\documentclass[journal]{IEEEtran}
\usepackage{wrapfig}

\usepackage{longtable} 
\usepackage{soul}
\usepackage{xcolor}
\definecolor{pinkhl}{RGB}{255, 182, 193}
\definecolor{cyanhl}{RGB}{0, 255, 255}
\definecolor{greenhl}{RGB}{204, 255, 0}
\newcommand{\hlpink}[1]{{\sethlcolor{pinkhl}\hl{#1}}} 
\newcommand{\hlcyan}[1]{{\sethlcolor{cyanhl}\hl{#1}}} 
\newcommand{\hlyellow}[1]{\hl{#1}} 
\newcommand{\hlgreen}[1]{{\sethlcolor{greenhl}\hl{#1}}} 
\renewcommand{\hlyellow}[1]{{\sethlcolor{white}\hl{#1}}}
\renewcommand{\hlgreen}[1]{{\sethlcolor{white}\hl{#1}}}
\renewcommand{\hlcyan}[1]{{\sethlcolor{white}\hl{#1}}}
\renewcommand{\hlpink}[1]{{\sethlcolor{white}\hl{#1}}}
\renewcommand{\hl}[1]{#1}

\usepackage{xcolor,soul,framed} 
\usepackage[table]{xcolor}
\usepackage{colortbl}
\colorlet{shadecolor}{yellow}
\usepackage[pdftex]{graphicx}
\graphicspath{{../pdf/}{../jpeg/}}
\DeclareGraphicsExtensions{.pdf,.jpeg,.png}
\usepackage[utf8]{inputenc}
\usepackage[english]{babel}
\usepackage[T1]{fontenc}
\usepackage[cmex10]{amsmath}
\usepackage{amssymb}
\usepackage{array}
\usepackage{mdwmath}
\usepackage{mdwtab}
\usepackage{eqparbox}
\usepackage{url} 
\usepackage{booktabs}
\usepackage{tabularx}

\begin{document}
\bstctlcite{IEEEexample:BSTcontrol}
    \title{\hlyellow{Active Electronic Terahertz Imaging for Industrial} \hlyellow{Applications: From Hardware to the Paradigm Shift} \hlyellow{by Artificial Intelligence}}

  \author {Aparajita~Bandyopadhyay, Hui~Yuan,
      Willie~J.~Padilla,~\IEEEmembership{Fellow,~IEEE,}\\
      David~J.~Brady,~\IEEEmembership{Fellow,~IEEE,}
      and~Hartmut~G.~Roskos

  \thanks 
  {A. Bandyopadhyay, H. Yuan and H. G. Roskos are with Goethe University Frankfurt, Germany (e-mail: Bandyopadhyay@physik.uni-frankfurt.de, also loveofconcept@gmail.com; yuan@physik.uni-frankfurt.de; roskos@physik.uni-frankfurt.de).}
  \thanks{W. J. Padilla is with Duke University, Durham, USA (e-mail: willie.padilla@duke.edu).}
  \thanks{D. J. Brady is with The University of Arizona, Tucson, USA (e-mail: djbrady@arizona.edu).}
  \thanks{This work has been submitted to the IEEE for possible publication. Copyright may be transferred without notice, after which this version may no longer be accessible.} }

\maketitle

\begin{abstract}
\hlyellow{Imaging with terahertz (THz) radiation (0.3--10~THz) benefits from a unique combination of attributes: penetration through dry, non-polar packaging materials; variations of dielectric functions to provide contrast between different materials; the existence of spectral fingerprint resonances for some classes of materials; non-ionizing photon energies that are safe for use around humans; and -- viewed from the low-frequency side -- an extension of the capabilities of established microwave radar to higher frequencies and thus to substantially better spatial resolution, at wavelengths which still permit direct measurement of the complex-valued radiation field.} \hlyellow{This review concentrates on active THz imaging with electronic sources combined with power detectors or coherent receivers -- the system class most likely to deliver fast (ideally real-time), cost-effective and deployable solutions for a wide range of industrial applications such as quality control, non-destructive testing, security screening and ranging for situational awareness. Such systems should be deployable on robotic and emerging autonomous platforms, for robotic machine vision and automotive sensing. Passive radiometric, astronomical and opto-electronic continuous-wave and broadband spectroscopic (time-domain) imaging are treated only where they inform this focus.} \hlyellow{We review the state of the art of compact semiconductor detector arrays, of imaging modalities ranging from focused-beam raster and frequency-modulated continuous-wave (FMCW) architectures to coherent Fourier-plane acquisition, and of augmentation techniques including near-field methods, compressive sensing (CS) and metasurface-based illumination control. Particular attention is paid to the rapidly growing role of Artificial Intelligence (AI): from Convolutional Neural Networks (CNN) and physics-informed deep learning for phase retrieval and image reconstruction, to Large-Language-Model (LLM)-driven agentic frameworks for autonomous system design.} \hlyellow{The long-standing bottlenecks of THz imaging -- acquisition speed, resolution, contrast and cost -- are re-examined in the light of these innovations, and a reference-anchored technology roadmap is derived which projects an order-of-magnitude reduction in the measurement requirements of THz imaging through end-to-end co-optimization of programmable THz hardware with physics-informed, heterogeneous agentic AI, further augmented by quantum-classical hybrid computing and edge inference. Converging with strategies from the infrared (IR) and  visible (VIS) imaging frontiers, active THz imaging is poised for decisive deployment across industrial disciplines within the next decade.} \\
\hlyellow{A separate list of all technical abbreviations used in this review is provided at the end of the article for quick reference.}
\end{abstract}

\begin{IEEEkeywords}
\hlyellow{Active THz imaging, industrial applications,} compact semiconductor detector arrays, THz imaging modalities, THz Fourier imaging, THz metasurfaces, compressive sensing, \hlyellow{physics-informed deep learning,} LLM-driven agentic AI, technology roadmap.
\end{IEEEkeywords}

\IEEEpeerreviewmaketitle


\section{Introduction}

\IEEEPARstart{T}{he} \hlgreen{THz frequency range, sandwiched between the microwave and IR regions of the electromagnetic spectrum, has for many decades been an important resource of scientific research. The technological exploitation of THz radiation for industrial and consumer applications, on the other hand, has long lagged behind. While this is not uncommon for any new technology, there have been specific hindrances for the THz spectral range. It being situated between microwaves and optics with its radical differences in technologies is already a clear indication that -- while borrowing as much as possible from either spectral range -- something new had to be invented to arrive at a reliable, cheap, manufacturable, standards-compliant and  system-level-robust device basis capable to provide the required performance for the desired practical applications. Possible application areas have long been identified, but they also meet physical constraints which have grounded many dreams: For THz communications, atmospheric absorption and propagation effects complicate real-world links; for inspection and diagnostic purposes, the limited or non-existent transmission of materials such as water, biological tissue, carbon-fiber-reinforced polymers and metals reduces the range of applications; for the inspection of small features with THz radiation, the wavelength imposes spatial-resolution constraints. And often, there are already solutions from other technologies for the task to be addressed, such that a THz alternative does not provide enough value at a competitive price to warrant the investment. Despite these restrictions, THz radiation has gradually made inroads into specialized applications, the most prominent ones being its being used for non-destructive testing (NDT) of car paints and pipe coatings and its consideration for ultrahigh-bandwidth communication. This success and the rapid development of a solid technological foundation give hope that this is just the beginning of a much wider exploitation of THz radiation in the coming years \mbox{\cite{Sun2025}}. An indication for the increasing maturity of the field is the abundance of existing review articles and books which cover the diverse aspects of recent developments. Here is a -- certainly incomplete -- list of important publications:}  \cite{Ferguson2002,Siegel2002,Tonouchi2007,Chan2007,Thomson2007,Lee2009,Jepsen2011,Zouaghi2013,Saeedkia2013,Carpintero2015,Song2015,Yang2016,Dhillon2017,Mittleman2018,Farrah2019,Naftaly2019,Tao2020,Ellrich2020,Saha2020,Valusis2021,Malhorta2021,Pavlidis2021,Akyildiz2022,Taday2022,Das2022,Sri2022,You2022,Ghzaoui2023,Leitenstorfer2023,Koch2023,LiX2023,Anitha2023,Jiang2024,Preu2025,Liu2025,Nongkseh2025,Zhang2025,Zogra2026}. 

\hlgreen{The present review concentrates on the status of \textit{imaging} with THz radiation and on an exploration of how the applied technologies may develop with regard to hardware and imaging concepts in light of the tempestuous progress of AI which revolutionizes photography and video imaging in the VIS spectral range and has just begun to have its impact on THz imaging. With a view towards potential larger-volume applications of THz imaging, we have restricted the scope of this publication further to active imaging with electronic radiation sources and power detectors, respectively receivers for coherent imaging. To put our publication into perspective with existing reviews of THz imaging,}
Table~\ref{table:review_positioning} \hlyellow{situates the present review within the landscape of those review articles, comparing the ambit of each prior review with what it offers to the reader, and thereby delineating the topical areas addressed here. }\\

\begin{table}[!t]
\centering
\scriptsize
\caption{\hlyellow{Positioning of the present review among published THz imaging reviews cited in this article}}
\label{table:review_positioning}
\renewcommand{\arraystretch}{1.3}
\begin{tabularx}{\columnwidth}{|>{\raggedright\arraybackslash}p{1.45cm}|>{\raggedright\arraybackslash}X|>{\raggedright\arraybackslash}X|}
\hline
\textbf{Review (year)} & \textbf{Ambit and scope} & \textbf{What it offers the reader} \\ \hline
Chan et al. \cite{Chan2007} (2007) & \hlyellow{Field-wide survey, pulsed/TDS-centric} & \hlyellow{Foundational account of THz imaging principles, early modalities and applications} \\ \hline
Jepsen et al. \cite{Jepsen2011} (2011) & \hlyellow{THz spectroscopy and imaging} & \hlyellow{Broadband time-domain techniques with materials and spectroscopic emphasis} \\ \hline
Rogalski and Sizov \cite{Rogalski2011} (2011) & \hlyellow{THz detectors and focal-plane arrays} & \hlyellow{Component-level review of detector physics, sensitivity limits and FPA technologies} \\ \hline
Mittleman \cite{Mittleman2018} (2018) & \hlyellow{Field-wide, historical} & \hlyellow{Twenty-year perspective on the evolution of THz imaging research} \\ \hline
Farrah et al. \cite{Farrah2019} (2019) & \hlyellow{Radioastronomy, earth observation,  instrumentation} & \hlyellow{A review of missions and instrumentation, and technology development trends} \\ \hline
Naftaly et al. \cite{Naftaly2019} (2019) & \hlyellow{Industrial applications} & \hlyellow{Focus on TDS for a broad range of applications} \\ \hline
Ellrich et al. \cite{Ellrich2020} (2020) & \hlyellow{Industrial applications} & \hlyellow{Automotive and aviation industries} \\ \hline
Tao et al. \cite{Tao2020} (2020) & \hlyellow{Industrial applications of optoelectronic systems} & \hlyellow{Non-destructive testing of materials} \\ \hline
Valu\v{s}is et al. \cite{Valusis2021} (2021) & \hlyellow{Field-wide} & \hlyellow{Components and imaging modalities. Challenges and trends discussed for each technology} \\ 
\hline
Bandyopadhyay and Sengupta \cite{Bandyopadhyay2022} (2022) & \hlyellow{THz spectral imaging} & \hlyellow{Concept, applications and implementation issues of spectral-imaging systems} \\ \hline
Taday et al. \cite{Taday2022} (2022) & \hlyellow{Applications in industry and medicine} & \hlyellow{TDS applications} \\ \hline
Li et al. \cite{LiX2023} (2023) & \hlyellow{High-throughput THz imaging} & \hlyellow{Hardware and computational-imaging routes to imaging speed; comparison of sensor-array technologies} \\ \hline
Anitha et al. \cite{Anitha2023} (2023) & \hlyellow{Technology trends and applications} & \hlyellow{Broad application-wise survey across usage domains} \\ \hline
Nongkseh et al. \cite{Nongkseh2025} (2025) & \hlyellow{Security screening} & \hlyellow{Comprehensive review of concealed-object detection from THz images} \\ \hline
Liu et al. \cite{Liu2025} (2025) & \hlyellow{Non-destructive testing and imaging} & \hlyellow{Use of TDS for thermal barrier coatings} \\ \hline
Zografopoulos et al. \cite{Zogra2026} (2026) & \hlyellow{Imaging techniques and applications} & \hlyellow{TDS systems} \\ \hline
Tian et al. \cite{TianReview2026} (2026) & \hlyellow{Field-wide, modality-organized survey} & \hlyellow{Comprehensive coverage from CW holography, ptychography and tomography to near-field nanoscopy, with an application panorama} \\ \hline
\textbf{This work} (2026) & \hlyellow{Active electronic imaging for industrial deployment, paradigm change by AI} & \hlyellow{Hardware--computation co-design thread from detector arrays and modalities to physics-informed and agentic AI, culminating in a reference-anchored technology roadmap} \\ \hline
\end{tabularx}
\end{table}

\hlgreen{Before addressing technical aspects of THz imaging, we want to comment on a choice: to define more closely the frequency range which we consider in the following. The literature defines two lower bounds for the THz frequency regime. Coming from the classical distinction of millimeter and sub-millimeter waves, the demarcation frequency is 0.3~THz which corresponds to vacuum radiation with a wavelength of 1~mm. Motivated by technological challenges of implementing electronic THz systems working above 0.1~THz, the boundary in the past has often been chosen to be at 0.1~THz. 
In this review, we adopt 0.3--10~THz as the THz frequency range per se, and treat 0.1--0.3~THz as a transition zone. This 'zoning scheme' becomes inevitable as the techniques and system concepts inherited from millimeter-wave radar are discussed where they inform THz imaging (as in Sec.~III-A and Fig.~3), while the forward-looking scope of this review -- and of the technology roadmap of Sec.~VIII -- begins at 0.3~THz. This choice is not of mere taxonomic value --- the transition zone, especially, 100--275~GHz is indeed a contested spectrum as per the ITU Radio Regulations which strictly allocate the band densely among incumbent services, with numerous sub-bands reserved for radio astronomy and for earth-exploration and space-research (passive) sensing. In fact, in several sub--bands, e.g., 200--209~GHz and 226--231.5~GHz, all active emissions are \textit{prohibited outright}} \cite{ITU_RR_2024}. \hlgreen{ 
Competitors for frequency bands are radiolocation and radio-navigation services, which are of strategic and defense significance, and land-mobile and fixed communication services in support of future wireless systems, for which WRC-19 additionally identified the 275--296, 306--313, 318--333 and 356--450~GHz ranges} (No. 5.564A of \cite{ITU_RR_2024}). \hlgreen{This is a regulatory certainty, and active imaging efforts concentrated in the transition zone would, therefore, have to either ensure electromagnetically shielded operation or secure spectrum in competition with established scientific, communication and strategic domains in the near future. 
Above 0.3~THz, by contrast, and particularly within the atmospheric windows discussed in Sec.~VII, spectrum access remains comparatively open.} \\

\hlyellow{As stated before and in} Table~\ref{table:review_positioning}, \hlyellow{the focus of this review is on \textit{active} THz imaging with \textit{electronic power detectors and coherent receivers}. }   
\hlyellow{In order to explain the first of these choices, that for active systems, we begin with a closer look at passive imaging. Passive systems (radiometers) record the thermal radiation naturally emitted or reflected by a scene. They are intrinsically covert and safe, and often employ polarization for the distinction of object features. At microwave frequencies, radiometers -- ground-based, air or spaceborne -- are widely used for astronomy, meteorology, earth observation, military surveillance and reconnaissance, but also for thermography, e.g., in medical diagnostics and non-destructive testing \mbox{\cite{Ulaby, Karmakar, Peichl2007, Ring2010, Vittucci2026}}. If temperature contrasts are small, e.g. only a few Kelvin, or if only minute amounts of radiation arrive from far distances as in astronomy, extreme demands are placed on detector sensitivity (noise-equivalent power, NEP). There are two established ways to reach the required sensitivity levels: resorting to cryogenically cooled superconducting detector technologies, and/or heterodyne detection. Radiometer systems with both superconducting and semiconducting sensors are also considered for security screening, especially stand-off person scanning \mbox{\cite{Appleby1999, Appleby2002, Appleby2007, Heinz2010, Rowe2016, LiLiLi2019, Luomahaara2021, Hadinejad2025}}. 
However, active imagers currently appear to be favored in security applications, probably due to an intrinsically better spatial resolution \mbox{\cite{Ahmed2021}}.}

\hlyellow{In this review, which deals with imaging at THz frequencies, we will touch radiometers only occasionally because they are -- with the main exception of the monitoring of specific gas species in atmospheric sensing and radioastronomy \mbox{\cite{Farrah2019}} -- not quite as broadly applied at frequencies above 0.3~THz as they are at microwaves. This may be attributed to two reasons: (i) The penetration depth of the radiation in the atmosphere, but also in human tissue and through fabrics, decreases with frequency, which reduces the benefit of using the THz part of the electromagnetic spectrum, except when spectral fingerprints of atomic and molecular species outweigh this deficit. (ii) Radiometers often amplify the incoming radiation before heterodyne detection \mbox{\cite{Wang2021}}, but the gain of amplifiers decreases with rising frequency which reduces the systems' sensitivity. At the present time, passive THz imaging continues to be explored (see, e.g., investigations with power detectors \mbox{\cite{Andree2024, Cibiraite2020}} and electrooptic upconversion \mbox{\cite{Santamaria2020}} at room temperature), which may lead to future industrial applications, but currently, it doesn't seem to have found as much attention for such applications as active imaging.} \\

\hlyellow{Active systems, in contrast, control the illumination and therefore the image contrast; they operate with orders-of-magnitude higher signal levels, relax the detector NEP requirements to values attainable with room-temperature semiconductor devices, and give access to the phase of the object wave via coherent detection. These properties align with the requirements of industrial inspection, non-destructive testing (NDT), security portals and automotive sensing. The remainder of this review therefore has its focus on active THz imaging.}\\

\hlgreen{The second of our choices, that is, the focus on electronic imaging systems, originates from the promise which such systems hold for fast (ideally real-time) operation, ease of deployment and cost-effectiveness - all of these considerations make them well suited for many industrial and consumer markets. Broadband spectroscopic (time-domain) imaging 
which has found industrial applications in non-destructive testing is treated only where it directly impacts this focus, being extensively covered by dedicated reviews} \cite{Chan2007, Jepsen2011, Mittleman2018, Naftaly2019,Tao2020,Ellrich2020,Valusis2021,Taday2022,Liu2025,Zogra2026} \hlgreen{and by companion invited papers of this Special Issue.} \\

\begin{figure*}[t]
\centering
\includegraphics[width=6in]{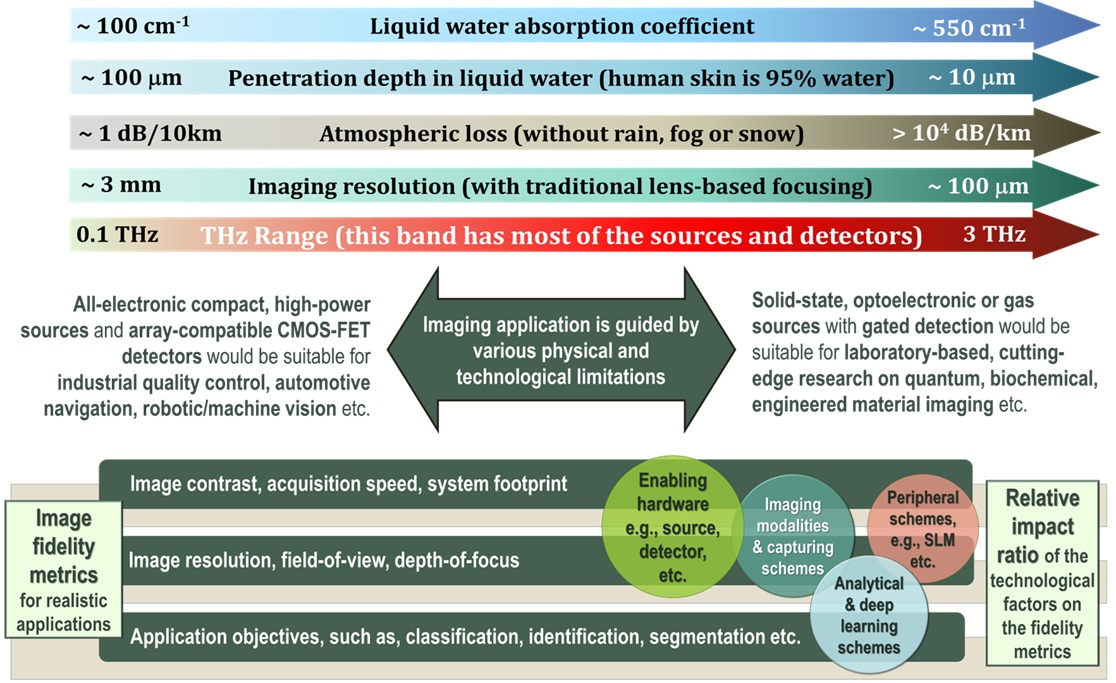}
    \caption{The physical and technological factors of the THz range determine THz imaging modality in a specific application domain. The image fidelity metrics, required for successful implementation of THz range in an application domain, are further impacted by various enabling factors, such as, hardware, combinatorial components and/or technique enhancement, as well as, computational framework. \hlgreen{The frequency span indicated in the top banner (0.1--3~THz) includes the 0.1--0.3~THz transition zone defined in Sec.~I.}\hlyellow{ Original artwork by the authors.}}
    \label{fig 1_Intro}
\end{figure*}

Fig.~\ref{fig 1_Intro} highlights key physical factors 
which determine the potential and \hlpink{limitations} of specific application domains of practical imaging modalities. A critical factor is the application-specific threshold for image fidelity, determined by key metrics such as contrast, resolution, acquisition speed, field of view, and depth of focus. These parameters ultimately dictate whether the required information can be extracted from the images. The imaging systems can be optimized with regard to these parameters\hlpink{; however, only} within the range allowed by the well-known fundamental physical constraints such as the absorption, reflection and scattering properties of the objects to be imaged. A central role is played by the wavelength of the radiation -- its large value ($\lambda$\;=\;1~mm at 300~GHz) introduces severe limitations for free-space THz imaging in comparison with VIS and IR radiation. Despite efforts with regard to algorithmic or optics-based \textit{free-space} super-resolution imaging \cite{Anitha2023}, the large THz  wavelength is responsible for pronounced diffraction and artifacts. It also restricts the pixel density of practical detector arrays resulting in limited spatial resolution. Combined with the still-limited availability of high-efficiency sources, detectors and modulation devices, this technique continues to suffer from limited contrast and lengthy acquisition times -- ultimately restricting its real-world applications \cite{Jepsen2011, Valusis2021}. On the other hand, diffraction can be a beneficial feature of THz imaging, as it encodes phase information in interference patterns, which, unlike in the VIS and NIR regimes, are conceptually straightforward \hlpink{to record and to exploit} for the extraction of phase information of the object wave needed for 2D and 3D computational image reconstruction: this aspect will be discussed in detail, in the context of Fourier Imaging. \\

\hlyellow{To appreciate the essential elements of THz imaging technology -- its evolution and its state of the art -- we adopt a simple organizing scheme that runs through this review: each imaging \textit{modality} is assessed together with the \textit{implementable hardware modules} that realize it and the \textit{computational model} that turns the recorded data into an image.} \hlpink{Within this scheme, we place specific emphasis on coherent Fourier imaging as a crucial modality with realistic application potential for 3D THz imagery in several emerging industrial domains that are also targeted by the IR and GHz spectral ranges} \cite{Bandyopadhyay2024}\hlpink{, such as self-driving automotive systems, robotics and situation-aware industrial quality control.} An expanded discussion is included on compact, semiconductor-based, array-compatible THz detectors, specifically those based on transistors such as \textit{TeraFETs} \cite{Oejefors2009,Lisauskas2009,Boppel2012,But2023a}; and the impact of the combinatorial execution of advanced computational image processing techniques\hlpink{, such as CS imaging} \cite{Valles2020} \hlpink{and SLMs} \cite{Chan2009}\hlpink{, combined with diverse DL schemes} \cite{LiX2023} \hlpink{based on AI and/or Machine Learning (ML) techniques -- from CNNs to multi-agentic frameworks} \cite{Wiecha2021} \hlpink{performing autonomous feature-space optimization based on LLMs.} \\

As described in Fig. ~\ref{fig 1_Intro}, these technological factors, starting from enabling hardware to imaging components and peripherals to computational framework; have perceptible and measurable impact in the resultant THz image fidelity required for a specific application field. \hlpink{Through the convergence of improved sensitivity, reduced acquisition times and enhanced analytical capabilities achieved over the years, THz imaging has emerged as an indispensable tool for twenty-first-century science and technology;} and Table~\ref{table:First report} highlights some of the first reports of experimentally achieved milestones in THz imaging over the years.\\

\begin{table*}[t]
    \centering
    \caption{First report of various experimental THz Imaging Schemes with specifications}
    \scriptsize
    \begin{tabular}{|>{\centering\arraybackslash}m{2.3cm}|>{\centering\arraybackslash}m{3.5cm}|>{\centering\arraybackslash}m{2.2cm}|>{\centering\arraybackslash}m{0.9cm}|>{\centering\arraybackslash}m{1.4cm}|>{\centering\arraybackslash}m{1.2cm}|>{\centering\arraybackslash}m{1.2cm}|>{\centering\arraybackslash}m{1.1cm}|}
        \hline
        \textbf{Imaging Techniques} & \textbf{Source Type} & \textbf{Detector Type} & \textbf{Freq} & \textbf{Spatial Resolution} & \textbf{Dynamic Range} & \textbf{Acquisition Time} & \textbf{Year [Ref]} \\
        \hline
        Far-Infrared/Sub-Millimeter Wave CW Imaging & HCN laser at 337 $\mu$m (10 mW) and an FIR waveguide laser at 554 $\mu$m (0.5 mW), 890 $\mu$m (0.2 mW), and 1020 $\mu$m (0.1 mW) & Liquid-He-cooled GaAs detector & 300\,GHz--1\,THz & Low resolution (mm-cm) & Not specified & Raster scan & 1976 \cite{Hartwick:76}\\
        \hline
        THz Time-Domain Imaging (TDI) & Photoconductive antenna (PCA) driven by femtosecond (fs) laser & Photoconductive receiver & 0.1--3\,THz & Sub mm & Not specified & $> 60$\,min & 1995 \cite{Hu:95} \\
        \hline
        Near-Field THz imaging & PCA driven by fs laser & Photoconductive receiver & 0.1--2\,THz & $\lambda$/4 & Not specified & $\sim15$\,min & 1998 \cite{Hunsche1998} \\
        \hline
        THz CW imaging (optoelectronic photomixing) & PCA driven by beating of two-color Ti:Sapphire lasers & LT-GaAs Photomixers & 1\,THz & $\sim 250\,\mu$m & SNR $> 10$ & 200 ms per pixel & 2002 \cite{10.1063/1.1469679} \\
        \hline
        THz SNOM & PCA driven by fs laser & Electro-optic sampling & 0.1--2.5\,THz & $\sim 150\,$nm (or $\sim\lambda/1000$) & Not specified & Less than 10 mins & 2003 \cite{10.1063/1.1616668} \\
        \hline
        Compressive Sensing THz imaging & Broadband PCA & Broadband PCA & Not specified & Not specified & Not specified & Not specified & 2008 \cite{Chan2008}\\
        \hline
        THz FMCW full-field radar imaging & Schottky diode based frequency-multiplier with 30 GHz bandwidth & Schottky diode mixer & 0.662--0.691\,THz (swept) & Lateral 1 cm, depth res 5.2 mm & Absolute SNR $\sim 65$\,dB & Frame rate of 1 Hz/pixel & 2011 \cite{Cooper2011} \\
        \hline
        THz imaging with Spatial Light Modulator (SLM) & Hg-Arc lamp with long-pass filter blocking $< 65\,\mu$m (or above 4.6 THz) & Liquid-helium cooled silicon bolometer & 0.2--4.6\,THz & 1.5 mm to $328\,\mu$m & Not specified & 31.5 to 511.5\,s (depends on total pixels) & 2013 \cite{Shrekenhamer2013} \\
        \hline
        Ultrafast THz scanning tunneling microscope imaging & Large-area aperture PCA driven Ti:Sapphire regenerative amplifier & Electro-optic detection with ZnTe crystal & 2.59\,THz & $\sim2\,$nm & Not specified & $\sim 15$\,min & 2013 \cite{Cocker2013} \\
        \hline
        THz QCL feedback interferometric imaging & THz-QCL with 50\% duty-cycle as emitter & THz-QCL with 50\% duty-cycle as receiver with self-mixing & 2.59\,THz & $\sim 250\,\mu$m & Not specified & 19\,min & 2013 \cite{Rakic2013} \\
        \hline
        THz digital holographic imaging & CO\textsubscript{2} laser with a wavelength of 118.83\,$\mu$m & Pyroelectric detector array & 2.52\,THz & $\sim 158\,\mu$m & Not specified & $< 1$\,min & 2015 \cite{Rong2015} \\
        \hline
        THz computational single-pixel imaging & Optical rectification of ZnTe crystal & Electro-optic detection with ZnTe & 0.75\,THz & $\sim 100\,\mu$m & Not specified & $500$\,ms & 2016 \cite{Stantchev2016} \\
        \hline
        THz Ptychographic imaging & CO\textsubscript{2}-laser-pumped CW far-infrared gas laser & Uncooled microbolometer array (480$\times$640 pixels) & 3.1\,THz & Lateral $\sim 300\,\mu$m and depth res $\sim 5\,\mu$m & Not specified & Not specified & 2018 \cite{Valzania2018} \\
        \hline
        THz 3D Reconstructed Fourier Imaging with TeraFET detector & Frequency multiplier chain & Antenna-coupled CMOS-TeraFET (heterodyning) & 0.3\,THz & Lateral resolution approx 2 mm & $> -60$\,dB & Several mins with raster scan & 2019 \cite{Yuan2019} \\
        \hline
    \end{tabular}
    \label{table:First report}
\end{table*}

In more recent times, the integration of advanced computational methods \cite{Hunt2013,Zhu2017}, particularly deep learning and AI-driven approaches \cite{Lin2018,Su2023}, promises to address historical limitations in THz image acquisition speed and data processing. As hardware becomes more accessible through Complementary Metal-Oxide-Semiconductor (CMOS) integration and manufacturing advances, and as algorithms become more sophisticated, THz imaging is poised to transition from specialized research applications to mainstream industrial, clinical and other emerging technology use, such as, unmanned automotive, robotics etc. This technical review on THz imaging, therefore, is not only an account of the present state-of-the-art and the challenges that the frontier of \hlpink{this} field is facing, but also an exploratory outlook based on the emerging trends: not only in the THz range, but also being developed in the neighboring spectral range of IR and GHz. Driven by the acute \hlpink{demands} of several application domains which require fusion of sensing technologies, such as healthcare \cite{Gezimati_healthcare_2023}, security \cite{Hadinejad2025}, multi-spectral machine vision \cite{Tokizane2021,Hung2024}, augmented reality  \cite{Guillet2024}, situation-aware decision analytic platform \cite{Fan2024} etc.; it is expected that THz imaging will \hlpink{take a major} development leap in the near future. Propelled by increasing adaptation of AI/ML and computational frameworks and innovative image augmentation schemes, this field of THz imaging is also likely to benefit from inspiration from \hlpink{the} much more advanced imaging in the IR and VIS spectral ranges. Two guiding principles are essential for THz imaging: first, leveraging AI/ML to shift complexity from hardware to software; second -- and equally vital -- ensuring the highest possible quality of physically recorded data, as these form the basis for all downstream AI/ML processing.  \hlyellow{The extended sections at the end of this review -- on outlook and perspectives, and on a reference-anchored technology roadmap -- are meant to provide a discourse and directional markers along these two guiding principles.}

\section{Present State-of-the-Art of THz Detectors}

\subsection {THz detectors (single pixel)}

THz technology has many types of detectors to offer for imaging purposes, and to cover them all is beyond the scope of this article. For an overview, we can refer to excellent review articles \cite{Rogalski2011, Sizov2018}. We limit ourselves here to a few types of detectors which we deem to be of broad relevance for cost-effective and high-speed active industrial imaging. If broadband spectral information is needed, then the imaging systems of choice will be (raster-scanning) optoelectronic ones, either operated in the time domain \cite{Jepsen2011} or the frequency domain \cite{Schwenson2025}, and the detectors will be photomixers providing time-of-flight information, respectively access to both amplitude and phase of the signal for a derivation of the complex-valued dielectric function, already covered extensively in several seminal reviews \cite{Mittleman2018,Valusis2021,Preu2025}. The detectors of choice considered here for imaging have to fulfill, most importantly, the requirement of a high sensitivity, as more often than not one will encounter situations with low radiation power arriving from the scene \cite{Spiegel2010, Petkie2008}. \\

The sensitivity is best specified by the \textit{optical} NEP, i.e., the radiation power yielding unity signal-to-noise ratio (SNR) for a 1-Hz detection bandwidth. The specifier  \textit{optical} denotes that the power is measured in front of the detector, unlike the \textit{electrical} or \textit{intrinsic} NEP, which relates to the power estimated to reach the active region of the detector \cite{Bauer2019, Javadi2021}. The comparison of published NEP values is often made difficult by the lack of information about how the values were derived. NEP values can be specified for \textit{direct} detection (units pW/$\sqrt{\mathrm{Hz}}$), measuring the radiation power, as well as for \textit{coherent} (hetero- or homodyne) detection (units pW/Hz or pW/$\sqrt{\mathrm{Hz}}$ \cite{Sizov2010}), measuring the electric radiation field by wave-mixing processes in receivers. If sensitivity were the only condition, one would preferentially work with cryogenically cooled detectors. Semiconductor power-detecting devices reach optical NEP values around 1~pW/$\sqrt{\mathrm{Hz}}$ \hlcyan{when cooled towards the boiling temperature of liquid He (e.g., photoconductive detectors based on Ge:Ga at 4.2~K, measured with broadband radiation around 3~THz with typical laboratory background radiation \mbox{\cite{GeGa-QMC}}, devices based on the photoelectric tunable-step (PETS) effect at 9~K and 1.9~THz \mbox{\cite{Michailow2022, Chen-Michailow2024}}, or Si CMOS TeraFETs at 20~K and 540-580~GHz \mbox{\cite{Holstein2025a, Holstein2026, Klimenko2012}}).
These values pale in comparison with 
the performance reached with superconductor devices, e.g. transition-edge bolometers and kinetic inductance detectors (KIDs), which exhibit NEP values in the range of 10$^{-18}$-10$^{-19}$~W/$\sqrt{\mathrm{Hz}}$ \mbox{\cite{Irwin1995, Suzuki2014, Day2003, DeVisser2014}}. 
At deep sub-1~K, even single-GHz/THz-photon-detection sensitivity has been reached with devices based on superconductors \mbox{\cite{Kokkoniemi2017, Echternach2018, PhysRevApplied.20.014003, Paolucci2021, Pankratov2025}}, semiconductor quantum-dots \mbox{\cite{Hashiba2006, Komiyama2011} and nanobolometers \cite{Karasik2007}}. }

Depending on the application, additional crucial aspects of relevance are total costs, robustness, size and operational complexity. These together usually enforce room-temperature operation, and have led to a trend to work with on-chip solutions instead of hollow-wave-guide-coupled ones because of the easier packaging. Given these constraints, the three types of detector devices, which appear to dominate present developments, are bolometers, transistor-based devices, and diode ones. The latter two work (mainly) on rectifying principles \cite{Sakhno2013, 
Andree2022, ludwig2024model}, the former on thermal effects \cite{Mather1982}.

A common feature is that all three lend themselves for the realization of multi-pixel arrays (MPAs) discussed in the following Sec.~\ref{MPAs}.  
Differences arise with regard to the frequency response. While all three can principally cover the entire frequency range from microwaves to 10~THz \cite{Bauer2014, Ahmad2019, Jang2019, Yadav2023, Zhu2023, Yadav2024, Ludwig_diss2024, Hirakawa2025, Holstein2026a}, with some restrictions especially at high frequencies in the reststrahlen band dependent on the specific material system used, transistors and diodes need to be embedded in antennas for the desired frequency band, whereas bolometers can be designed either with antenna or antenna-free with a large-area absorber. They also differ with regard to their applicability as receivers in heterodyne detection. Transistor and diode receivers, due to their high intrinsic speed, can operate at high intermediate frequency (IF) up to the GHz range \cite{Glaab2010, Boppel:12, Lisauskas2013mix, Grzyb2015, Lin2020, Feng2022, Ding2023}, whereas bolometers with their slower, thermal-diffusion-dominated response are restricted to much lower IF \cite{Huhn2013}. 

The most common diode-type detectors are based on Schottky diodes \hlpink{\mbox{\cite{Liu2010, Hoefle2014, Mehdi2017, Ito2017}}}. Very high performance is also achieved \hlpink{with heterostructure backward diodes \mbox{\cite{Rahman2018, Shi2022}}}, and with double- and triple-barrier resonant tunnel diodes (RTDs) \cite{Takida2020, Clochiatti2025}. In the case of transistor-type detectors, both field-effect transistors (TeraFETs) \cite{Knap2004, Oejefors2009, Lisauskas2009} and heterobipolar transistors (HBTs) \cite{Hadi2013} are broadly used. The former are commonly operated at zero gate bias voltage which ensures a low noise floor determined by Johnson-Nyquist thermal noise of the channel \cite{Tauk2006, Lisauskas2013}). HBTs, on the other hand, are current-biased and exhibit 1/f-noise which is, however, of minor significance when the radiation is chopped at 1-10~kHz \cite{Andree2024}. 

Diode-type detectors are commonly fabricated from III-V semiconductors. Transistor-type detectors, on the other hand, are implemented in both III-V technology (HEMTs) \cite{Bauer2019, Zhu2023, Yadav2024, Fatimy2006, Boppel2016, Hou2017, Qin_GaN2017, Sun2019} and Si technology (CMOS and BiCMOS) \cite{Boppel2012, hillger_terahertz_2019, Ikamas2018}. All of these are fabricated in common semiconductor foundries, no extra technology is required. This can also be the case with bolometers \cite{Corcos2015, Grant2013}, but usually additional fabrication steps (e.g., MEMS fabrication or deposition of unconventional absorber layers) are required  \cite{gou2017spiral, Huhn2013, Simoens2013,  Hirakawa2023}; however, when built on a Si platform, they are still compatible with CMOS integration \cite{Lisauskas2013}. 

Regarding the sensitivity of the detectors in the frequency range 0.3-0.8~THz, best values of the optical NEP lie in the 5-30-pW/$\sqrt{\mathrm{Hz}}$ window for all of them. Recent collections of NEP data including comparisons of frequency trends are found in the following papers \cite{Bauer2019, hillger_terahertz_2019,  Ferreras2021, ludwig2024terahertz,
Holstein2026, Holstein2026a}. The NEP values of the rectifying devices universally increase with frequency. Ref.~\cite{Holstein2026a} finds that -- below roughly 1~THz -- diode-type detectors (typically embedded in hollow waveguides) appear to be slightly more sensitive than TeraFETs (typically with substrate-lens-coupling of the THz radiation). Above 1~THz, TeraFETs exhibit lower values of the optical NEP than Schottky diodes, because the TeraFETs exhibit a much weaker sensitivity roll-off with frequency, allowing for a patch-antenna-coupled TeraFET with superstrate radiation coupling to still achieve an optical NEP of 43~pW/$\sqrt{\mathrm{Hz}}$ at 2.45~THz. 

To conclude this section on single-pixel THz detector, we summarize the discussion with few final remarks. (i) A  field of current exploration is the exploitation of the nonlinear characteristics of radiation sources for the \textit{coherent} detection of back-coupled radiation by the sources themselves via the \textit{self mixing} effect. This saves the use of a separate detector. Coherent self-mixing has been applied for imaging in the far-field \cite{Dean2013, Wienold2016, Cai2025} and the near-field \cite{Reichel2021} using THz quantum cascade lasers (QCLs). The data acquisition speed is limited by raster-scanning. 
Self-mixing is also possible with transistors \cite{Hadi2017, But2023} and diodes \cite{Manh2018, Duelme2019}, which, while not much explored at present, is expected to have a promising  future. (ii) While the detector devices discussed above all have been fabricated from established crystalline material systems, we expect that the material basis will be expanded in the near future by new materials such as carbon nanotubes, graphene and related 2D van-der-Waals materials as well as topological materials \cite{Tarasov2007, Vicarelli2012, Zhu2013, Koppens2014, Zak2014, Qin2017, Suzuki2018,
Castilla2019, Viti2020, Kokkoniemi2020, Seifert2020, asgari_chip-scalable_2021, Viti2021, MIAO2023112, Jumaah2023, HE2024118886, Thomson2024, Delgado2024, ludwig2024terahertz, Xia2025, Wei2025}. 
Challenges related to material uniformity, process standardization, and pixel-to-pixel reproducibility remain to be solved.
(iii) Before considering detector arrays in the following Sec.~\ref{MPAs}, we want to point out gesture-based human-machine interaction as a possible application scenario for few-detector systems which may merit future investigation. Gesture recognition requires sufficient scattering of the radiation by skin (or gloves) to unambiguously identify a finger or a hand. At microwave frequencies, skin reflects mainly in a specular manner, making gesture recognition difficult. With increasing frequency, however, scattering becomes more significant, and thus gesture identification may become possible.   
(iv) Coming back to detector technology, it is an interesting and entirely open question whether THz detectors could be realized using printed materials. While printing is an established approach for the fabrication of DC electronic circuits, it is usually assumed that, for high-frequency signals, the losses are too high and the mobility of the charge carriers in the conductors too low. It may be worth to carefully re-examine the options. Today, printing is possible on different geometrical scales down to nanometers and many different materials including metals can be printed \cite{Ghosh2025}. Possibly, conceptually simple detector structures such as planar diodes could be of interest \cite{Xu-Song2008, Minkevicius2011, Sangare2013}. And at least the distributed resistive mixing effect of THz rectification in field-effect devices is not directly dependent on carrier mobility (albeit indirectly via impedance matching limitations of power coupling) \cite{Boppel2012}. If successful, a printing technology possibly would open a much less expensive way to build size-scalable MPAs.  

\begin{table*}[t]
    \centering
    \caption{State of the art of THz cameras:}
    \scriptsize
    \begin{tabular}{|>{\centering\arraybackslash}m{2.3cm}|>{\centering\arraybackslash}m{1.8cm}|>{\centering\arraybackslash}m{1.8cm}|>{\centering\arraybackslash}m{2cm}|>{\centering\arraybackslash}m{2cm}|>{\centering\arraybackslash}m{2cm}|>{\centering\arraybackslash}m{1.5cm}|>{\centering\arraybackslash}m{1.2cm}|}
    \hline
    \textbf{Camera} & \textbf{Pixel count} & \textbf{Pixel size} & \textbf{Sensing area} & \textbf{Technique} & \textbf{NEP} & \textbf{Frequency} & \textbf{Framerate}\\
    \hline
    \rowcolor{blue!30}
    i2S/TZ cam\cite{i2s} & 320 $\times$ 240 & 50\,$\mu$m & 16 $\times$ 12\,mm$^2$ & Micro-bolometer & $<$30\,pW @ 2.5\,THz & 0.4--3\,THz & 25\,fps\\
    \hline
    Swiss THz/RIGI\cite{swiss} & 80 $\times$ 80 -- 1920 $\times$ 1080 & $>$15\,$\mu$m & Max. 28.8 $\times$ 16.2\,mm$^2$ & Micro-bolometer & $<$1.5\,pW/Hz @ 4.6\,THz & $<$1\,THz -- 18\,THz & 10, 30, 60\,fps\\
    \hline
    INO/MICROXCAM-384I\cite{ino} & 384 $\times$ 288 & 35\,$\mu$m & 13.4 $\times$ 10\,mm$^2$ & Micro-bolometer & 0.32\,pW/$\sqrt{\text{Hz}}$ @ 0.4\,THz & 0.09--20\,THz & 50\,fps\\
    \hline
    TeraSense\cite{terascense} & 16 $\times$ 16, 32 $\times$ 32, 64 $\times$ 64 & $>$1500\,$\mu$m & 24 $\times$ 24, 48 $\times$ 48, 96 $\times$ 96\,mm$^2$ & GaAs FET & 1\,nW/$\sqrt{\text{Hz}}$ & 50--700\,GHz & 50\,fps\\
    \hline
    FBH\cite{FBH} & 12 $\times$ 12 & 380\,$\mu$m & 4.56 $\times$ 4.56\,mm$^2$ & AlGaN/GaN & 57\,pW/$\sqrt{\text{Hz}}$ @ 0.5\,THz & 0.1--1.5\,THz & 100\,fps\\
    \hline
    Ohio HELIOS\cite{ohio} & 80 $\times$ 64 & 100\,$\mu$m & 0.84 $\times$ 0.64\,mm$^2$ & Diode & -- & 0.6--1.2\,THz & --\\
    \hline
    Ticwave\cite{ticwave} & 32 $\times$ 32 & 1000\,$\mu$m & 3 $\times$ 3.2\,mm$^2$ & CMOS FET & 10\,pW/$\sqrt{\text{Hz}}$ @ 0.32\,THz & 0.3--1.1\,THz & 54\,fps\\
    \hline
    \end{tabular}
\label{table:THz camera}
\end{table*}

\subsection{THz multi-pixel arrays (MPAs)}
\label{MPAs}

From the perspective of practical THz imaging systems, the transition from single-pixel detectors to scalable MPAs represents a critical enabling step toward real-time, high-throughput, and application-relevant imaging. 
\hlcyan{With regard to sensitivity, the best-performing MPAs have been realized with superconducting detectors. Especially KIDs are well suited for MPA implementation and are employed as the sensor elements of astronomical observation instruments with up to 17.5~kpixels \mbox{\cite{Farrah2019, Meeker2018, Wilson2020, Reyes2026, Lourie2018}}. 
Superconducting MPA technology has also been applied to demonstrate passive body scanners for security applications \mbox{\cite{Heinz2010, Rowe2016, Luomahaara2021}}.} 

\hlcyan{It is to be expected that the need for cryogenic cooling will limit systems based on superconducting detector elements to high-value applications. For most use-cases, only room-temperature operation is considered.} The most mature and widely adopted THz MPAs of that kind are presently based on thermal and semiconductor electronic detection mechanisms. Thermal MPAs are dominated by microbolometer arrays \cite{Oda2010, fan2020broadband}, which typically operate in the 1--10~THz range, with pixel counts reaching up to 320$\times$240 and pixel pitches of 20--30~$\mu$m, as well as pyroelectric arrays \cite{5615383,hack2016comparison} that cover a broad THz spectral range with typical formats of 160$\times$160 pixels and pixel sizes of 80--100~$\mu$m. Owing to their wide use in the IR spectral range, these platforms benefit from relative technological maturity and commercial availability. Limitations of presently commercialized devices exist typically with regard to their sensitivity at frequencies below 1~THz, for which the sensors seemingly have not been optimized (cp., however, \cite{Oda2010}). Heterodyne operation is fundamentally constrained by the read-out bandwidth, which is limited by thermal response times. 

Electronic THz MPAs, based on field-effect transistors (FETs) and diodes, provide an alternative pathway that is inherently compatible with scalable integration and high-speed read-out. Such arrays have been realized using CMOS \cite{9365832, Zda15, Boukhayma2016, 
Yokoyama2019, Kanazawa2019, Hoogelander2023, Liu2024}, SiGe BiCMOS \cite{8310362,Hoogelander2026}, III--V compound semiconductor technologies (e.g., GaN, GaAs and InP) \cite{9567181, Trichopoulos2013, Luo2018}, as well as MEMS-based processes \cite{8538071}. These detector technologies can cover the sub-THz to THz frequency range and typically offer superior temporal response compared to thermal detectors, with reported array sizes on the order of 1-10$\times 10^3$ pixels (e.g., 32$\times$32 \cite{6363486} and 64$\times$64 \cite{Wang2026}) and pixel dimensions in the few-hundred-micrometer range. Importantly, their electronic nature makes them attractive for integration with on-chip signal processing and computational imaging architectures.

Table~\ref{table:THz camera} summarizes representative commercially available THz MPAs. A common trend across these platforms is the trade-off between pixel count, sensing area, and achievable sensitivity, which ultimately limits spatial resolution, field-of-view (FoV), and scalability. The long wavelength of THz radiation implies that achieving pixel counts comparable to VIS or IR cameras would require wafer-scale fabrication with stringent uniformity control, resulting in prohibitive cost and yield challenges. Moreover, the reliance on direct power detection restricts these MPAs to focal-plane-array (FPA) operation, which inherently constrains image resolution and FoV in conventional imaging geometries.

At present, essentially all commercially available THz cameras operate in power-detection mode \cite{i2s,swiss,ino,terascense,FBH,ohio,ticwave}, employing focal-plane imaging (FPI) architectures in which the MPA serves as a passive FPA. While this approach benefits from conceptual simplicity and technological maturity, it fundamentally limits the accessible imaging modalities and achievable image fidelity \cite{Zatta2021a, Zatta2021}.

For all types of FPAs, the finite sensing area relative to the THz wavelength, combined with diffraction-limited spot sizes, constrains the achievable FoV and spatial resolution \cite{Echternach2018,PhysRevApplied.20.014003, 5615383,hack2016comparison, Zatta2021, 6105155}. These constraints motivate the exploration of alternative imaging modalities and hybrid hardware--computational approaches that can decouple image quality from raw pixel count, a theme that becomes central in the discussion of coherent and computational THz imaging techniques in the following sections.

\begin{figure*}[ht]
\centering
\includegraphics[width=7in]{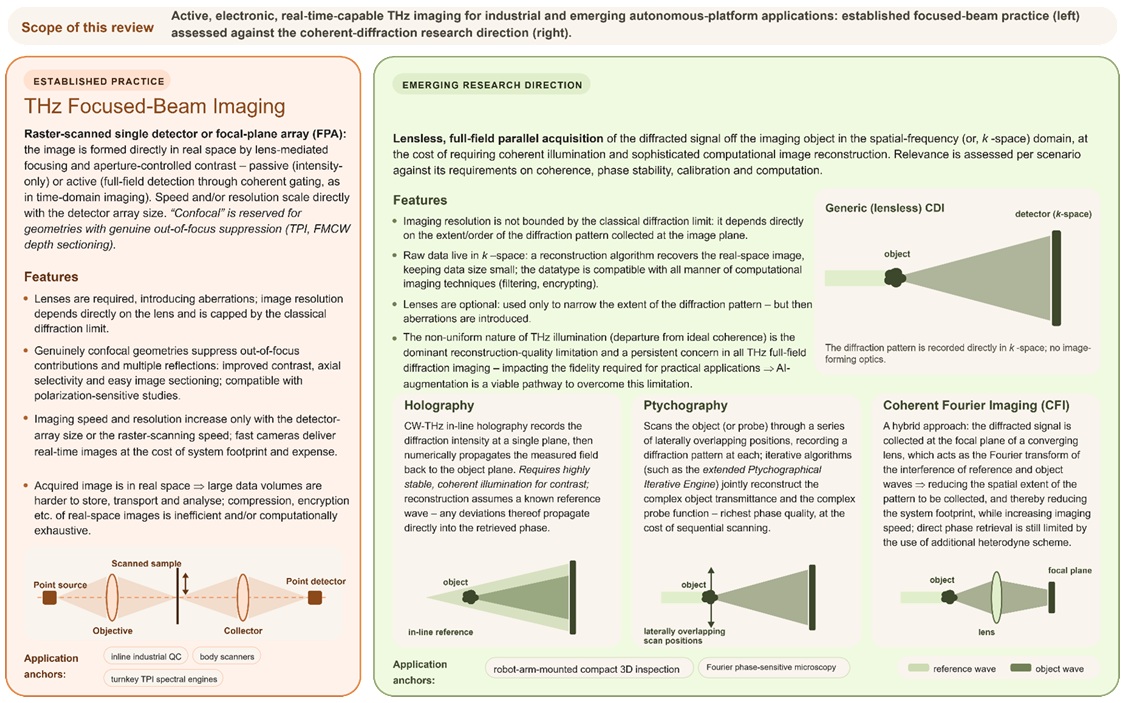}
    \caption{\hlyellow{A consolidated chart of various THz imaging schemes, both in direct-space and \textit{k}-space with corresponding salient features and application anchors for the imaging technique. Original artwork by the authors.}}
    \label{New_SecIII_fig.jpg}
\end{figure*}

There, another central topic will be the challenges associated with the measurement or computational retrieval of the phase distribution of the object wave. Heterodyne detection is a standard way to determine the phase information. While straightforward to implement with single detectors, it is challenging to extend the approach to FPAs. Often, one uses an externally generated local-oscillator (LO) wave -- i.e., a reference wave --, frequency-locked to the emitter used for object illumination and slightly frequency-shifted relative to it, and couples the LO wave quasi-optically onto the receiver array. In order to recover the phase of the object wave, the IF signal generated in each receiver must be read out with a high sampling rate for subsequent external digital signal processing. As the need for high-bandwidth sampling of the IF waveform requires direct and continuous access to each receiver and impedes multiplexing, it most of all factors limits the number of receivers which can be implemented in the FPA. A case in point is radioastronomy (where system costs are not a prime limiting factor). For gas spectroscopy at frequencies of 0.5-5~THz, FPAs are commonly based on superconducting hot-electron bolometers (HEBs) or superconductor-insulator-superconductor (SIS) mixers \cite{Tretyakov2025}, the IF lies typically in the range 0.5-12~GHz. The implemented FPAs contain only between a few up to several tens of mixers \cite{Graf2015}. Interestingly, at this scale, their number is not restricted by the required LO power, which -- for HEBs -- is on the order of 1~$\mu$W \cite{Risacher2016}. 

In contrast, semiconducting mixers operated at room temperature, as one would employ them for more mundane terrestrial applications, require much higher power levels for best performance \cite{Feng2022}. 
We want to make a case, however, that optimal operation condition may not always be necessary to achieve sufficiently good phase measurements.  

In the imaging studies of \cite{Boppel:12, Glaab2010}, it was explored how much LO power per receiver (in this case CMOS TeraFETs, with the object-illuminating radiation as well as the LO radiation at about 600~GHz) is necessary to achieve an improvement of heterodyne detection over direct dynamic range by 10~dB (for a 20-ms lock-in time constant) and at the same time provided access to the phase information of the object wave. Hence, while more LO power would undoubtedly result in a larger dynamic range and less phase noise, one wins in sensitivity and general access to the phase even at low power levels. There is quite some room for optimization for the specific heterodyne imaging purposes. We will come back to related issues in the Outlook under topic \textit{3D imaging with heterodyne cameras}.

\section{\hlyellow{Active} Terahertz Imaging Modalities}
\label{Chap3}
The achievable performance of a THz imaging system is governed not only by the choice of source or detector technology, but critically by the choice of imaging modality and the associated image data acquisition, storage, transport and analysis strategy  \cite{Wang2018},\cite{Ning2019}. The real interest of THz imaging lies in the fact that it is compatible with quantitative phase imaging  \cite{Peiponen2013}; catalyzing three largely parallel but increasingly convergent approaches: \hlyellow{THz Time-Domain Pulsed Imaging (TPI) and THz FMCW imaging, which acquire the image in real space by focused-beam raster or focal-plane-array (FPA) detection -- with genuinely confocal implementations additionally providing out-of-focus suppression --} and a gamut of techniques which can collectively be called Coherent Diffraction Imaging (CDI) techniques \cite{Rodenburg2008}, namely, THz digital holography \cite{Heimbeck2020}, THz Fourier Imaging \cite{Yuan2019} and THz Ptychography \cite{Valzania2018}.  Fig.~\ref{New_SecIII_fig.jpg} shows a consolidated chart of these various imaging techniques with corresponding salient features and very basic schematics.

\subsection{Far-field focused-beam imaging and THz-FMCW imaging} 

In the THz regime, where detector pixel counts are inherently limited due to the large wavelength and prohibitive cost of fabrication of large-area MPAs, and where diffraction effects are pronounced and limit the pixel pitch, image formation may rely either on multi-pixel focal-plane detection or on single-pixel detection combined with spatial, angular, or temporal scanning. Consequently, the imaging modality defines how spatial information is encoded, how acquisition time scales with resolution, and how effectively computational reconstruction can compensate for hardware constraints \cite{Dolganova2018},\cite{Bandyopadhyay2022}. TPI, grounded in THz time-domain spectroscopy, retrieves both amplitude and phase spectra via Fourier transformation of the measured electric field transient. Its spectral richness is unmatched, but its reliance on raster scanning and mechanical time delays makes it inherently slow. Acceleration strategies \cite{Tsubouchi2020},\cite{Kanda2017} — galvanometer scanning, two-dimensional electro-optic sampling, asynchronous optical sampling — partially address speed, yet the fundamental serial architecture remains a bottleneck of this conventional approach. In parallel, the use of electronic FPAs in \hl{focused-beam}  THz imaging has been explored to improve acquisition speed through spatial parallelism. To enhance signal coupling efficiency in such systems, hyper- or hemispherical silicon lenses are frequently integrated with individual pixels, often in combination with free-space optical components\cite{6557453,8048280,7404034,Zda15,Kutaish2024}. These substrate lenses effectively increase the numerical aperture (NA) and improve impedance matching, but at the cost of increased optical complexity and a reduced effective FoV. Moreover, the limited pixel count and relatively large pixel pitch of electronic MPAs impose fundamental constraints on spatial sampling density and achievable resolution. On the other hand, the confocal geometry inherently suppresses out-of-focus contributions and multiple reflections, resulting in improved contrast and axial selectivity, which has made it a widely adopted technique in laboratory-scale THz imaging systems; and even beyond, especially with commercial TPI-based turn-key spectral engines providing less than microsecond per pixel formation \cite{Arbab2025, Bandyopadhyay2026}.

THz-FMCW imaging traces its roots to millimeter-wave radar technology, which has long underpinned important commercial applications such as body scanners, package and letter inspection machines and inline quality control systems in production lines  \cite{terascense, RohdeSchwarz, Leidos, Nuctech1, Huebner, Orteh, CamTHz, Basler, Terakalis, Sikora}. FMCW techniques are commonly real-time capable and able to produce 3D images. Unlike radar cross-section measurements of an object at a certain distance, the THz-FMCW technique employs a confocal sectioning approach for depth discrimination \cite{DiLazaro2018}. In this approach, signals from specific axial positions within the sample are isolated temporally or spectrally -- either through sweeping the source or through \hlpink{time-domain} phase compensation \cite{Liang2025} -- while lateral resolution remains governed by optical focusing and scanning \cite{Hu2022}. \hlpink{In this technique, THz waves whose frequency is swept linearly in time are transmitted toward the target. Upon reflection from the target, the returning signal is mixed with a copy of the transmitted wave, and a beat signal is produced. The linearity of the frequency sweep ensures that the beat frequency is constant during the overlapping part of the sweep.} The time measured by the imager is determined by the frequency of the beat signal obtained at the IF port of the mixer; and is directly proportional to the difference between the delays caused by propagation on two paths: illumination and reflection. Thus, the measured beat frequency depends on the delay, which in turn depends on the distance to the target; or essentially the profile of the target can be determined by measuring the beat frequency. High bandwidths up to 300 GHz may allow sub-millimeter range precision. As a consequence, it may require from milliseconds to seconds per A-scan (essentially a 1D range profile providing structural information of a sample at a single transverse location) and several seconds for full 2-D electronic sweep of the scenery with bulky RF front-end, which is prone to phase noise and mixer non-linearity, as well as antenna side-lobes \cite{Tian2026}.

\begin{figure}
    \centering
    \includegraphics[width=0.7\linewidth]{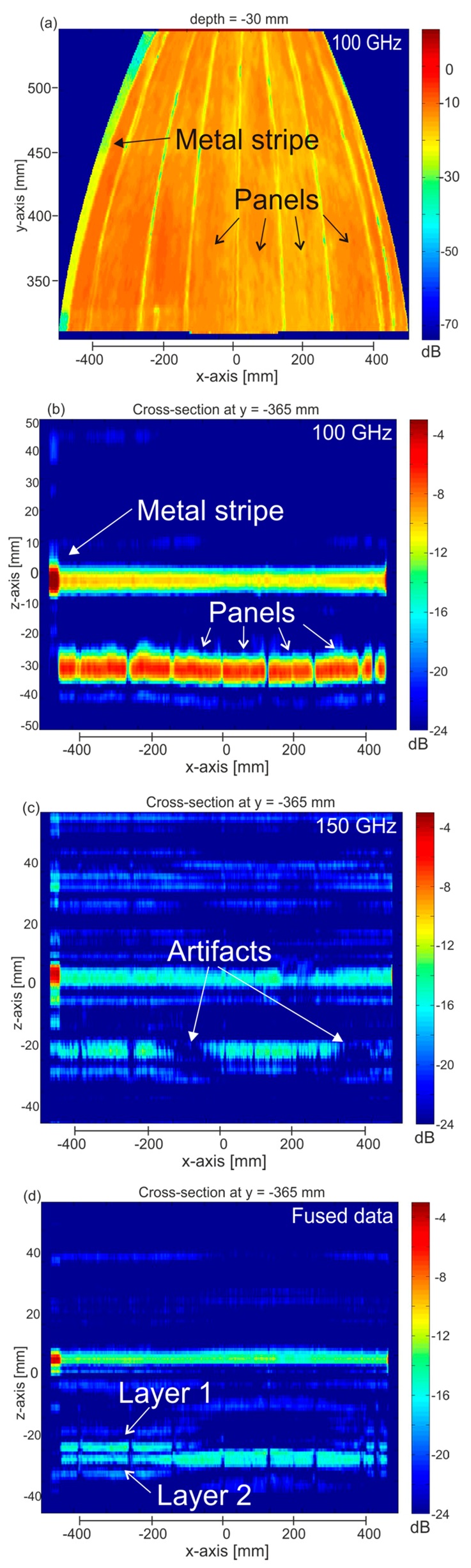}
    \caption{{\hlyellow{Three-dimensional FMCW THz inspection of an aircraft radome inside a manufacturing machining centre.
(a)~Reconstructed back surface at a depth of $-30$\,mm measured with the
100\,GHz unit (40\,GHz modulation bandwidth), showing the individual GFRP
panels and the metal alignment stripe; (b)~corresponding depth cross-section
at $y=-365$\,mm; (c)~the same cross-section acquired with the 150\,GHz unit
(60\,GHz bandwidth), exhibiting reduced penetration depth and artefacts;
(d)~fusion of both bands to a synthetic bandwidth of 100\,GHz centred at
120\,GHz, resolving the delamination of the GFRP panels from the carbon
mandrel (Layers~1 and~2). }\hlgreen{The measurement frequencies fall within the 0.1--0.3~THz transition zone discussed in Sec.~I.}\hlyellow{ Reproduced from
\mbox{\cite{friederich2018radome}}, \copyright~2018 the authors, licensed
under CC~BY~4.0.}}}
    \label{fig:radome_fmcw}
\end{figure}

 Recent studies \cite{Taba2022} highlight how high-speed transistors and harmonic oscillators have paved the way for scalable THz FMCW radar-on-chip solutions, revolutionizing security screening, NDT, and biomedical sensing. \hlpink{SiGe-based 220--320 GHz harmonic-oscillator and multiplier-chain architectures have been demonstrated to achieve superior linearity and reduced phase noise} \cite{Taba2022}. Concurrently, synthetic aperture radar (SAR) advancements \cite{Batra2021} and computational imaging frameworks \cite{Pan2020} have redefined 3D reconstruction fidelity and image interpretability. Modern FMCW imaging leverages adaptive modulation, compressive sensing (CS), and deep learning-based reconstruction to transition from range detection to real-time volumetric imaging and intelligent material diagnostics. These innovations have led to chip-scale FMCW arrays supporting portable THz imaging modules. \hlpink{SAR enhances FMCW imaging through aperture synthesis, improving lateral resolution without large antennas.} The back-projection (alternatively, back-propagation) algorithm reconstructs object profiles by integrating phase-aligned echoes. \hlpink{It has been demonstrated that SAR-based FMCW imaging in the 0.22--1.1~THz range can achieve sub-millimeter 3D imaging resolution} \cite{Batra2021}\hlpink{, beneficial for NDT and concealed-object detection.} \hl{A representative industrial realization is shown in Fig.~}\ref{fig:radome_fmcw} \hl{:
a two-band FMCW sensor head combining transceivers operating from 70 to
110~GHz and from 110 to 170~GHz was integrated into the machining centre of
an aircraft-radome production line and scanned spiralwise over the rotating
part }\cite{friederich2018radome}\hl{. The individual
glass-fibre-reinforced-plastic (GFRP) panels and the metallic alignment stripe
are resolved both in the reconstructed back-surface image and in the depth
cross-section, while fusion of the two bands into a synthetic modulation
bandwidth of 100~GHz sharpens the depth resolution to about 1.5~mm in air
and exposes a delamination of the panels from the carbon mandrel that neither
band resolves on its own.}

 Regardless of whether single-pixel scanning or FPA-based detection is employed, far-field \hl{focused-beam} THz imaging remains fundamentally constrained by diffraction-limited focusing and by the direct mapping between detector sampling density and image resolution. Increasing resolution typically requires either tighter focusing, which rapidly encounters wavelength-scale limits, or finer spatial scanning, which increases acquisition time. As a result, \hl{focused-beam} imaging architectures intrinsically couple image fidelity to hardware complexity and measurement duration, posing challenges for real-time, wide-FoV, and high-throughput applications. However, some of the specific scenarios of diverse application domains are expected to drive development, especially in FMCW THz imaging through (1) advancements in fabrication of BiCMOS and CMOS processes, achieving f\textsubscript0 values approaching 500 GHz, enabling fully integrated THz transceivers with enhanced range resolution, compactness, and power efficiency in recent times \cite{Li2017}, \cite{Dobroiu2022}; and (2) combining the principles of photonics and electronics, especially with respect to beam-scanning techniques \cite{Berkel2024}; depth super-resolution \cite{Reyes2025}; high-performance integrated lens antennas \cite{Loncarevic2025}; algorithms and analysis for curved SAR \cite{Gezimati2023} and even sensing precision approaching quantum noise limit \cite{Lu2025FMCWLiDAR} to circumvent some of the lingering issues of THz imaging.

\subsection {THz Coherent Diffraction Imaging}

\hlyellow{Before turning to lensless techniques, the relative maturity of the modalities deserves an explicit statement: focal-plane, scanned-beam, time-domain, radiometric, FMCW and SAR systems constitute today's established practice, whereas CDI -- holography, ptychography and Fourier imaging -- is a research direction whose relevance we assess for specific scenarios, such as compact robot-arm-mounted 3D inspection and laboratory NDT, balanced against its requirements on coherence, phase stability, calibration and computation. We nevertheless retain a detailed treatment of CDI deliberately: it is the segment of the field where the hardware-computational co-design advocated in this review has the largest projected impact.} 

\hlyellow{In contrast to the focused-beam imaging modalities described so far,} where the resulting image is acquired in the real space using a combination of lens (imparting aberration) and scanning opto-mechanics with single or detector array; CDI techniques acquire the image without lenses as it avoids aberrations and achieves the highest possible resolution, even beyond the classical diffraction limit, through the extent of the orders of the diffraction pattern collected in the Fourier space \cite{Rodenburg2008}. Since intensity-only detectors (FPA or traditional camera, mostly pyroelectric detector array) are used for most of the CDI based techniques, the imaging modality becomes insensitive to phase information. To reconstruct the entire scenery, therefore, the phase of the complex-valued scattered wave must be recovered from the intensity-only diffraction pattern, which produces the so-called "phase problem" \cite{Maiden2009}. In essence, the lensless CDI schemes consist of the experimental acquisition of an interference pattern followed by numerical phase retrieval from the same based on specific retrieval algorithms, such as, Iterative Phase Retrieval (IPR) \cite{Brault2025}, Transport of Intensity Equation (TIE) \cite{Rong2021} etc. - where the engineering trade-off is the spectral richness of THz TPI for spatial parallelism and phase recovery (usually without mechanical time-delay scanning) \cite{Rong2015}. A persistent limitation of all lensless CDI approaches is the quality of the coherent beam. High-power, long-wavelength coherent sources, such as far-infrared gas lasers produce spatially non-uniform illumination. Normalization by a reference background suppresses some of this non-uniformity, but residual structure in the probe persists and degrades phase accuracy. Holographic reconstruction assumes a known reference wave; deviations from this assumption propagate directly into the reconstructed phase. To take care of this weakness in the in-line holography, Ptychography offers the richest phase quality with inherent probe estimation but requires sequential scanning \cite{Valzania2018}. In recent times, the TIE approach has become the standard as it is computationally the cheapest and most suitable for weakly absorbing, extended samples under stable illumination \cite{Wang2023}.
\vspace{12pt}
\subsubsection{\textbf{THz Holography and Ptychography}}

Digital holography is a full-field coherent diffraction imaging technique which uses a reference signal to retrieve the phase -- it has very broad applications in bio-medicine, microscopy,
commercial electronics, entertainment, manufacturing, augmented and virtual reality as well as security and defense, etc \cite{Javidi2021}. In contrast, in Ptychography, the imaging phase is retrieved through a computational algorithm (usually iterative in nature) from a series of intensity patterns -- patterns which are recorded either by displacing the object to various positions relative to an illumination field; or modulating the illumination field relative to the object \cite{Pfeiffer2018}. In recent years, both these techniques have started to impact the long-wavelength of the THz region as these approaches can effectively overcome the space–bandwidth product bottleneck and diffraction limit simultaneously \cite{Hack2014, Ma2025b, Heimbeck2020, Constable2024, Mukherjee2026,Valzania2018}.

The core issue addressed by these techniques is the fact that semiconductor detectors measure the irradiance of the electric field, not the coherent field itself. For the past 75 years, phase retrieval and holography have been the primary mechanisms for recovering coherent fields from irradiance measurements. Both of these techniques solve the ``quadratic form'' problem of recovering a signal $x$ from the squared magnitude of an affine transformation $g=|Hx +r|^2$. Holography uses a reference to linearize the quadratic form problem. As computational imaging has developed and neural estimators have emerged, linearization has been \hlpink{replaced} by more \hlpink{aggressive computational} methods, such as ptychography. However, traditional methods are not orthogonal; emerging systems may use diverse strategies such as coded diffusers or masks, structured space-time illumination, and adaptive sampling. Such methods enable feature-specific and compressive sampling strategies, which we consider in the next section on augmentation techniques.

Both holography and phase retrieval may recover $x$ from $g$ with quantum-limited precision from appropriately coded measurements. In both cases, 4 or more measurements per reconstructed complex value are required~\cite{schulz2021photon}. Measurement strategy is designed around the nature of available reference signals, sensor arrays and optics. For example, Fourier diffraction imaging is a relatively slow point scan system with a complex detector. Holography and phase retrieval may alternatively combine larger arrays of simple detectors and optics to recover the field.

Ptychography is a self-referencing phase retrieval technique based on windowed measurements in image space and Fourier space. Advantages of ptychography are that the system forward model is relatively simple and easily calibrated and that only local coherence is required in the illuminating field. THz imaging is following the optical Ptychography development path with a time lag of roughly five to seven years \cite{Su2023}, recapitulating in the long-wavelength regime the same sequence of innovations: lensless holography \cite{Georges2022HolographyInvisible} → Ptychography with probe estimation \cite{Mukherjee2026} → probe optimization via divergent illumination \cite{Rong2021THzPtychography} → numerical aperture extension via extrapolation \cite{Xing2022THzPhaseRetrieval} → non-interferometric alternatives (TIE) \cite{Rong2021} → post-reconstruction deep learning \cite{xiang2025reconstruction} → and, prospectively, integrated physics-guided networks and learned sensing \cite{Chen2025UncertaintyAwareFP}. The most significant unexplored intersection is the application of optical Ptychography's coded-detection architecture; particularly the thickness-independent imaging model and its micron-step continuous scanning capability — to the THz frequencies \cite{Mukherjee2026}. At THz range, however, the coded surface is required to be engineered from THz-compatible materials with wavelength-scale features, presenting a fabrication challenge analogous to (but solvable by) the disorder-engineered surfaces \cite{Wang2023}. 

The evolutionary trajectory is indicative: THz computational imaging is converging toward a hybridized framework where physically motivated forward models (angular spectrum propagation, Fresnel diffraction) are optimized using either classical iterative algorithms or gradient-descent-based deep learning, probe estimation and illumination optimization are integrated rather than separated, and resolution enhancement through divergent illumination and numerical extrapolation are now becoming standard \cite{Zhang2026DualBandMetasurface}, rather than experimental. The insights developed in optical Fourier Ptychography — aberration recovery, multiplexed illumination, multi-aperture parallelism, learned sensing — represent a paradigm shift that the THz imaging domain is only beginning to follow \cite{LiX2023}.
\vspace{12pt}

\begin{figure*}[ht]
    \centering
    \includegraphics[width=1\linewidth]{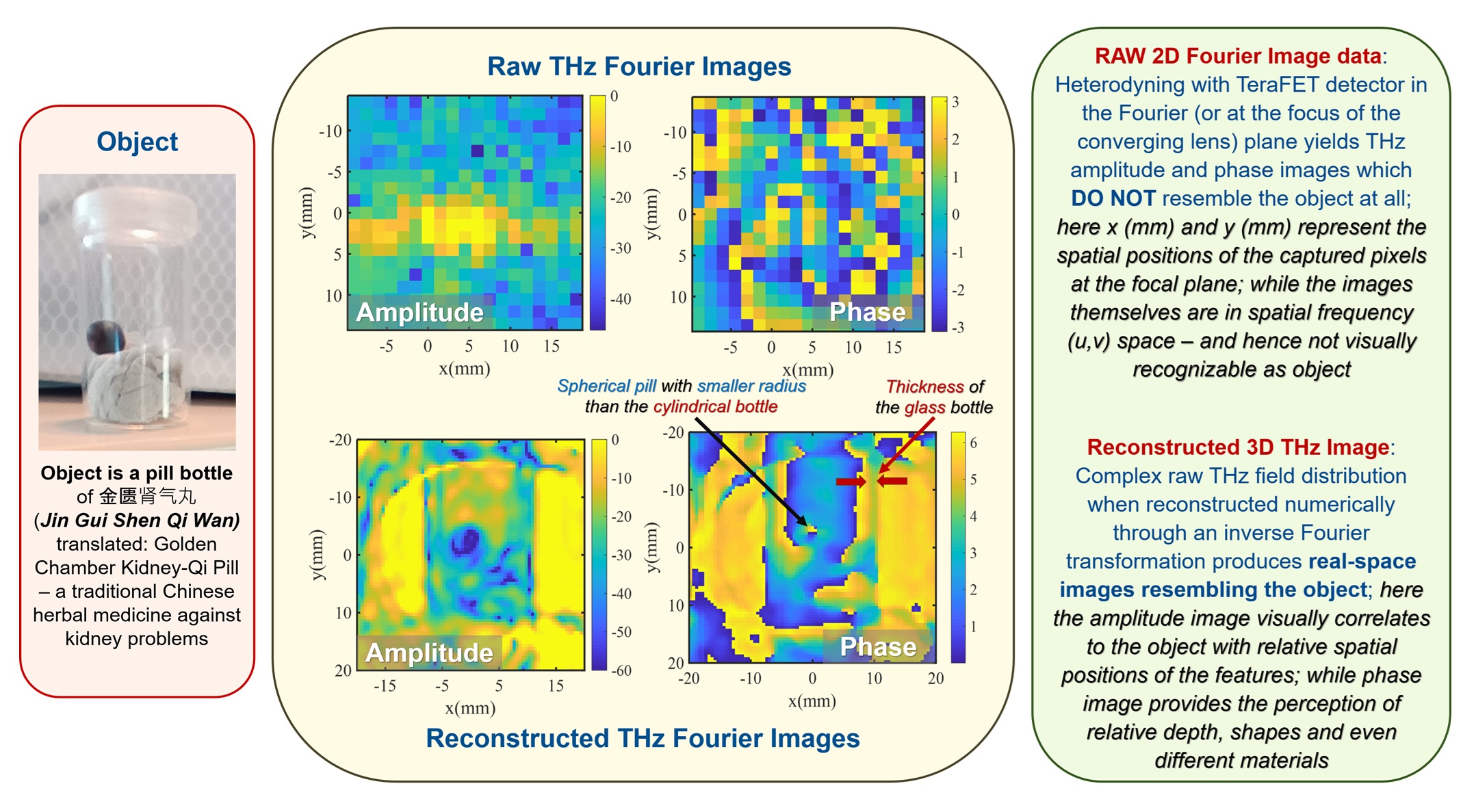}
    \caption{{\hlyellow{
Experimental illustration of how coherent Fourier imaging renders a
three-dimensional THz image. \textit{Left:} the object, a closed glass pill
bottle containing a single spherical pill. \textit{Top centre:} raw amplitude
and phase recorded by heterodyne detection with a CMOS TeraFET in the focal
(Fourier) plane of the converging lens. The axes $x$ and $y$ denote the
physical positions of the sampled pixels in the focal plane, while the data
themselves reside in spatial-frequency $(u,v)$ space and are therefore not
visually recognizable as the object. \textit{Bottom centre:} after numerical
inverse Fourier transformation of the complex field, the reconstruction
returns to real space. The amplitude image reproduces the relative spatial
arrangement of the object features, while the phase image conveys relative
depth, shape and material contrast, so that the smaller radius of the pill
compared with the cylindrical bottle and the finite wall thickness of the
glass both become directly readable. Acquisition at 300\,GHz with a step of
1.5\,mm over [20$\times$20] pixels at an object distance of 6\,cm. Original
artwork by the authors.}}}
    \label{fig:fourier3D}
\end{figure*}

\subsubsection{\textbf{THz Fourier Diffraction Imaging}}
Fourier diffraction imaging can be regarded as the far-field limit of CDI \cite{Yuan2019, Yuan2023, Yuan2025, Guerboukha:18,10.1063/1.5094728, Kumar:25}, where the detected field is located in the focal plane, where beam cross-section reaches its minimum in the optical system. In the description of the framework of the Fraunhofer diffraction theory for a 2D scene, the field distribution in the focal plane is proportional to the Fourier transform of the scenery under imaging, and distance information is contained in a pre-factor reflecting the change of the wave's phase during propagation to the lens. When the complex field is recorded coherently, both amplitude and phase information can be measured, allowing the object distribution to be reconstructed numerically through an inverse Fourier transformation. \hlyellow{Figure}~\ref{fig:fourier3D} \hlyellow{illustrates this two-step process on a concrete measurement: the raw focal-plane data bear no visual resemblance to the object, whereas the reconstruction recovers both its real-space geometry and, through the phase, its depth structure.}

\begin{figure*}[ht]
\centering
\includegraphics[width=7in]{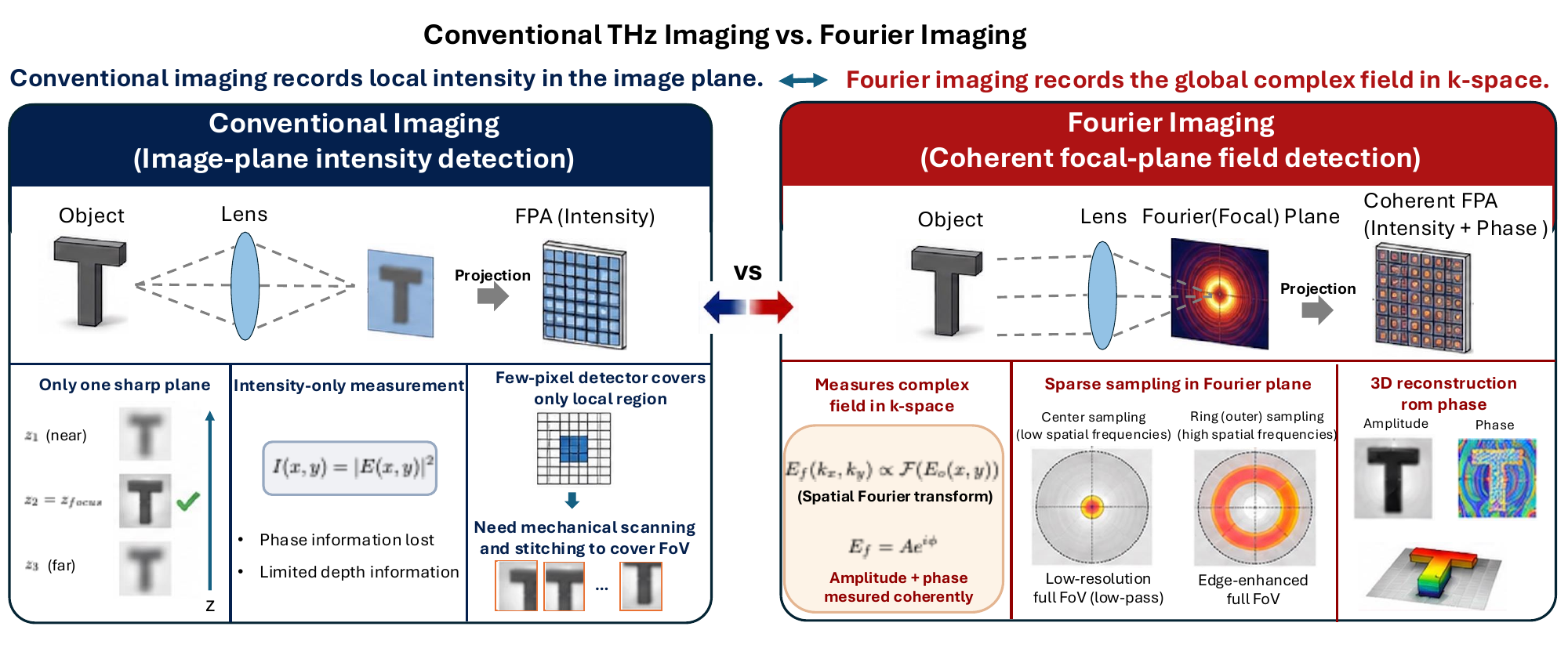}
    \caption{Conventional THz imaging measures intensity in the image plane, while Fourier imaging records the complex field in the Fourier plane, enabling full-FoV and phase-sensitive 3D imaging. \hlyellow{Original artwork by the authors.}}
    \label{conventional_vs_fourier}
\end{figure*}

In the THz regime, Fourier diffraction imaging has attracted significant attention due to the availability of coherent detection techniques such as heterodyne receivers. Several research groups have demonstrated THz Fourier imaging systems capable of reconstructing object distributions from measurements performed in the focal plane \cite{Zha10,Yuan2023,Yuan2025}. In contrast to ptychographic approaches\cite{Mukherjee2026}, where only the intensity in the Fourier plane is measured and phase information must be recovered through iterative phase retrieval algorithms using multiple Fourier spectrum intensity measurements under varying illumination positions, coherent Fourier imaging directly measures the complex field distribution in the focal plane. This enables a more straightforward reconstruction process and allows access to additional information such as phase-based depth cues. 

Fourier diffraction imaging offers several advantages for THz systems compared with conventional image-plane intensity imaging, as schematically illustrated in Fig.~\ref{conventional_vs_fourier}. (i)In Fourier imaging, the measured amplitude and phase distribution in the focal plane is mathematically linked to the object through spatial Fourier transformation \cite{goodman2005introduction}. Consequently, even sparse sampling in the Fourier plane still contains information about the entire FoV. Sampling close to the optical axis mainly captures low-spatial-frequency information and therefore provides a low-resolution view of the entire scene, whereas sampling farther away from the optical axis emphasizes high-spatial-frequency object features such as edges \cite{Yuan2019, Yuan2023}. The spatial resolution is therefore determined by how far the Fourier spectrum is sampled away from the optical axis, while the FoV is primarily determined by the spatial sampling interval (pixel pitch) of the detector array rather than by the physical detector aperture. 
In contrast, conventional image-plane imaging with only a few detector pixels covers only a limited spatial region and generally requires mechanical scanning and image stitching to recover the full FoV. (ii) Moreover, coherent Fourier imaging directly measures both amplitude and phase information, enabling depth reconstruction and three-dimensional imaging capabilities that are generally inaccessible in conventional intensity-only imaging systems, where only one image plane remains sharply focused for a given object distance. 
(iii) This Fourier-space representation also makes the required detector area physically scalable for practical THz camera implementations. Following Ref.~\cite{Yuan2019}, the illuminated Fourier-plane area can be estimated as $A_f \approx \pi(fD/2d)^2$. When the product $fD$ scales with the object distance $d$, i.e., when the object-space numerical aperture is kept constant, the illuminated Fourier-plane area remains approximately unchanged. For representative parameters of $D=15$~cm, $f=25$~cm, and $d=1$~m, this corresponds to approximately 11~cm$^2$, comparable to the sensor size of modern full-frame CMOS cameras. Assuming a detector pitch on the order of the wavelength, a coherent THz focal-plane array operating at 600~GHz would require only a few thousand pixels, suggesting that near-diffraction-limited large-FoV coherent THz imaging systems may become technologically feasible using scalable CMOS fabrication technologies. 
For comparison, a conventional intensity-recording FPA of similar size and pixel pitch could also provide near-diffraction-limited large-FoV imaging, but without direct access to phase and depth information. Interestingly, for large object distances, the image plane approaches the Fourier (focal) plane geometrically. For the representative parameters above, the image plane is located 33.3~cm from the lens, corresponding to a separation of only 8.3~cm from the Fourier plane, while at an object distance of 5~m this separation decreases further to only 1.63~cm. This suggests that future THz imaging systems may potentially integrate or switch between conventional image-plane imaging and coherent Fourier imaging within closely related hardware architectures, provided that coherent phase detection is available. From a broader perspective, Fourier optics provides a natural framework for optical analog information processing and computational imaging. Beyond image reconstruction, the Fourier plane can be used for wavefront manipulation, spatial-frequency encoding, and optical-computational processing. Recently, these capabilities have attracted increasing interest in the context of physics-informed neural networks and optical analog computing. In particular, diffractive deep neural networks (D$^2$NN) utilize cascaded diffractive layers to modulate the amplitude and phase of propagating waves, enabling task-specific optical neural networks implemented directly through free-space diffraction \cite{Lin2018}. Such systems exploit Fourier or Fresnel propagation to perform large-scale optical transformations at the speed of light, highlighting the potential of Fourier-space processing for future intelligent THz imaging systems. Looking forward, several emerging directions may further enhance the capabilities of Fourier diffraction imaging systems. One promising trend is the integration of large-scale coherent detector arrays into THz cameras, enabling parallel acquisition of Fourier-plane data and significantly improving imaging speed. More broadly, Fourier diffraction imaging can be viewed as a physically realizable far-field platform for modern computational imaging concepts under aperture-limited boundary conditions. Within this framework, techniques such as coherent diffraction imaging (CDI), holography, ptychography, CS, structured illumination, and super-oscillatory wavefront engineering can all be interpreted as different strategies for sampling or manipulating the Fourier spectrum of the object wave \cite{10.1063/1.5094728,Yuan2023,Fang:25,guerboukha2020superresolution,Kumar:25}. Combined with recent advances in AI-assisted phase retrieval and physics-informed reconstruction \cite{xiang2024amplitude, xiang2025hybrid}, these developments suggest that Fourier diffraction imaging may evolve into a unified platform integrating optical hardware, computational imaging, and intelligent wave-based information processing.

\section{Advanced State-of-the-art image augmentation techniques}

In this section, we discuss some of the techniques which \hlpink{have been implemented in recent times} to circumvent some of the lingering issues of THz imaging\hlpink{, such as image resolution, speed of image acquisition, and the limited range of frequency-selective THz modulators and lenses.} 

\subsection{To augment the THz image resolution: nearfield technique}

THz imaging with deep sub-wavelength resolution can be achieved in various ways. One approach is to detect the THz signal via techniques such as electrooptic sampling where the spatial resolution is determined by the cross-sectional area of the VIS or NIR laser beam used for the nonlinear read-out of the THz signal in the electrooptic crystal \cite{Pfeifer1998, Pfeifer1996}. This technique is not per se a near-field technique, but can be used as such \cite{Seo2007}. In similar spirit, use of THz guiding structure has been employed to exploit modal electric field localization to beat the classical diffraction limit \cite{Aalam2024}. In contrast, a true-near-field approach, but with limited sub-diffraction \hlpink{resolution}, employs photoconductive microprobes for the localized launching and detection of THz waves \cite{Wolfgang2024, Hoof2021}. The most commonly applied THz near-field technique, however, with an excellent spatial resolution in the tens-of-nm range, is scattering-type near-field optical microscopy (s-SNOM). Since its beginnings 
\cite{Huber2008, Wiecha-Intech2021}, it has experienced broad application 
\cite{Soltani2020, ZhangZL2023} in parallel with continuing technical refinement even with regard to spatial resolution, which now reaches down to 1~nm  \cite{Shiotari2025} (demonstrated for wavelengths in the VIS, not the THz range), only surpassed by THz-assisted tunneling imaging and spectroscopy \cite{Siday2024}. For many practical THz inspection applications, it would be important to have cost-effective s-SNOM machines available. With the radiation source and the detector (notably cryogenically cooled ones) being major cost factors, application-specific solutions, operating over a limited target frequency range, are to be found. A first step in this direction has been the demonstration that low-cost TeraFET detectors, operated at room \hlpink{temperature}, are suited for s-SNOM imaging \cite{Wiecha-nano-2021} and allow to determine the carrier density of moderately doped semiconductors \cite{Wiecha-Si-2021}. Work to also replace the multiplier-chain-type electronic THz emitters used in these experiments by electronic THz oscillators (such as Colpitts CMOS, SiGe or InP oscillators) emitting directly at the target frequencies is underway. We finally point out the development of Si microelectronic circuits for biomedical near-field imaging \cite{Grzyb2017} and chemical sensing \cite{Chernyadiev2024}. They do not employ a probe tip like s-SNOM for enhanced spatial resolution, but the electric field confinement in the gap region of split-ring resonators and at antennas.
\subsection {To augment THz imaging speed: Compressive sensing}

Image compression is possible because images are more efficiently represented as feature maps rather than raw pixel sequences. CS extends this efficiently into the measurement regime by recognizing that one may directly measure features instead of na\"ively sampling Cartesian pixels. Various strategies for compressive sampling have emerged over the past two decades. A major theme focuses on quasi-random projections on the pixel basis\cite{donoho06, candes06, eldar2012compressed}. This approach rests on three pillars: sparsity, incoherence, and convex optimization. Sparsity requires that the signal of interest, while high-dimensional in its native representation, admits a concise description when expressed in an appropriate basis (e.g., wavelet, discrete cosine transform, or gradient domain for images). Incoherence demands that the measurement matrix, which maps the signal to observations, exhibits low correlation with the sparsity basis. And finally, and most importantly, the recovery algorithms which reconstruct the sparse signal from incomplete measurements with high probability. These recovery algorithms are deeply rooted in the advanced mathematical strategies to solve unknown variables in large scale representations, namely Non-deterministic Polynomial time (NP). NP relates to CS primarily through the computational complexity of recovering sparse signals. Specifically, to find the absolute best sparse representation from a small number of measurements, also known as L0-minimization, is an NP-hard problem, meaning it is computationally infeasible to solve optimally for large datasets. In CS scheme, therefore, it is overcome by using convex optimization (or commonly known as, L1-minimization) or greedy algorithms to find a ‘nearly optimal solution’ in polynomial time. This approach is most directly represented in single pixel cameras~\cite{Chan2008}.

CS theory revolutionized signal processing by demonstrating that signals possessing sparse representations can be accurately reconstructed from far fewer measurements than dictated by the Nyquist-Shannon sampling theorem. For THz imaging, where acquisition speed, hardware complexity, and data volume present significant challenges, CS offers practical algorithm-driven pathways for dramatic acquisition time reduction without compromising image resolution.

Another branch of CS focuses on the mismatch between the embedding dimensionalities of measurements and objects, specifically the fact that measurements are typically drawn from 2D planes but objects span 3D space, time, color and polarization. Snapshot compressive imaging (SCI)~\cite{yuan2021snapshot} \hlpink{seeks to efficiently recover high-dimensional objects from low-dimensional measurements. While improving measurement efficiency is a common goal of both approaches, single-pixel and SCI systems seek to increase the entropy of measurement projections through physical coding rather than to obtain fully random incoherent projections.} In both SCI and single pixel compression, convex optimization has increasingly been replaced by Bayesian and neural estimators.

Compressive holography, consisting of the reconstruction of 3D objects from 2D holograms\hlpink{,} is the SCI approach of greatest utility in THz systems~\cite{cull2010millimeter}. Surfaces embedded in 3D are naturally sparse. For 2D THz scenes, several natural sparsity-inducing representations exist. Most objects exhibit edge sparsity—they consist of smooth regions separated by discontinuities, leading to sparse gradients. Spectroscopic THz images often show spectral sparsity, with each pixel's spectrum containing only a few resonances against a smooth baseline. Spatial-spectral decomposition may reveal that object properties vary slowly in space but rapidly in frequency, or vice versa. Exploiting appropriate sparsity priors enables reconstruction from random or optimized subsets of Fourier components, spatial samples, or time-domain points. Practical THz-CS implementations take various forms: Frequency-domain under-sampling (such as in THz-FMCW) acquires only a subset of spectral points during swept-frequency measurements, with missing frequencies recovered through CS reconstruction. Spatial under-sampling randomly samples pixel positions rather than performing exhaustive raster scans (such as in TPI or even THz Fourier Imaging), reducing acquisition time proportionally. Combined spatial-spectral CS acquires partial data in both domains simultaneously, exploiting joint sparsity for maximum compression.

Reconstruction algorithms for THz CS range from basic basis pursuit (L1 minimization via linear programming) to sophisticated variants incorporating total variation (TV) regularization, wavelet-based denoising, or learned sparsifying transforms. Iterative hard thresholding and proximal gradient methods offer favorable computational complexity for real-time applications, while alternating direction methods of multipliers (ADMM) enable incorporation of multiple constraints and priors. Recent deep learning approaches could ‘learn’ optimal measurement matrices and reconstruction operators jointly, achieving superior performance compared to hand-designed CS systems.
Key challenges include robustness to model mismatch (when true sparsity deviates from assumptions), computational demands of reconstruction for high-dimensional data, and difficulty establishing rigorous performance guarantees for practical systems with approximate rather than exact sparsity. Despite these limitations, CS provides a framework for THz imaging system design, enabling informed trade-offs between measurement count, reconstruction quality, and computational resources. As THz sources and detectors improve, it is expected that, CS will increasingly migrate from academic curiosity to practical implementation in commercial systems.

\subsection {Metastructures for THz optics - from metalenses to SLMs} 

\textit{Properties of metamaterials for THz optics}
\vspace{1mm} 

Electromagnetic metamaterials are composite media fashioned from metals, dielectrics, semiconductors, and sometimes other, more exotic, materials such as superconductors, liquid crystals, phase change materials, etc. Metamaterials may be broadly classified into two categories depending on which material platform is used to support their electromagnetic response: metals or dielectrics \cite{chen2016review}. The principle idea of metamaterials is that they can achieve precise values of the electric permittivity ($\epsilon$) and the magnetic permeability ($\mu$) through their geometry rather than through their chemical composition \cite{capolino2009theory, padilla2006spectroscopy}. Most often a unit-cell is fashioned using patterned conductors or shaped dielectrics and then this unit-cell is replicated over 1D, 2D, or 3D space \cite{capolino2009theory}.

The electromagnetic response obtained from metamaterials is resonant and has a strong dependence on frequency $(\epsilon(\omega),\mu(\omega))$ often described by a Lorentz oscillator model with parameters: oscillator strength $\omega_p$, resonance frequency $\omega_0=2\pi c/\lambda_0$, and damping frequency $\gamma$. It is desirable to make the metamaterial dimensions and the unit-cell lattice parameter ($a$) as sub-wavelength as possible ($a/\lambda_0 \ll 1$) to accurately describe them with $(\epsilon(\omega),\mu(\omega))$ parameters, as well as to avoid diffractive and photonic band-gap effects \cite{capolino2009theory}. If metamaterials are not too sub-wavelength they may exhibit spatial dispersion, often described by the wavevector ($k$) which results in an additional electromagnetic response dependence $(\epsilon(\omega,k),\mu(\omega,k))$ \cite{fan2022active}.

As metamaterials are composite media, the materials used to form them may serve multiple functions, i.e. metamaterials are multi-functional \cite{fan2022active}. For example, in metal-based designs the patterned conductors which give rise to $(\epsilon(\omega),\mu(\omega))$ may also be used to function as leads in a two-terminal device, such as a diode. Using the multi-functional properties of metamaterials a number of devices have been demonstrated including modulators, spatial light modulators (SLMs), and even frequency and direction agile wavefront manipulator \cite{zhou2023realtime, fan2022active}. For example, nearly 100\% of reflection modulation has been demonstrated for radation above 1~THz with metamaterials using graphene-on-silicon as active material \cite{Xia2025}.

Another realizable feature afforded from the composite nature of metamaterials is that of spatial dependence \cite{chen2016review}. That is, since metamaterials are modular, each location in a periodic lattice can be populated with a specific unit-cell design. In the simplest case, the modular property of metamaterials may be used to obtain graded designs over a plane (2D), i.e. gradient index optics $n(\omega,\mathbf{r})$, where $n=\sqrt{\epsilon}$ is the refractive index and $\mathbf{r}$ is the position vector. \cite{smith2005gradient} In a more extreme case termed \textit{transformation optics} both $\epsilon(\omega,\mathbf{r})$ and $\mu(\omega,\mathbf{r})$ are tailored over a volume (3D) to obtain, e.g. an electromagnetic cloak \cite{pendry2006controlling, schurig2006metamaterial, zhang2019transformation}. Other extreme cases involve non-graded or discretely varying architectures, where unit-cells are distributed according to pseudo-random, or digital sequences rather than continuous spatial gradients. These unit-cell configurations are utilized to control arbitrary electromagnetic phenomena, including diffuse scattering through inverse-designed quasi-random arrays \cite{rozman2024deep}, arbitrary multi-beam synthesis and radar cross-section (RCS) reduction via digital coding metamaterials \cite{cui2014coding}, spatially multiplexed architectures for decoupled multi-wavelength or polarization control \cite{arbabi2015dielectric}, and the reconstruction of complex wavefronts in metasurface holography \cite{ni2013metasurface}.

The properties of metamaterials which make them ideal for THz optics are summarized as: \cite{o2007properties}

\begin{itemize}
    \item The electromagnetic response is determined by geometrical parameters \cite{pendry1999magnetism, capolino2009theory}.
    \item Independent and direct control of $\epsilon$ and $\mu$ \cite{capolino2009theory}.
    \item Multifunctional properties \cite{fan2022active}.
    \item The electromagnetic response is determined by the `meta-atom', not the array \cite{chen2016review}.
    \item Electromagnetic similitude, i.e. metamaterials can be scaled throughout the spectrum \cite{fan2022active, padilla2022imaging}.
\end{itemize}

Although the above intrinsic properties of metamaterials indeed enable unprecedented control over electromagnetic radiation, they introduce a large number of geometrical parameters and material choices that must be determined. If this space is too large and not guided by theory or intuition, it may be infeasible to find an optimal solution in a reasonable amount of time. It is this challenge that has led to a paradigm shift toward data-driven computational frameworks.

\vspace{1mm}
\section{Artificial Intelligence in THz Imaging Systems}
\label{AI}
\vspace{1mm}

\subsection{Introduction to Artificial Intelligence for THz Imaging}

Artificial intelligence -- the primary term used to describe approaches ranging from deep learning to large language models (LLMs) -- has impacted nearly every field of science and engineering. Advancements in AI are fundamentally distinct and more significant than progress in domain-specific research. That is, because algorithmic optimization is inherently domain agnostic, progress in AI directly enhances capabilities across all scientific fields. This universality is underpinned by the \textit{universal approximation theorem} \cite{hornik1989multilayer, cybenko1989approximation} which demonstrates that in a feedforward neural network, a single hidden layer consisting of a finite number of neurons can approximate any continuous compact function to any desired degree of accuracy.

One common way to define artificial intelligence in relation to deep learning and machine learning is with a set theoretical description,
\setlength{\abovedisplayskip}{3pt}
\setlength{\belowdisplayskip}{3pt}
\begin{equation}
    \text{DL} \subset \text{ML} \subset \text{AI} \,,
\end{equation}
\noindent where DL is deep learning and ML is machine learning. Thus AI is the overarching set which enables a machine to mimic cognitive functions or optimize a specific objective. AI therefore not only encompasses systems that learn stochastically, i.e. machine / deep learning, but also those that are rule-based and deterministic. Machine learning deviates from classical AI by using stochastic probabilistic reasoning and statistical optimization rather than deterministic rules. Algorithms used in the ML paradigm automatically adjust their internal parameters to minimize a mathematically defined \textit{loss function} to extract and generalize structural patterns from empirical data. \cite{mitchell1997machine} Deep learning is the innermost subset within the AI superset and further specializes the statistical approach of ML. Here a large number of specialized parameters are used and the architecture specifically employs layers of hidden neurons. This hidden layer approach allows the network to automatically learn the optimal representation of the data thereby circumventing the need for manual feature engineering of ML. \cite{lecun2015deep} Although \textit{artificial intelligence} is often used as an umbrella term, the field is currently dominated by deep learning architectures which have enabled the most impactful systems including LLMs for natural language processing \cite{vaswani2017attention}, deep convolutional neural networks (CNNs) for high-dimensional image reconstruction, \cite{krizhevsky2012imagenet} and generative diffusion models \cite{ho2020denoising}.
\hl{The application of artificial intelligence to terahertz (THz) imaging is an active area of modern research and demonstrates a shift from traditional hardware and system optimization to advanced computational frameworks. As discussed, progress in THz imaging is limited by several challenges including long-wavelength diffraction limits, phase ambiguities, and prolonged acquisition times. AI architectures offer a path to mitigate or solve some of these challenges and some preliminary studies have indeed shown promise. For example, deep learning has been increasingly used across the entire THz imaging pipeline. AI excels at learning patterns in high-dimensions and extracting features, permitting them to recover complex optical fields, extract sub-wavelength features, and autonomously optimize measurement configurations. These capabilities establish AI as a key component of modern THz imaging systems which may be broadly applied in domains of: computational image reconstruction, spatial classification, super-resolution imaging, and end-to-end learned sensing. A comprehensive summary of these key developments and representative algorithmic architectures is presented in Table {\ref{tab:ai_thz_summary}}.}

\subsection{\protect\hl{Computational Image Reconstruction and Phase Retrieval}}
\hl{The acquisition and determination of complex-valued THz fields through use of intensity-only detectors is an ill-posed inverse problem. Common modalities subject to such difficulties include digital holography, Fourier imaging, and ptychography, and have traditionally relied on numerical approaches such as iterative phase retrieval (IPR) {\cite{Maiden2009}} or the transport of intensity equation (TIE) {\cite{Rong2021}} to recover the phase. These classic iterative techniques are based on solid theoretical underpinnings, however they suffer from computational complexities, require precise calibration procedures, and are highly sensitive to spatial non-uniformities.}

\hl{Artificial intelligence has been used to mitigate some of these challenges and good results have been demonstrated using deep learning non-iterative phase retrieval techniques. For example, in one study a convolutional neural network (CNN) was developed to learn a direct mapping from the $k$-space irradiance distribution to the recovered amplitude and phase distributions in real space. {\cite{xiang2024amplitude}} Although a disadvantage in this non-iterative approach is the upfront cost of training and verifying a deep learning model, the image formation stage is significantly accelerated compared to traditional techniques. Thus, in various industrial settings the potential for real-time image reconstruction may outweigh the development time. It is worth noting, however, that purely empirical data-driven CNNs often lack generalization outside of their specific training scenarios.}

\hl{It has long been known that the incorporation of domain knowledge into machine / deep learning models significantly mitigates ill-posedness and enhances model generalization. {\cite{abu1990learning}} In modern research this approach can be used for inverse scattering and phase retrieval and may be generally termed physics-informed deep learning (PIDL) {\cite{xiang2024amplitude}}. For example, various forward physics based models -- such as the Fresnel approximation of angular spectrum theory {\cite{goodman2005introduction}} -- may be incorporated into the loss function or into the network architecture {\cite{deng2025physics}}. Thus the network is forced to obey electromagnetic theory permitting PIDL models to achieve accurate holographic and tomographic reconstructions. In one study the authors used a two-step process in which they first used supervised pre-training followed by unsupervised PIDL refinement. {\cite{xiang2024amplitude}} This permitted the model to suppress standing-wave signatures and multiple-reflection artifacts, effectively recovering the phase which would be challenging for a classical linear method {\cite{xiang2024amplitude}}.}

\subsection{\protect\hl{Spatial Classification and Semantic Segmentation}}

\hl{Once the complex THz field is acquired or reconstructed a common desire is to perform classification and / or segmentation of spatial features in the image. Although the THz region is rich in spectral features stemming from intermolecular vibrational modes, phonon resonances, and hydrogen bonds, traditional methods such as principal component analysis \mbox{\cite{jolliffe2016principal,shlens2014tutorial}} or partial least squares discriminant analysis {\cite{fordellone2018partial}} do not perform well in high-dimensions, with low signal-to-noise, or in non-linear scattering regimes.}

\hl{Artificial intelligence may be used to address some of the above limitations and CNNs are particularly well suited for this task as they can perform efficient pixel-wise semantic segmentation across multi-spectral THz datasets {\cite{gezimati2024terahertz}}. For example, in biomedical research deep learning models applied to THz images can delineate boundaries between different biological structures. {\cite{cong2023biomedical}} The successful determination of malignant carcinomas entails differentiating it from surrounding healthy tissue by analyzing subtle variations in hydration levels and structural density, and may be performed with classification networks. Other approaches incorporate various signal processing techniques including the wavelet synchro-squeezed transformation combined with transfer learning to determine spatial features from noisy THz time domain spectra data {\cite{liu2022deep}}.}

\hl{Other areas of investigation have used spatial classification frameworks in non-destructive testing (NDT) and security screening. Deep learning models can autonomously classify concealed objects and detect structural defects including sub-surface delaminations, and potentially voids and porosity, fiber misalignments, and matrix cracks. This is accomplished by mapping local anomalies observed in the refractive index and absorption coefficient. In one study, the authors utilized a neural network architecture that integrated a multi-layer perceptron (MLP) and a complex-valued CNN, and a rapid detection of signatures within tomographic volumes outperformed that achievable with conventional techniques. {\cite{gezimati2024terahertz}} }

\subsection{\protect\hl{Sub-Diffraction Super-Resolution and Denoising}}

\hl{The relatively long wavelength of terahertz radiation limits spatial resolution to a few hundred microns at 1.0 THz, while high-speed acquisition of images inherently limits the signal-to-noise-ratio (SNR). Early work on super-resolution imaging (an inverse ill-posed problem) used multi-frame interpolation methods to increase image resolution using robust $L_{1}$-norm regularization constraints {\cite{farsiu2004fast}}. Although these conventional methods achieved good results, the increase in resolution was often at the expense of high-spatial frequency information under low SNR scenarios, which resulted in suboptimal image reconstruction. More recent artificial intelligence approaches have been shown successful as robust estimators which are capable of producing super-resolution images even with corrupted or diffraction-limited observations.}

\hl{Artificial intelligence approaches have also shown success in image denoising using both supervised and unsupervised GANs paired with deep convolutional frameworks. These systems have demonstrated an ability to decouple stochastic noise from deterministic object scattering. In one study, the authors developed models termed CycleGAN and Pix2pix which demonstrated an ability to reconstruct high-fidelity, denoised THz images from highly degraded datasets. {\cite{dutta2022deep}} Their approach was shown to preserve edge discontinuities and structural integrity in NDE and does not require low noise target datasets for the unsupervised training stage {\cite{dutta2022deep}}.}

\hl{Artificial intelligence frameworks have also shown an ability to recover sub-diffraction spatial image features through operating as generative interpolators. Here the model projects low-resolution diffraction limited THz images to high-dimensional feature space where specialized networks use sub-pixel convolutions and attention mechanisms to map these features to high-resolution output images. {\cite{ruan2022efficient}}. In another study, researchers used angular spectrum propagation as a physics constraint in an untrained neural network model to perform sub-wavelength single-pixel imaging with low sampling ratios {\cite{zhu2025deep}}. This artificial intelligence approach to super-resolution imaging effectively achieves a larger numerical aperture which extends the accessible spatial-frequency bandwidth of the system.}

\subsection{\protect\hl{End-to-End Learned Sensing and Hardware-Algorithm Co-Design}}

\hl{The aforementioned techniques apply artificial intelligence for post-processing only and therefore are limited. That is, traditionally the design of terahertz imaging systems is segmented into two separate efforts - the design of the optical components and the development of image reconstruction algorithms. This approach is sub-optimal and a more holistic approach may achieve higher performance with better imaging results. For example, in an end-to-end learning sensing paradigm, the physical optical path may be considered as an analog computational layer, and thus be incorporated into the optimization process of a complete terahertz imaging system utilizing artificial intelligence {\cite{Lin2018}}. In one study, researchers developed an end-to-end system where physical modulation layers (inverse-designed metamaterials or coded apertures) were parameterized and trained jointly with a digital backend neural estimator. Output errors were backpropagated along both the digital layers and the physical propagation model, permitting optimized co-design of a THz focal-plane array imaging system. {\cite{LiX2023}}.}

\hl{Agentic frameworks have been demonstrated that are able to tailor programmable metasurfaces to self-optimize a THz imaging system {\cite{xu2024physics}}. These artificial intelligence agents autonomously reconfigure hardware through continuous evaluation of measurements compared to predefined imaging objectives. In this study, the agentic system tuned the spatial phase distribution or the structural illumination pattern to adapt to the specific scattering environment of the target. This agentic framework is thus an example of a closed-loop autonomous imaging pipeline that demonstrates image formation through active learning of the wave-matter interactions.}

\begin{table*}[htbp]
\centering
\caption{\hlpink{Summary of Artificial Intelligence Applications and Developments in THz Imaging}}
\label{tab:ai_thz_summary}
\renewcommand{\arraystretch}{1.3}
\begin{tabular}{>{\raggedright\arraybackslash}p{3.5cm} p{6.5cm} >{\raggedright\arraybackslash}p{4.5cm} >{\raggedright\arraybackslash}p{1.5cm}}
\toprule
\textbf{Application Domain} & \textbf{Key Developments \& Capabilities} & \textbf{Representative Architectures} & \textbf{References} \\
\midrule
Computational Image Reconstruction \& Phase Retrieval & Direct non-iterative mapping from $k$-space to real space; mitigation of ill-posed inverse problems; suppression of standing-wave and multiple-reflection artifacts via domain knowledge integration. & Convolutional Neural Networks (CNNs), Physics-Informed Deep Learning (PIDL) & \cite{xiang2024amplitude, deng2025physics} \\
\midrule
Spatial Classification \& Semantic Segmentation & Delineation of biological structures and malignant carcinomas; autonomous anomaly detection of concealed objects and structural defects (e.g., sub-surface delaminations, voids) in NDT. & CNNs, Multi-Layer Perceptron (MLP) coupled with complex-valued CNNs, Transfer Learning & \cite{gezimati2024terahertz, cong2023biomedical, liu2022deep} \\
\midrule
Sub-Diffraction Super-Resolution \& Denoising & Decoupling of stochastic noise from deterministic scattering; sub-wavelength single-pixel imaging; recovery of sub-diffraction spatial features at low sampling ratios by extending accessible spatial-frequency bandwidth. & Generative Adversarial Networks (CycleGAN, Pix2pix), Sub-pixel CNNs, Untrained Neural Networks & \cite{dutta2022deep, ruan2022efficient, zhu2025deep} \\
\midrule
End-to-End Learned Sensing \& Co-Design & Joint optimization of physical modulation layers and digital estimators; autonomous hardware reconfiguration via active learning of wave-matter interactions in a closed-loop pipeline. & End-to-End Neural Estimators, Agentic Frameworks & \cite{Lin2018, LiX2023, lu2025agentic, lupoiu2025multiagentic, huang2025mcpenabled} \\
\bottomrule
\end{tabular}
\end{table*}

\section{Artificial Intelligence for Photonic Design}

The realization of advanced THz optics and systems is hampered by the lack of suitable materials, the complexity and vastness of the design space, and the inefficiency of computational tools. As a result, AI approaches are increasingly used for the design of photonic systems including metamaterials, photonic crystals, frequency selective surface, and plasmonics -- collectively termed \textit{artificial electromagnetic materials} (AEMs).

\subsubsection{Forward Deep Learning -- Surrogate Modeling}

In THz artificial electromagnetic material disciplines, the goal of the forward model is to make predictions of the electromagnetic scattering properties given a complete description of the physical system---the AEM geometry, desired frequency operational range, constituent materials, and embedding environment \cite{Malof2023}. Existing AEMs have demonstrated novel and surprising phenomena, and yet it is likely that even more exotic and applicable electromagnetic properties can be achieved with sufficiently-complex AEM designs \cite{khatib2021deep}. Unfortunately, manipulating AEMs of this complexity is largely beyond the reach of current design techniques including theory and conventional computational electromagnetics. This design bottleneck is perhaps the most important open problem in AEM research, limiting progress across the research community \cite{khatib2021deep}. 

The advent of artificial intelligence---deep neural networks (DNNs) in particular---has transformed traditional research methods across many disciplines \cite{deng2025physics}. DNNs are empirically derived systems that use large quantities of data to learn patterns that are fundamental to a process \cite{deng2025physics, khatib2021deep}. One notable application is that of forward modeling, where DNNs have been found to accurately predict the properties of metamaterials based upon their design \cite{lu2025agentic, li2024machine}. In this paradigm, researchers develop a deep neural network that solves the forward problem of metasurface design---given any geometry ($\mathbf{x}$), predict the corresponding S-parameters ($\mathbf{s}$), i.e., $\mathbf{x} \rightarrow \mathbf{s}$ \cite{nadell2019deep}.

Deep learning networks can be highly accurate, achieving an average mean square error on the order of $10^{-3}$, and are over five orders of magnitude faster than conventional electromagnetic simulation software \cite{nadell2019deep, deng2022benchmarking}. This extraordinary speed permits the evaluation of thousands of candidate designs in a fraction of a second, bringing complex architectures within reach. Despite the tremendous recent progress, however, DNNs are ``black box'' models that efficiently interpolate between training samples, and generally only approximate the true underlying physical relationships \cite{khatib2022learning}. 

As a consequence, neural surrogate models, which require significant quantities of training data, suffer from poor generalization outside the training domain \cite{khatib2022learning, peng2024transfer}. Recent research has shown that some of the limitations of conventional DNNs can be mitigated by employing fundamental physical knowledge to guide or constrain the DNN training process, or the network architecture, in different ways \cite{khatib2022learning}. These types of models are typically described as ``physics-informed'' or ``physics-guided'' neural networks \cite{khatib2022learning}, and represent an important step to combining deep neural networks with the laws of physics to yield models that are both fast and physically rigorous.

\subsubsection{Inverse Design of THz Artificial Electromagnetic Materials}

Due to the curse of dimensionality, the forward paradigm is unlikely to be a viable long-term solution for complex artificial electromagnetic material (AEM) research \cite{khatib2021deep}. That is, the design space may be so vast that even a fast forward method, such as evaluating a deep learning surrogate model, is infeasible. Fortunately, there is a fundamentally different approach—termed inverse design—which theoretically avoids some of the computational challenges \cite{khatib2021deep}. In AEM disciplines, the inverse problem is, given some desired electromagnetic scattering, to determine the AEM geometry which produced it \cite{Malof2023}. 

Although machine learning has recently been applied to the design of metasurfaces with impressive results, the much more challenging task of finding a geometry that yields a desired spectra remains a significant hurdle \cite{deng2021neural}. This difficulty arises because inverse problems are often ill-posed \cite{deng2022deep}, meaning the conditions of existence, uniqueness, and continuity are violated -- termed the well-posed Hadamard conditions \cite{deng2021neural}. 

To address this, deep inverse models (DIMs) have been specially designed to solve ill-posed inverse problems \cite{ren2022inverse}. DIMs have achieved impressive results, often surpassing capabilities possible with other approaches \cite{deng2022deep}. A variety of deep-learning-based inverse design architectures, such as tandem neural networks, generative adversarial networks (GANs) \cite{liu2018_nanolett}, variational autoencoders (VAEs) \cite{ma2019_advmater}, adversarial autoencoders (AAEs) \cite{makhzani2015_aae}, the neural-adjoint (NA) method \cite{deng2021neural}, and invertible neural networks (INNs) \cite{ardizzone2019_iclr}, have been successfully adopted for metasurface design and have demonstrated excellent performance \cite{xu2024physics}.

While no single model performs consistently the best across every metric, comprehensive comparisons indicate that the NA approach achieves the best overall performance across multiple AEM problem settings \cite{ren2022inverse, ren2020benchmarking}. The NA method is capable of finding accurate solutions to ill-posed inverse problems, where conventional methods struggle \cite{deng2021neural}. Crucially, since the NA method finds the globally optimal solution even in its worst-performing cases, results suggest that the NA always (or nearly always) finds the globally optimal solutions, even for highly complex problems \cite{deng2021neural}. 

Interestingly, this suggests that the main obstacle of custom design is no longer the challenges associated with the inverse model, but rather the space over which we choose to search for designs \cite{deng2021neural}. To mitigate this, the NA method can intelligently grow the search space to include designs that increasingly and accurately approximate the desired scattering response \cite{deng2021neural}. With its exceptional computational speed, high accuracy, and potential use in active learning, the NA method has an impressive future in not only thermal emitters \cite{yang2023normalizing}, but also any THz AEM inverse problem \cite{deng2021neural, rozman2024deep}.

\subsubsection{Automated Agentic Design and Autonomous Discovery}

\hl{ChatGPT {\cite{achiam2023gpt}}, Gemini {\cite{anil2023gemini}}, LLaMA {\cite{touvron2023llama}}, and Claude {\cite{anthropic2024claude}} are often designated as LLMs, although their underlying architectures are more accurately categorized as foundational models (FMs) {\cite{bommasani2021opportunities}}. LLMs are open-loop and reactive sequence generation systems. An agentic system, on the other hand, is a closed-loop and proactive iterative system. In contrast to LLMs, agentic systems are stateful, i.e. they maintain memory and track global task progression over time. {\cite{lu2025agentic}} Agentic frameworks interact with the external environment, execute code, invoke APIs (application programming interfaces), and can autonomously evaluate feedback from these processes to compare against their objectives and subsequently adjust their actions. Although the terminology within the AI community has not coalesced around an accepted acronym, here we term these agentic systems as artificial intelligent agents (AIAs).}

\hl{Relatively recently, AIAs have been applied to the physical modeling and design of artificial electromagnetic materials. For instance, researchers have demonstrated multi-agentic frameworks that leverage trained, fixed forward surrogate models to enable real-time, autonomous freeform metasurface design {\cite{lupoiu2025multiagentic}}. Other systems have utilized the model context protocol (MCP) to enable LLMs to interface directly with differentiable solvers to initialize fixed inverse design workflows {\cite{huang2025mcpenabled}}.}

\hl{In one study an AIA demonstrated the completion of a complex workflow for the deep inverse design of an all-dielectric metamaterial. {\cite{lupoiu2025multiagentic}} To achieve the target electromagnetic scattering state, the AIA needed to complete the following multi-stage steps: understand the problem, dataset generation, forward surrogate training, and inverse model optimization. After the agent successfully generated the forward surrogate model, it automatically turned to solving the inverse problem by using a deep inverse model {\cite{ren2020benchmarking}}. This work showed that the AIA is able to autonomously plan, store memory, use tools, adaptively iterate, and achieve its goal with no humans in the loop. This autonomous optimization framework serves as a direct analog for the complexities inherent in THz image reconstruction, establishing a foundation for the agentic workflows that will be explicitly exploited in Sec. {\ref{sec:emerging_strategies}}.}

In the future, the paradigm of artificial intelligent agents will transition from autonomous optimization to independent idea generation and execution of novel hypothesis. Advanced AIA could parse the literature, generate quantitatively novel ideas, identify gaps in current understanding, formulate research objectives, carry out experiments, and publish novel scientific discoveries. The shift toward autonomous scientific discovery will be disruptive in many areas, including terahertz science and technology. 

It is important to stress that success of AIAs is not accomplished through memorization and generalization of large datasets, as in deep learning, but rather through its emergent capacity to parse text, interpret structured data, and interface with computational engines. Further, since these agentic frameworks exist as digital entities, a successful agent demonstrating efficient and novel capability for scientific discovery can be readily and instantly replicated. Cloning allows for parallel deployment of thousand of specialized identical agents all concurrently working across vast domains of science. This capability decouples the rate of scientific discovery from human limits, enabling the systematic advance of electromagnetic topologies and the realization of novel THz optics at unprecedented scale.

\section{Outlook and perspectives}
Having presented key aspects of the current state-of-the-art of THz imaging, we now turn our attention to the (near) future and try to assess potential for major developments and breakthroughs, being fully aware that our statements are undoubtedly influenced by subjectivity. \hlpink{Let us begin with the impression that, with the increasing success of THz imaging in the market, the current trend away from purely focused-beam imaging modalities will continue.}  

\subsection{Hardware developments} 
\subsubsection{\textbf{Sub-1-THz regime}}

The frequency regime below 1~THz will be dominated by  electronic solutions for rapid imaging as long as spectral scanning over a large frequency range is not required. 

\textbf{\textit{THz cameras (power detection):}} With regard to hardware, it is foreseeable that THz cameras with ten thousand or -- depending on radiation frequency -- \hlpink{several tens of thousands of} pixels will become available as market opportunities arise; quite possibly with a strong share of CMOS-based cameras \cite{6363486, Zda15, ticwave} in addition to microbolometer arrays \cite{Oda2010, Nguyen2012, ino}. 
For CMOS cameras, one can expect substantial improvements (notably with regard to signal integration and the reduction of read-out noise) to be implemented in the near future by the adaptation of existing circuit solutions from VIS/IR cameras \cite{Liu2024}. 

\textbf{\textit{THz optics:}} In order to make best use of the available radiation power and to achieve a good spatial resolution, lens-based THz imaging modalities tend to have lenses with large diameter. This makes it challenging to minimize lens aberrations. Classical optics for the visible spectral range has solved this problem by the use of performance-optimized lens groups. It seems to be conceptually straightforward to adapt such lens-combination concepts also for THz optics, especially to deal with astigmatism hampering wide-angle imaging and impossible to correct in all directions by aspherical lens design alone. However, as in the case of VIS/NIR cameras, it is mandatory to suppress the spurious reflections occurring at each surface of the lenses with anti-reflection (AR) coatings. For some THz optical components, AR coatings are commercially available \cite{Tydex}. Further research on suitable materials and technologies is ongoing \cite{Gatesman2000, Chen2009, Yao2018, Soeda2019, Shevchick2023, Song2025}. The implementation of AR materials on lenses, the fusion of lenses to lens groups, and their assembly to composite optical systems still are in their infancy. The field of VIS/NIR optics continues to provide rich inspiration, exemplified by lens designs which provide flat focal areas for large MPAs and zoom lens designs that allow continuous adjustment of magnification and FoV in imaging systems  \cite{Wang2026}.     

\textbf{\textit{THz emitters:}} The main limiting factor for THz imaging, however, has always been the lack of sufficient radiation power. If there is still a remnant of the famous \textit{THz gap}, then it is here. For real-time imaging, tens or hundreds of mW of power are required which seems to preclude optoelectronic sources for deployable mobile systems. Benefiting from advances in semiconductor device technology for wireless communications, more and more powerful sources become available at the transition region from millimeter to submillimeter waves \cite{Makhlouf2023}. Commercially, one finds sources for 300~GHz based on Schottky-diode technology, which provide 40~mW of cw power \cite{ACST}.  But these microelectronic sources continue to exhibit a strong power \hlpink{roll-off} when scaled towards high frequencies, before lasers such as QCLs take the power lead above 2~THz. A game-changing development seems to be on-chip coherent power combination. This development occurs across technology platforms, but we want to highlight recent advances achieved with THz oscillators employing RTDs as gain medium. Despite recent breakthroughs with regard to the emitted power from a single RTD oscillator (1~mW in the 0.6-0.7-THz range) \cite{Tanaka2025}, RTD oscillators tend to provide low power, on the order of tens of $\mu$W in the 0.4-0.9-THz window. Researchers tried for a long time to achieve coherent power combination, but failed. In 2023, scientists at Canon Inc. finally succeeded to coherently couple 36 oscillators on-chip, achieving an output power of 10~mW at 0.45~THz \cite{Koyama2022}, and these chips can be combined (incoherently) for even \hlpink{higher} radiated power. Modified on-chip coupling schemes were implemented subsequently by other teams 
\cite{Tai2023, Meng2024} at even higher frequencies. These advances, in combination with  progress regarding wave amplification with RTD waveguides  \cite{Greenberg2026}, have begun to make the 0.4-0.9~THz range accessible for imaging applications, especially as the beam quality of 2D emitter arrays is very high \cite{Koyama2022}. 

We also note, that other candidates for powerful emission in this frequency regime are vacuum-tube emitters \cite{Thumm2020}. Usually, these devices are large and heavy, but there were comparatively compact backward-wave oscillators (BWOs) on the market covering a large frequency regime extending beyond 1~THz. However, they had only a limited lifetime and were difficult to repair. Compact vacuum-tube sources seem to have largely disappeared from view in the THz imaging community, although work on traveling-wave tubes (e.g. as amplifiers for communication purposes \cite{Ulisse2022}) continues. A number of years ago, the fabrication technology for a promising palm-sized traveling-wave-tube emitter for 650~GHz was reported \cite{Lueck2011}. Its compactness was described to be due to diamond structures for the waveguide. The company Teraphysics Inc. behind this development shows on their website a palm-sized device, now as amplifier (and apparently no longer as an emitter) designed for output powers from tens to hundreds of Watts  \cite{Teraphysics}. It would be very interesting to obtain and explore such devices for imaging purposes.

\textbf{\textit{Imaging at high frequencies for improved spatial resolution:}} 
That the frequency range 0.4-0.9 THz has gained in practical interest, can be seen from the multitude of applications which Canon Inc. is exploring with their emitter device. Among these are stand-off security screening, package and luggage inspection, NDT and wireless communication \cite{Canon2023, yet2}. An obvious benefit as compared with millimeter waves is the much better spatial resolution of imaging at these high frequencies. Especially for stand-off security scanning, this had been recognized as crucial already two decades ago \cite{Appleby2007}. During a first wave of research on such security scanners (which ultimately did not lead to systems in the market), systems were explored for heterodyne imaging for example at 675~GHz (including FMCW for depth resolution)  
\cite{Cooper2011} and 812~GHz \cite{Friederich2011}. While one can make a case that for object identification through garments it may be better to employ lower frequencies towards 300~GHz \cite{Heinz2010}, the identification of objects in free space over many meters should clearly benefit from higher frequencies as long as they are within the atmospheric windows of low THz absorption by water vapor (the windows being 350$\pm$22~GHz, 410$\pm$25~GHz, 670$\pm$43~GHz, 850$\pm$55~GHz, with rising background attenuation for bands at higher frequencies \cite{Sheikh2016}). As sub-1-THz technology becomes more advanced, attempts to develop distance scanners based on sub-1-THz radiation are being renewed for various application scenarios. Perhaps the most ambitious endeavor at this time may be the development of an imaging radar technology at 500+~GHz for cars by the company TeraDar Inc. at Boston, USA, aiming at an angular resolution similar to that of Lidar \cite{Teradar1}. It purportedly uses silicon transistor technology and it is claimed that it can meet the auto industry's 300-meter distance requirement \cite{Teradar2}. If this technology would make it successfully to the market, one could speak of a true "killer application" for THz radiation above 300~GHz, \hlpink{for which the THz community has been searching and hoping for decades.}

\textbf{\textit{3D imaging with heterodyne cameras:}} 
The technology of TeraDar Inc. is 
based on classical active radar concepts such as SAR and FMCW, the latter because distance determination is of paramount significance for their goals. 
However, SAR and SAR/FMCW imaging systems tend to be  bulky because of the inherent distributed nature of the emitter and detector arrangements. While these ensure a relatively large effective numerical aperture, i.e., detection of object waves arriving at the detectors over a fairly large angular range, which is beneficial especially when taking images of weakly scattering objects, more compact alternative imaging systems with lenses or other ways of beam collection may be preferable for applications with only limited available space for the device. A typical situation could be the mounting of a THz imaging system on the arm of a robot. 
Such possible alternatives could be lens-based imaging systems, among them Fourier Imaging systems for 3D vision, see \hlpink{Sec.~III-B}. The latter employs a FPA for detection of the radiation, mounted in the focal plane of a lens where the radiation is concentrated the most along the entire beam path. The scene is illuminated with a single emitter, but the use of distributed emitters is feasible, although, in the latter case, evaluation of the coherent signals and image reconstruction remain to be demonstrated. 

As stated above, Fourier Imaging is a holographic technique and requires the phase information of the object wave for image reconstruction. The phase is obtained by heterodyne detection. As discussed in Sec.~\ref{MPAs}, heterodyne detection is straightforward to achieve with single detectors (mixers), but FPAs, which one needs in order to become real-time-capable, are challenging. One has three basic options to proceed. First, to employ an externally generated and radiatively injected LO wave, and to read out the full temporal waveforms of the IF signals. As of Sec.~\ref{MPAs}, this approach limits the pixel number of such arrays to less than 100 due to the huge data rates to be read out. This is the established approach. 

We suggest here a second, novel option. As before, it is based on the use of an external LO beam, but the idea is now to process the phase information at each pixel and read out a quasi-DC signal. For this to work, it is necessary to develop the ability to process the IF signal on-chip at each receiver, and to read out only two quasi-DC signals which represent the amplitude and phase. This can be achieved in principle with integrated I\&Q converters. The concept has been implemented in the NIR spectral regime with photonic circuits \cite{Khachaturian2021}. The challenge for an electronic implementation, especially embedded in a receiver array, is the required chip area. A 3D architecture, with the I\&Q converters placed underneath the receiver antennas, would be desirable. Another possibility to detect the phase could be an adaptation of interferometric heterodyne phase sensing, which is used for optical wavefront characterization with multi-pixel arrays \cite{Furth2011}. Both approaches are on the forefront of device research, and large heterodyne arrays suitable for real-time imaging will \hlpink{take} a long time to become available, although they have the potential to bring forth THz heterodyne imaging substantially. 

Even further in the future is the third option, which combines pixel-wise phase processing with the generation of LO power on chip or on the carrier board of the receiver units. The advantage of local LO power generation and distribution is the availability of, in principle, higher power levels per receiver. However, heat generation as well as space constraints and losses of power distribution are major limiting factors. We are not aware of publications on implementations of the third option. Local LO power generation on its own, both at fundamental and sub-harmonic frequencies, has been reported in a number of publications for individual receivers and  small arrays of them  \cite{Guerra2013, 
Grzyb2015, Hu2019,  Zhang_Rennings2022}.

\textbf{\textit{Substrate lenses, superstrate lenses and micro-lens arrays for FPAs:}} A common challenge of all THz detectors is the proper loss-reduced in-coupling of the THz radiation. Losses arise not only from reflections at surfaces and absorption in materials, but -- in the case of antenna-integrated detectors -- also by a mismatch of the wavelength and the size of antenna. Antennas are designed for their effective refractive index, which for surface-mounted antennas is an average of the indices of air and the substrate material. The radiation, however, when impinging from air, covers a spot which even for diffraction-limited focusing is larger than the effective cross-section of the antenna because of the refractive-index mismatch. In order to reduce the radiation spot size, it is common to employ substrate and superstrate (solid-immersion) lenses 
\cite{Rutledge1982, Filipovic1993, Jepsen1995, Suszek2015, Yuan2023b, Krysl2024, Chen_lens2025, 
Holstein2026a}. For FPAs with their densely packed detectors, the mismatch also leads to a spread ("spilling") of the radiation to neighboring pixels.
While for intensity recording, the spilling effect reduces spatial resolution, for Fourier Imaging it additionally leads to a distorted recording of the phase of the object wave, which can lead to errors in the image reconstruction. Ways to cope with this challenge are the use of a single large sub-/superstrate lens on the FPA with the risk of strong aberrations \cite{ Trichopoulos2013, Zatta2021, Zatta2021a, ticwave}, the application of superstrate micro-lens arrays  \cite{Park_microlens2013,
Guo_lens2015, Delpino2017, 
Zhang_Llombart2024} or possibly of suitable near-field diffraction elements. There is still a clear need for continued optimization of the radiation coupling into FPAs (optimizing the FPA and the optics jointly) for an improved image quality. 

\textbf{\textit{3D imaging with incoherent radiation:}} 
3D imaging can also be achieved with incoherent radiation. An interesting approach is light-field imaging. Its concept is rooted in the integral imaging concept of Gabriel Lippmann from 1908 who introduced the idea to employ a microlens array (lenticular array) in front of a photographic film in order to derive the directions of the light beams arriving from the objects in addition to the  usual intensity and the greyscale/color distribution \cite{Lippmann1908}. Cameras with the ability to determine beam directions are termed plenoptic cameras; they can employ microlens arrays \cite{Dansereau2013} or other ways to derive beam directions \cite{Adelson1992}. After many decades of development, VIS/NIR plenoptic cameras have begun to enter the market and are being employed for special applications such as fluid velocity field measurements \cite{Liu_Lightfield2024, Niu2025}, improved digital refocusing \cite{Ng2005} or in astronomy for the measurement of the phase distribution of wavefronts propagating through the atmosphere \cite{Rodriguez2012}. The microlens array can be fabricated in the form of  metalens array \cite{Lin2019}. At THz frequencies, the low pixel count of available FPAs makes the use of microlens arrays on the FPA impractical. Instead, plenoptic functionality was demonstrated with a single large Si substrate lens on a 32$\times$32-pixel CMOS FPA \cite{9365832, Jain2016}. As 3D image reconstruction can be computationally demanding, CS strategies are being explored to minimize data acquisition requirements \cite{Kutaish2024}. Light-field techniques have strong potential, but require -- as other THz imaging techniques -- larger FPA pixel numbers to succeed. 

Yet another approach to 3D imaging with incoherent radiation is stereoscopic imaging -- inspired by biological vision, not only of humans, but most remarkably of birds of prey -- which represents a natural way of performing 3D imaging via triangulation. 
Indeed, one finds stereoscopy mainly investigated in the context of passive microwave imaging  \cite{Luethi2005, Yeom2011}. Compared with SAR/FMCW techniques, however, the charm of stereoscopic imaging lies in the potential use of incoherent radiation with additional constraint that the source is powerful to achieve required level of scene illumination and the availability of pairs of FPAs with sufficient pixel numbers. At the present time, limited availability of powerful THz sources and a lack of suitable FPAs may have been the primary reasons as to why stereoscopic imaging has been largely overlooked in the THz community. Given that stereoscopic imaging is conceptually capable of real-time processing, it certainly merits greater research attention in the future.

\textbf{\textit{Super-resolution imaging:}} Sub-THz imaging systems possess several important advantages for practical applications compared with imaging above 1~THz, including the availability of compact high-power electronic sources operating at room temperature, as well as generally lower atmospheric attenuation and material absorption, which enable improved penetration through optically opaque or scattering materials \cite{Makhlouf2023,Heinz2010, Sheikh2016}. These properties are particularly attractive for industrial applications. However, the longer wavelength fundamentally limits the spatial resolution in far-field imaging systems due to diffraction, since the accessible spatial-frequency bandwidth is constrained by the NA and system geometry. As a result, deeply subwavelength object features are strongly attenuated or completely lost. Consequently, super-resolution imaging has become a key research direction for future sub-THz systems, aiming to combine the penetration capability and practical advantages of long-wavelength operation with spatial resolutions approaching those of higher-frequency THz systems. Similar diffraction-related challenges have driven the rapid development of super-resolution and computational imaging techniques in VIS optics \cite{1203207, Zheng2014}, many of which are now beginning to migrate into the THz domain.

Current THz super-resolution approaches mainly rely on two conceptual categories. The first represents spatial-bandwidth-extension techniques, which utilize subwavelength modulation structures or metamaterials to convert evanescent high-spatial-frequency information into propagating waves detectable in the far-field \cite{10.1063/1.5094728, 
guerboukha2020superresolution, Alekseyev2011, Lu_Superres2012, Stantchev2017, Zhao2021, Chen_Superres2019, Vazquez2024, Zeng2025}. Although such near-field approaches can achieve deep-subwavelength resolution (values as low as $\lambda/100$ have been reported), they require close proximity between the object and modulation structure in order to achieve the spatial-frequency shifting into the detectable Fourier-space region. The second category relies on the 
redistribution, selective enhancement, or computational recovery of spatial-frequency information already accessible in the far-field. The radiation field is manipulated through various forms of Fourier-space engineering, including filtering processes, phase reshaping, structured illumination, synthetic apertures, and super-oscillatory wavefront engineering
\cite{Abramova2023, Cecconi2024, Li_Superres2024, Fleming2025, Zhao_Superres2026}. Unlike near-field modulation techniques, these methods generally do not physically extend the accessible spatial-frequency bandwidth beyond the conventional diffraction limit. Although the achievable resolution enhancement of such far-field techniques is generally moderate, they are  compatible with stand-off imaging geometries and practical industrial inspection scenarios. 

The achievable degree of super-resolution is usually accompanied by increased acquisition time and computational complexity, since additional measurements are required to recover inaccessible spatial-frequency information. Therefore, practical implementations must balance imaging speed, signal-to-noise ratio, and resolution enhancement depending on the application requirements. In this context, recent developments in computational imaging, diffractive deep learning, and physics-informed neural networks are particularly promising. By jointly optimizing the physical modulation process and reconstruction algorithms, AI-assisted approaches may significantly reduce the required measurement number while preserving subwavelength features. Combined with ongoing advances in THz spatial modulators, detector arrays, and high-speed computational hardware, these developments suggest that practical real-time THz super-resolution imaging systems could become feasible for future industrial inspection, security screening, and biomedical applications.

\subsubsection{\textbf{Above-1-THz regime}}

The region above 1~THz is physically distinct from the sub-1-THz band in commercially important ways \cite{Blackburn2022}. The water absorption coefficient rises from approximately 100~cm$^{-1}$ at the low end of the THz band to around 550~cm$^{-1}$ at 3~THz, while the penetration depth in liquid-containing biological tissue falls from roughly 100~$\mu$m to approximately 10 $\mu$m at the upper end of the THz range. Counterintuitively, this surface-sensitivity becomes a commercial advantage rather than a limitation in specific application domains enabling surface-specific chemical imaging of thin films, polymer coatings, and pharmaceutical tablet coatings etc. In these domains the imaging objective stresses on real-time acquisition and where superficial penetration combined with molecular spectroscopic fingerprinting is most valuable to suppress image ambiguity arising out of substantial volume penetration of the imaging illumination \cite{Zheng2014}. At the same time, the imaging resolution achievable above 1~THz improves markedly relative to lower-frequency THz systems. The free-space wavelength at 1~THz is 300~$\mu$m; at 3~THz it falls to 100~$\mu$m. Traditional lens-based focusing yields diffraction-limited spot sizes of the same order, which is already superior to any microwave, millimeter-wave or sub-1~THz imaging systems. Ptychographic reconstruction further improves on the above \cite{Li2022}: the divergent-illumination extrapolation approach achieved lateral resolution of approximately 1.5$\lambda$ and the subwavelength longitudinal-shift method demonstrated full-field imaging with sub-wavelength resolution without requiring near-field proximity of the detector. At 1 THz these system specifications correspond to spatial resolutions in the 150–300 $\mu$m range achievable with far-field optics – a very important imaging performance that no other competing far-field THz modality currently matches with this approach \cite{Wang2025}.

The emergence of THz Fourier Ptychographic Microscopy (THz-FPM), therefore, marks a pivotal inflection point in coherent THz imaging in recent years, especially above 1 THz range \cite{Kumar2025, Mukherjee2026}. Since the foundational FPM architecture has established that angular multiplexing in the Fourier domain could synthesise gigapixel-class resolution beyond the physical aperture of any single detector, the concept has migrated across the electromagnetic spectrum with remarkable fidelity \cite{Wang2025}. Recent demonstration of proof-of-concept resolution enhancement through multi-angle plane-wave illumination without hardware modification of the detector has opened up a credible engineering roadmap toward clinical and industrial deployment of this technology \cite{Zheng2014, Horstmeyer2015, guerboukha2020superresolution, Brault2025}.

\subsection{Emerging data acquisition and image processing strategies}
\label{sec:emerging_strategies}

\textbf{\textit{Fourier Imaging  with intensity-only measurements and AI-enabled phase retrieval:}} As we have seen, FPAs for coherent detection with large dynamic range (ensured by sufficiently high LO power per pixel) will not be available anytime soon commercially, at least not with pixel numbers as one would like them to have for real-time imaging. For that reason, the idea has come up to derive the phase information of the object wave not by heterodyne techniques, but from interference patterns present in the intensity profiles measured by a power-detecting FPA. The question arose whether -- for the sake of a simplified hardware implementation -- this could be achieved without the use of a reference wave. The retrieval of the phase without a well-defined reference wave is  a notoriously ill-posed inverse problem \cite{Zhang2024, Xiang2026}, and, given the state-of-the-art of established mathematical techniques, would have had virtually no chance of success. A novel way to tackle the problem has emerged with the advent of AI. Before addressing lens-based Fourier Imaging, the potential of AI-enabled phase retrieval was tested for lens-less in-line holography of 2D objects in transmission mode. First, successful phase retrieval and image reconstruction was demonstrated with both unsupervised and supervised DL \cite{xiang2024amplitude}. As sufficiently large THz image databases are still lacking for the pre-training of CNNs, emulated complex-valued 2D THz images were generated from a publicly accessible database of 70\,000 digitized photographs of handwritten digits. A key element for the successful phase retrieval was the simulation of the diffraction process with wave propagation models, either rooted in the Fresnel diffraction picture or the angular spectrum theory \cite{goodman2005introduction}. The simulation codes were combined with the CNN algorithms to an integrated physics-informed deep learning (PIDL) framework. The best image reconstruction was achieved with a combination of supervised DL followed by an unsupervised DL process. This sequential approach allowed to take care of unwanted standing-wave and spurious interference effects: As the pre-training of the supervised DL process included only forward propagation of the wave and no reflections, the processing of the measured diffraction images (which displayed a strong influence of unwanted standing-wave effects) via the pre-trained PIDL turned out to disregard the spurious interference signatures. This came at \hlpink{the} price of a reduction of reconstruction fidelity. The subsequent processing of the reconstructed image via unsupervised PIDL mended the errors to a considerable degree. Related interference cleaning effects were reported in the literature, e.g. for the case of biological samples on microscope slides, where CNN-based analysis eliminated self-interference effects incurred by the partial coherence of the visible-range light source
\cite{Rivenson2018}. This cleaning is an extremely useful feature because it solves otherwise intractable interference problems. 

Raising the stakes, subsequent studies tested the additional determination of the object distance \cite{xiang2025hybrid} as well as the reconstruction of 2D objects placed one behind the other with a lateral offset, thus one partly blocking the radiation impinging onto the other 
\cite{xiang2025reconstruction}. In the former case, supervised PIDL made distance determination possible, but the additional degree of freedom raised the requirements for the size of the dataset for pre-training. For future work, it is encouraging that also unsupervised PIDL with a multi-head CNN could be applied successfully, albeit with the need for additional information input about the scene. In this specific case, the scene was recorded twice, partly masked at different regions. In the future, one could use other approaches such as controlled variations of the wavelength, polarization or phase of the THz radiation, by the introduction of well-known diffraction elements, or by an on-axis shift of the FPA. Also the second problem, the reconstruction of partly obscured 2D objects, required additional information to be fed to the CNNs. Successful reconstruction was achieved by pre-training of the CNNs with synthetic images augmented by self-training (self-learning) with a limited amount of unlabeled measured data.

Encouraged by these results of inline holography, the next step was to make the transition to Fourier Imaging. There are significant differences between the two, conceptual and practical ones. On the conceptual side: While diffraction patterns in inline holography remain localized in the vicinity of the object's projection onto the detector, diffracted waves in Fourier Imaging are spread across the focal plane. And on the practical side: As THz lenses -- compared to optics for the VIS -- typically have large diameters and are \hlpink{strongly} curved, there may arise a need to take lens aberrations into account in the PIDL. This remains an issue to be addressed, possibly on the basis of approximate treatments as in \cite{Llombart2015, Dabironezare2021, Zhang_Llombart2024}. Based on conceptual arguments, there initially arose strong doubts whether phase retrieval is actually possible at all in the case of Fourier Imaging. The argument comes from the mathematics of Fourier transformation. If the mathematical descriptor of a planar object is Fourier transformed, then the information on the longitudinal or transverse position of the object resides entirely in the phase spectrum and none is in the amplitude spectrum. One can speak of complete amplitude-phase separation. This would imply for Fourier Imaging that it is impossible to retrieve phase information from the amplitude or intensity distribution in the focal plane. It turns out that this conclusion does not fully hold in practical systems. In realistic imaging configurations, various system-dependent effects influence the measured intensity distribution and its relation to the underlying wave field.
These effects may not be fully described by commonly used simplified propagation models -- such as the Fresnel or Fraunhofer approximations -- and therefore need to be considered differently in the reconstruction process. With appropriate modeling of such practical system characteristics, phase retrieval and image reconstruction can be achieved in Fourier imaging configurations (publication in preparation \cite{Yuan2026}).

The big challenges now are to extend the analysis to more complex scenes -- especially opaque and transparent 3D ones -- and to explore reflection-mode imaging which for many possible application scenarios may be the more relevant mode of operation. Many physics issues will arise -- such as the question of image reconstruction in the case of predominant specular reflections, or the treatment of internal reflections in transparent 3D objects. While the kind of PIDL approach which was taken until now may still be able to address these issues -- albeit foreseeably with a far larger effort regarding the implementation of the wave propagation model and the preparation of pre-training image datasets -- it may turn out to be a better approach to explore agentic AI approaches and follow similar paths as described in Sec. VI for the case of inverse metamaterial design \cite{lu2025agentic}. Here an agentic framework selectively used an increasing amount of data (from a pre-existing dataset) to train a forward model. Starting from a small initial number of elements, the \textit{Planner} LLM of the framework initiated and controlled the iterative growth of the dataset only to a size sufficient to reach the pre-set validation mean-square error (MSE). In this elegant way, the user does not have to make a more-or-less blind guess about how large his dataset should be. 
Considering THz imaging, one could proceed in a related, framework-controlled way of an iterative build-up of the image database, keeping in mind, however, that the generation of images for the pre-training will be numerically more costly. As much as possible, existing images should be used. In the near future, this will not be possible, because there are not enough available for pre-training yet, especially not 3D ones. A large number of synthetic images need to be produced, either from scratch based on mathematical object models, or by the conversion of VIS or IR images to artificial THz images, taking care in both cases to stay as close to the THz physical reality as possible. As successful \hlpink{image datasets} become available, they can be re-used for training tasks. 

The second advantage of the agentic approach is that the \textit{Inverse Designer} LLM autonomously modifies the CNNs in order to improve the code and thus to reduce the MSE. Regarding Fourier Imaging, the question is how far this can be pushed in the future. For example, an aspect where autonomous action could be very useful, is the correction for aberrations induced by the lens system. It would ease the burden on the design of the applied wave propagation model drastically if aberrations could be handled by a CNN and wouldn't have to be included in the model. Concerning such questions, one can only speculate at this time about possible answers. While agentic AI opens tremendous potential, it remains to be seen how far it can augment not only THz Fourier Imaging, but any form of THz imaging.

\subsection{Lessons from IR/VIS: data acquisition and image processing strategies} 
Computational imaging, consisting of joint optimization of physical measurement and image data processing, has emerged as structured discipline over the past half century~\cite{Brady2025}. Computational imaging first emerged in radar and X-ray systems, where data pixel densities and data loads are modest, but has increasingly informed visible and infrared imaging system design. Today dedicated image processing hardware and neural co-processors are essential components of mobile devices and head-mounted displays. Due to the more modest market size, these innovations have been slower to reach THz imaging systems.

Improved multisensor data fusion may be the most dramatic impact of computational imaging on THz system design. Traditional imaging systems build a physical forward model of the form $g=Hf$ describing the relationship between the object-generated field $f$ and the field measurements $g$. Image formation consists of mathematical inversion of this forward model. Emerging neural estimators, however, may invert forward models that combine diverse physical phenomena, such as a simultaneous visible, infrared, millimeter wave, THz and acoustic data set, to obtain a maximum likelihood estimate of an object. Such a system might use THz signals for atmospheric penetration and ranging, but infrared signals for cross-range imaging. 

Neural estimators also play a dramatic role in measurement system control. In IR-VIS systems the traditional control problem is pan-tilt-zoom, focus and exposure. An imaging system has many capture parameters, the photographer adjusts these parameters to capture an image. Automatic neural systems replace the photographer with systems optimized to capture the full ``light field.'' Rather than setting the system to a particular focus, an automated system may dynamically adjust focus, exposure and zoom to capture all-in-focus images with high \hlpink{dynamic} range and resolution~\cite{wang2021deep}. 

This same approach applies to THz imaging systems. A system has a range of illumination patterns and sensor receiver patterns. Rather than blindly adjusting these parameters to scan a scene, artificially intelligent control systems may dynamically adjust the capture system to maximize object information. These systems may also coordinate across sensor modalities. 

\section{A Reference-Anchored Technology Roadmap}
\label{sec:roadmap}

\begin{figure*}[ht]
\centering
\includegraphics[width=5in]{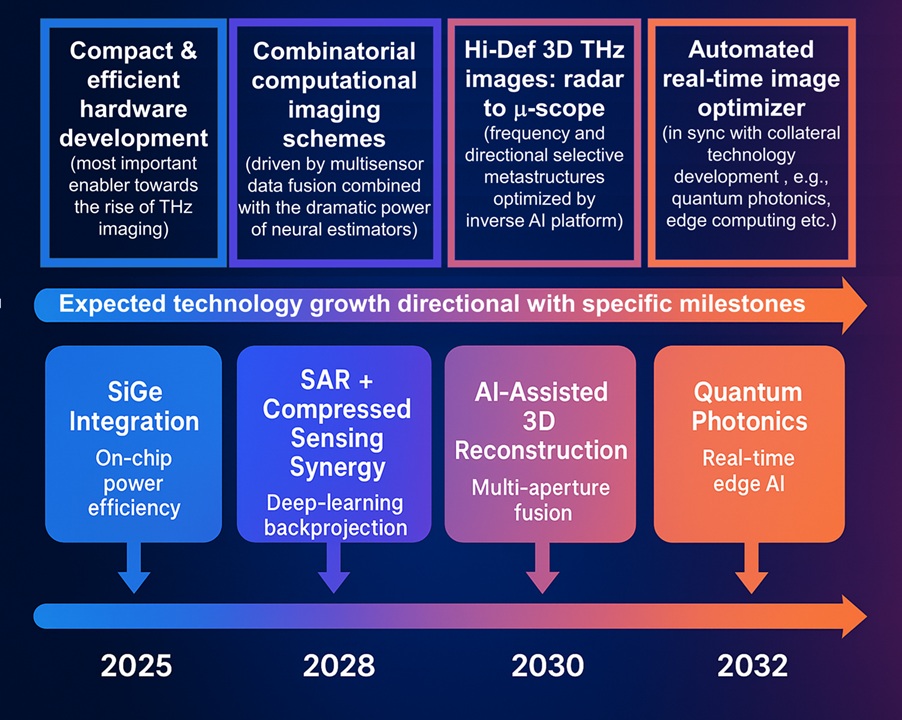}
    \caption{\hlyellow{Reference-anchored projection of the expected growth trajectory of active electronic THz imaging for industrial applications within the next decade. Milestones, dating methodology, readiness levels and uncertainty bounds are substantiated in Sec.}~\ref{sec:roadmap}\hlyellow{. Original artwork by the authors.}} 
    \label{Roadmap.jpg}
\end{figure*}

\subsection{Substantiation of the technology milestones}
\label{sec:roadmap-milestones}

\hlyellow{The technology trajectory summarized in Fig.}~\ref{Roadmap.jpg} \hlyellow{is not proposed as a universal roadmap for the THz imaging domain as a whole -- no single trajectory can serve applications as disparate as astronomical radiometry, biomedical diagnostics and industrial inspection. It is, instead, a reference-anchored projection for the system class on which this review concentrates: active, electronic, real-time-capable imaging for industrial quality control, security screening and emerging autonomous platforms. Each milestone block corresponds to a technology thrust with existing chip- or system-level demonstrations in the literature cited in the preceding sections, and each date marks the window in which we expect that capability to transition from laboratory demonstration to application-specific system integration (approximately technology readiness levels, TRL, 5--6). Below we substantiate each block, name its application drivers, and state the assumptions and uncertainty behind the timing.}

\hlyellow{The first block, SiGe/CMOS integration with on-chip power efficiency (2025), reflects capabilities demonstrated today and therefore serves as the calibrated anchor of the timeline rather than a prediction. SiGe BiCMOS and CMOS processes with characteristic frequencies approaching 500 GHz have enabled fully integrated FMCW transceivers} \cite{Li2017, Dobroiu2022}\hlyellow{, 220--320 GHz radar-on-chip architectures with improved linearity and phase noise} \cite{Taba2022}\hlyellow{, and monolithic focal-plane arrays reaching kilopixel counts} \cite{Hadi2013, Zda15, hillger_terahertz_2019, Hoogelander2023}\hlyellow{. On the source side, Koyama et al. at Canon Inc.} \cite{Koyama2022} \hlyellow{coherently combined 36 resonant-tunneling-diode oscillators on-chip, obtaining 10 mW at 0.45 THz, extended by the coupling schemes of Van Mai et al.} \cite{Tai2023} \hlyellow{and Meng et al.} \cite{Meng2024} \hlyellow{and the RTD waveguide amplification of Greenberg et al.} \cite{Greenberg2026}\hlyellow{. This baseline is already commercially load-bearing, as evidenced by the inspection, screening and 6G activities around the Canon platform} \cite{Canon2023, yet2} \hlyellow{and the 500+ GHz automotive imaging radar of TeraDar Inc.} \cite{Teradar1, Teradar2} \hlyellow{targeting the industry's 300-m range requirement.}

\hlyellow{The second block, SAR--compressive-sensing synergy with learned backprojection (2028), rests on ingredients each demonstrated in isolation: short-range SAR with sub-millimeter 3D resolution from GHz to THz frequencies} \cite{Batra2021}\hlyellow{, guided FMCW reflectometry} \cite{Pan2020}\hlyellow{, the compressive-sampling framework of Donoho} \cite{donoho06} \hlyellow{and Cand\`es et al.} \cite{candes06}\hlyellow{, compressive holography for 3D reconstruction from 2D measurements} \cite{cull2010millimeter}\hlyellow{, near-field compressive imaging down to} $\lambda$/100 \cite{Stantchev2017, Chen_Superres2019}\hlyellow{, large-FoV super-resolution CW compressive imaging} \cite{Fang:25}\hlyellow{, and deep-learning-assisted FMCW inspection of building materials} \cite{Tian2026}\hlyellow{. The 2028 placement reflects our assessment that the remaining step is not a missing component but the joint co-design of sampling hardware and reconstruction algorithms in one deployed system, for which the chip-scale FMCW platforms of the first block} \cite{Taba2022, Li2017} \hlyellow{provide the substrate. The drivers are non-destructive testing and inline process monitoring, where acquisition time, not sensitivity, is the binding constraint.}

\hlyellow{The third block, AI-assisted 3D reconstruction with multi-aperture and multi-sensor fusion (2030), is dated by an explicit methodology rather than intuition. As discussed in Sec.~III-B, Su et al.} \cite{Su2023} \hlyellow{have shown that THz computational imaging recapitulates the development sequence of optical computational imaging with a lag of roughly five to seven years. Applying this documented lag to capabilities established in the VIS/IR between 2023 and 2025 -- especially the uncertainty-aware Fourier ptychography and learned sensing of Chen et al.} \cite{Chen2025UncertaintyAwareFP} \hlyellow{and the end-to-end optimized diffractive networks of Lin et al.} \cite{Lin2018} \hlyellow{-- places robust THz counterparts of these research directions near the end of the decade. The THz-native groundwork exists: the physics-informed deep-learning studies of Xiang et al. demonstrated amplitude/phase retrieval for THz holography} \cite{xiang2024amplitude}\hlyellow{, depth estimation} \cite{xiang2025hybrid}\hlyellow{, reconstruction of partially obscured objects} \cite{xiang2025reconstruction} \hlyellow{and a unified treatment of the ill-posed inverse problem} \cite{Xiang2026}\hlyellow{, with the extension to lens-based Fourier imaging in preparation} \cite{Yuan2026}\hlyellow{; multi-sensor THz/IR/VIS fusion has been shown for inspection} \cite{Tokizane2021}\hlyellow{, tomography} \cite{Hung2024}\hlyellow{, concealed-object detection} \cite{Hadinejad2025} \hlyellow{and video-SAR} \cite{Fan2024}\hlyellow{. The quantitative target is the three- to ten-fold reduction in measurement number per unit reconstruction quality reported for learned illumination in the VIS/IR} \cite{Lin2018, Chen2025UncertaintyAwareFP}\hlyellow{, transferred to THz acquisition.}

\hlyellow{The final block, quantum photonics and real-time edge AI (2032), is deliberately labeled as the exploratory horizon and carries the largest uncertainty. Its inclusion is nevertheless grounded in measurable precedents already visible on the present research horizon: FMCW ranging approaching the quantum noise limit} \cite{Lu2025FMCWLiDAR}\hlyellow{, large-scale integrated photonic quantum computation} \cite{zhu2026photonic}\hlyellow{, quantum machine learning} \cite{peralgarcia2024qml}\hlyellow{, AI-assisted development of quantum hardware} \cite{alexeev2025aiqc}\hlyellow{, and photonic edge-AI accelerators} \cite{NeuraWave1}\hlyellow{. On the inference-cost side -- which, as elaborated in Sec.}~\ref{sec:roadmap-viability}\hlyellow{, will determine whether AI-heavy THz imaging is economically deployable -- Belcak et al.} \cite{belcak2025small} \hlyellow{and Sarker Bijoy et al.} \cite{sarker2025prost} \hlyellow{have shown that heterogeneous agentic architectures pairing a large orchestrator model with small specialized node models reduce memory bandwidth, latency and cost, complemented by CPU-based phase retrieval avoiding GPU-bound infrastructure} \cite{garisto2024cutting}\hlyellow{.}

\hlyellow{Two remarks on the validity of this technology roadmap are in order. First, the dates denote the earliest credible window for application-specific system-level demonstrations within the industrial scope of Sec.~I (estimated uncertainty of} $\pm$2 \hlyellow{years per block, growing toward the right of the timeline); they are not predictions of commercial maturity. Second, the trajectory is conditional on the stated application domains: passive radiometric, telescope-based astronomical and spectroscopic biomedical imaging follow different development logics and lie outside this projection. Within these bounds, every element of the figure is traceable to the demonstrated results cited here, and the figure should be read as a graphical summary of this analysis rather than an independent forecast.}

\subsection{Long-term viability and restructuring of AI augmentation}
\label{sec:roadmap-viability}

\hlyellow{In light of the prospective sweeping adoption of AI in THz imaging systems geared for industrial applications, a collateral trend deserves attention. The first phase of AI adoption in other industrial domains -- the implementation of large models across many workflows -- has remained largely exploratory, because the priority has been learning rather than optimization. The second phase of adoption is likely to be very different: many adopters will begin to ask a harder question -- which uses of AI actually create enough value to justify their cost of implementation? In the specific case of the computational THz imaging of tomorrow, GPU-hours in shared high-performance-computing centres may become a decisive metric. This shift will also change how end-users assess the potential application of THz imaging that depends heavily on AI adoption. Today, most AI infrastructure is still sold in terms of GPU-hours, which assumes that customers want access to homogeneous pools of expensive hardware. To keep the cost-prohibitive nature of this endeavor in check, phase retrieval may be executed on CPUs instead of GPUs} \cite{garisto2024cutting,Ding2023HybridGPU}\hlyellow{. A rapidly accelerating area of research involves the use of heterogeneous agentic architectures in which an LLM acts as the central orchestrator and small language models (SmLMs) serve as specialized node agents} \cite{belcak2025small}\hlyellow{. These agentic systems reserve the massive parametric models exclusively for the complex reasoning needed for orchestration and delegate more structured and repetitive tasks to SmLMs, which significantly reduces memory-bandwidth and computational requirements since SmLMs have a much smaller number of parameters. Heterogeneous agentic frameworks should therefore decrease latency and lower operational inference costs} \cite{belcak2025small, sarker2025prost}\hlyellow{.}

\hlyellow{In parallel to this present paradigm of AI infrastructure, it is also instructive to highlight the cross-pollination of two major aligned and overlapping technologies: the impact of AI techniques on developing and operating useful quantum computers (AI for quantum)} \cite{alexeev2025aiqc}\hlyellow{, and the slightly more speculative prospect of quantum computers enhancing AI techniques (often referred to as quantum for AI), combined with edge computing} \cite{NeuraWave1} \hlyellow{for real-time imaging applications} \cite{zhu2026photonic, peralgarcia2024qml}\hlyellow{. It is perhaps safe to say that, within the next decade, THz imaging enabled with automated real-time image-optimizing features -- especially for automated cooperative systems, such as collaborative robotics (cobotics) and automated transportation networks -- will combine several of these disruptive technology trends, as well as cross-over techniques from the IR and VIS imaging domains.}

\section{Conclusion}
\hlyellow{Active electronic THz imaging has come a long way across frequency bands, acquisition modalities and augmentation techniques; yet the systems demonstrated so far remain constrained by three inter-related bottlenecks: mechanical beam steering, which makes image formation slow; iterative phase retrieval, which is computationally heavy and expensive; and fixed optical hardware, which is spectrally narrow and inflexible in its spatial-frequency coverage.}

\hlyellow{This review has argued that these bottlenecks -- long treated as hardware problems -- are increasingly being dissolved at the interface between hardware and computation. The convergence of two transformative technologies, programmable THz metasurfaces and physics-informed deep learning, offers a concrete path: the end-to-end co-optimization of the illumination pattern and the neural-network reconstruction algorithm, in which the set of angular illuminations is not fixed by the experimenter's intuition but is learned from data to maximally reduce reconstruction uncertainty per unit measurement time. Early results in the VIS and IR ranges suggest that learned illumination can reduce the number of measurements required for a given reconstruction quality by a factor of three to ten} \cite{Lin2018, Chen2025UncertaintyAwareFP} \hlyellow{-- a dramatic speed-up with direct implications for real-time industrial THz imaging.}

\hlyellow{In line with the title of this review, our aim has therefore been twofold: to give an account of the present challenges of active THz imaging for industrial applications, and to substantiate -- through the reference-anchored roadmap of Sec.}~\ref{sec:roadmap} \hlyellow{-- the AI-driven computational route by which we expect these challenges to be resolved, in sync with collateral technology developments, over the coming decade.}

\section*{Acknowledgment}
\hlyellow{All figures -- except Fig. 3, an adapted version reproduced from} \cite{friederich2018radome} \hlyellow{-- in this article are original artwork designed and drawn manually by the authors; no generative-AI tools were used in their preparation.}

HGR acknowledges funding by the Deutsche Forschungsgemeinschaft, especially through recent grants RO 770\,/\,40, 43, 48, and 50, and by the Hessian LOEWE program.  AB, HY and HGR would like to thank Mark D Thomson for many helpful discussions. 

\ifCLASSOPTIONcaptionsoff
  \newpage
\fi
\bibliographystyle{IEEEtran}
\bibliography{IEEEabrv,Bibliography}

@IEEEtranBSTCTL{IEEEexample:BSTcontrol,
    CTLuse_article_number = "yes",
    CTLuse_paper = "yes",
    CTLuse_forced_etal = "no",
    CTLmax_names_forced_etal = "50",
    CTLnames_show_etal = "50",
    CTLuse_alt_spacing = "yes",
    CTLalt_stretch_factor = "4",
    CTLdash_repeated_names = "yes",
    CTLname_format_string = "{f.~}{vv~}{ll}{, jj}",
    CTLname_latex_cmd = "",
    CTLname_url_prefix = "[Online]. Available:"
 }

@book{Saeedkia2013,
  title={Handbook of Terahertz Technology for Imaging, Sensing and Communications},
  editor={D. Saeedkia},
  year={2013},
  publisher={Woodhead Publishing},
  isbn = {978-0-85709-235-9}
}

@book{Carpintero2015,
  title={Semiconductor TeraHertz Technology: Devices and Systems at Room Temperature Operation},
  author={G. Carpintero and E. Garcia-Munoz and H. Hartnagel and S. Preu and A. Raisanen},
  year={2015},
  publisher={Wiley-IEEE Press},
  isbn = {9781118920428}
  }

@book{Lee2009,
  title={Principles of Terahertz Science and Technology},
  author={Y.-S. Lee},
  year={2009},
  publisher={Springer},
  isbn = {978-1-4419-3491-8}
  }

@book{Das2022,
  title={Terahertz Devices, Circuits and Systems -- Materials, Methods and Applications},
  editor={S. Das and A. Nella and S. K. Patel},
  year={2022},
  publisher={Springer},
  isbn = {978-981-19-4107-8}
  }

@book{Pavlidis2021,
  title={Fundamentals of Terahertz Devices and Applications},
  editor={D. Pavlidis},
  year={2021},
  publisher={Wiley},
  isbn = {9781119460718}
  }

@book{Ghzaoui2023,
  title={Terahertz Wireless Communication Components and System Technologies},
  editor={M. {El Ghzaoui} and S. Das and T. R. Lenka},
  year={2023},
  publisher={Springer},
  isbn = { 9789811691843 }
  }

@book{You2022,
  title={Terahertz Technology},
  editor={B. You and J.-Y Lu},
  year={2022},
  publisher={Intech Open},
  isbn = {9922463511505561}
  }

@book{Song2015,
  title={Handbook of Terahertz Technologies: Devices and Applications},
  editor={H.-J. Song and T. Nagatsuma},
  year={2015},
  publisher={CRC Press},
  isbn = {9789814613088}
  }

@incollection{Zhang2025,
  title={Real-Time Terahertz Imaging},
  booktitle={High Throughput Imaging Technology},
  author={Y. Zhang and Y- T. Li},
  editor = {Z. J. Liu and Y. T. Li},
  pages = {311-318},
  year={2025},
  publisher={Springer},
  isbn = {978-981-96-1931-3},
  doi = {10.1007/978-981-96-1929-0}
  }

@book{Malhorta2021,
  title={Terahertz Antenna Technology for Imaging and Sensing Applications },
  author={I. Malhotra and G. Singh},
  year={2021},
  publisher={Springer},
  isbn = {978-3-030-68962-9}
  }

@incollection{Sri2022,
  title={Terahertz Imaging: Timeline and Future Prospects},
  author={G. Srivastava and S. Agarwal},
  booktitle={Terahertz Devices, Circuits and Systems -- Materials, Methods and Applications},
  editor = {S. Das and A. Nella and S. K. Patel},
  year={2022},
  publisher={Springer},
  doi = {10.1007/978-981-19-4105-4_16}
  }

@incollection{Saha2020,
  title={Advances in Terahertz Imaging},
  author={A. Saha},
  booktitle={Emerging Trends in Terahertz Solid-State Physics and Devices -- Sources, Detectors, Advanced Materials, and Light-matter Interactions},
  editor = {Biswas, A. and Banerjee, A. and Acharyya, A. and Inokawa, H. and Roy, J.},
  year={2020},
  publisher={Springer},
  doi = {10.1007/978-981-15-3235-1_10}
  }

@article{xiang2024amplitude,
  title={Amplitude/phase retrieval for terahertz holography with supervised and unsupervised physics-informed deep learning},
  author={Xiang, M. J. and Yuan, H. and Wang, L. X. and Zhou, K. and Roskos, H. G.},
  journal={IEEE Transactions on Terahertz Science and Technology},
  volume={14},
  number={2},
  pages={208--215},
  year={2024},
  publisher={IEEE}
}

@article{xiang2025hybrid,
  title={Hybrid multi-head physics-informed neural network for depth estimation in terahertz imaging},
  author={Xiang, M. J.  and Yuan, H. and Zhou, K. and Roskos, H. G.},
  journal={Computer Physics Communications},
  volume={312},
  pages={109586},
  year={2025},
  publisher={Elsevier}
}

@article{xiang2025reconstruction,
  title={Reconstruction of partially obscured objects with a physics-driven self-training neural network},
  author={Xiang, M. J. and Zhou, K. and Yuan, H. and Roskos, H. G.},
  journal={Optics Express},
  volume={33},
  number={10},
  pages={21482--21495},
  year={2025},
  publisher={Optica Publishing Group}
}

@article{Anitha2023,
  title={{THz} Imaging Technology Trends and Wide Variety of Applications: a Detailed Survey},
  author={Anitha, V. and Beohar, A. and Nella, A.},
  journal={Plasmonics},
  volume={18},
  number={},
  pages={441--483},
  year={2023},
  doi = {10.1007/s11468-022-01775-9}
}

@article{Grzyb2017,
  title={Solid-State Terahertz Superresolution Imaging Device in 130-nm {SiGe} {BiCMOS} Technology},
  author={Grzyb, J. and Heinemann, B. and Pfeiffer, U. R.},
  journal={IEEE Transactions on Microwave Theory and Techniques},
  volume={65},
  number={11},
  pages={4357--4372},
  year={2017},
  doi = {10.1109/TMTT.2017.2684120}
}

@article{Fleming2025,
  title={Virtually structured illumination for terahertz super-resolution imaging},
  author={Fleming, JP and Downes, LA and Girkin, JM and Weatherill, KJ},
  journal={Optics Express},
  volume={33},
  number={12},
  pages={26509-26516},
  year={2025},
  doi = {10.1364/OE.563675}
}

@article{Chen_Superres2019,
  title={Terahertz wave near-field compressive imaging with a spatial resolution of over $\lambda$/100},
  author={Chen, SC and Du, LH and Meng, K and Li, J and Zhai, ZH and Shi, QW and Li, ZR and Zhu, LG},
  journal={Optics Letters},
  volume={44},
  number={1},
  pages={21-24},
  year={2019},
  doi = {10.1364/OL.44.000021}
}

@article{Vazquez2024,
  title={Terahertz imaging super-resolution for documental heritage diagnostics},
  author={Vazquez, DA and Pilozzi, L and {DelRe}, E and Conti, C and Missori},
  journal={IEEE Transactions on Terahertz Science and Technology},
  volume={14},
  number={4},
  pages={455-465},
  year={2024},
  doi = {10.1109/TTHZ.2024.3410674}
}

@article{Zhao_Superres2026,
  title={Physics-enhanced neural network for terahertz rotating coherent forward-scattering full-field super-resolution imaging},
  author={Zhao, C. Y. and Wang, D. Y. and Zhao, J. and Wang, Y. X. and Rong, L. and Healy, J. L.},
  journal={Optics \& Laser Technology},
  volume={203},
  number={},
  pages={115563},
  year={2026},
  doi = {10.1016/j.optlastec.2026.115563}
}

@article{Li_Superres2024,
  title={Resolution enhancement in terahertz imaging with multi-wavelength information},
  author={Li, D. and Wang, X. K. and Zhang, Y.},
  journal={Optics Express},
  volume={32},
  number={19},
  pages={33369},
  year={2024},
  doi = {10.1364/OE.530610}
}

@article{Liu2010,
  title={A broadband quasi-optical terahertz detector utilizing a zero bias {Schottky} diode},
  author={Liu, L. and Hesler, J. L. and Xu, H. Y. and Lichtenberger, A. W. and Weikle, R. M.},
  journal={IEEE Microwave and Wireless Components Letters},
  volume={20},
  number={9},
  pages={504-506},
  year={2010},
  doi = {10.1109/LMWC.2010.2055553}
}

@article{Rahman2018,
  title={A {G-band} monolithically integrated quasi-optical zero-bias detector based on heterostructure backward diodes using submicrometer airbridges},
  author={Rahman, S. M. and Jiang, Z. G. and Shams, M. I. B. and Fay, P. and Liu, L.},
  journal={IEEE Transactions on Microwave Theory and Techniques},
  volume={66},
  number={4},
  pages={2010-2017},
  year={2018},
  doi = {10.1109/TMTT.2017.2779133}
}

@article{Shi2022,
  title={A 200 {GHz} fully integrated, polarization-resolved quasi-optical detector using zero-bias heterostructure backward diodes},
  author={Shi, Y. and Deng, Y. J. and Li, P. Z. and Fay, P. and Liu, L.},
  journal={IEEE Microwave and Wireless Components Letters},
  volume={32},
  number={7},
  pages={891-894},
  year={2022},
  doi = {10.1109/LMWC.2022.3155959}
}

@article{Meeker2018,
  title={Darkness: A microwave kinetic inductance detector integral field spectrograph for high-contrast astronomy},
  author={S. R. Meeker and B. A. Mazin and A. B. Walter and P. Strader and N. Fruitwala and C. Bockstiegel and P. Szypryt and G. Ulbricht and G. Coiffard and B. Bumble and G. Cancelo and T. Zmuda and K. Treptow and N. Wilcer and G. Collura and R. Dodkins and I. Lipartito and N. Zobrist and M. Bottom and J. C. Shelton and D. Mawet and J. C. van Eyken and G. Vasisht and E. Serabyn},
  journal={Publications of the Astronomical Society of the Pacific},
  volume={130},
  number={988},
  pages={065001},
  year={2018},
  doi = {10.1088/1538-3873/aab5e7}
}

@article{Reyes2026,
  title={{Amkid}: A large kid-based camera at the apex telescope},
  author={Reyes, N. and Weiss, A. and Yates, S. J. C. and Baryshev, A. M. and Cámara-Mayorga, I. and Dabironezare, S. and Endo, A. and Ferrari, L. and Görlitz, A. and Grutzeck, G. and Güsten, R. and Heiter, C. and Heyminck, S. and Hochgürtel, S. and Hoevers, H. and Jorquera, S. and Kovács, A. and Koopmans, D. and König, C. and Llombart, N. and Menten, K. M. and Murugesan, V. and  Ridder, M. and Schmitz, A. and Thoen, D. J. and {van der Linden}, A. J. and Wang, L. and Yurduseven, O. and Baselmans, J. J. A. and Klein, B.},
  journal={Astronomy and Astrophysics},
  volume={707},
  number={},
  pages={A294},
  year={2026},
  doi = {10.1051/0004-6361/202558596}
}

@article{Wilson2020,
  title={The {Toltec} camera: an overview of the instrument and in-lab testing results},
  author={G. W. Wilson and S. {Abi-saad} and P. Ade and I. Aretxaga and J. E. Austermann and Y. Ban and J. Bardin and J. A. Beall and M. Berthoud and S. A. Bryan and J. Bussan and E. {Castillo-Domínguez} and M. Chavez and R. Contente and N. W. {DeNigris} and B. Dober and M. Eiben and D. Ferrusca and L. Fissel and J. Gao and J. Golec and R. Golina and A. Gomez and S. Gordon and R. Gutermuth and G. Hilton and M. Hosseini and J. Hubmayr and D. Hughes and S. W. Kuczarski and D. Lee and E. Lunde and Z. Ma and H. Mani and P. Mauskopf and M. {McCrackan} and C. {McKenney} and J. {McMahon} and G. Novak and G. Pisano and A. Pope and A. Ralston and I. Rodriguez and D. {Sánchez-Argüelles} and F. P. Schloerb and S. M. Simon and A. Sinclair and K. Souccar and A. {Torres Camposana} and C. Tucker and J. Ullom and E. {Van Camp} and J. {Van Lanen} and M. Velazquez and M. Vissers and E. Weeks and M. S. Yun},
  journal={Proc. SPIE, Millimeter, Submillimeter, and Far-infrared Detectors and Instrumentation for Astronomy X},
  volume={11453},
  number={},
  pages={1145302},
  year={2020},
  doi = {10.1117/12.2562331}
}

@article{Lourie2018,
  title={Preflight characterization of the {BLAST-TNG} receiver and detector arrays},
  author={Lourie, N. P. and Ade, P. A. R. and Angile, F. E. and Ashton, P. C. and Austermann, J. E. and Devlin, M. J. and Dober, B. and Galitzki, N. and Gao, J. S. and Gordon, S. and Groppi, C. E. and Klein, J. and Hilton, G. C. and Hubmayr, J. and Li, D. and Lowe, I. and Mani, H. and Mauskopf, P. and {McKenney}, C. M. and Nati, F. and Novak, G. and Pascale, E. and Pisano, G. and Sinclair, A. and Soler, J. D. and Tucker, C. and Ullom, J. and Vissers, M. and Williams, P. A. and Paul A.},
   journal={Proc. SPIE, Millimeter, Submillimeter, and Far-infrared Detectors and Instrumentation for Astronomy IX},
  volume={10708},
  number={},
  pages={10708L},
  year={2018},
  doi = {10.1117/12.2314396}
}

@article{Luomahaara2021,
  title={A passive, fully staring {THz} video camera based on kinetic inductance bolometer arrays},
  author={Luomahaara, J. and Sipola, H. and Grönberg, L. and Mäyrä, A. and Aikio, M. and Timofeev, A. and Tappura, K. and Rautiainen, A. and Tamminen, A. and Vesterinen, V. and Leivo, M. and Gao, F. and Vasama, H. and Luukanen, A. and Hassel, J.},
   journal={IEEE Transactions on Terahertz Science and Technology},
  volume={11},
  number={1},
  pages={101-108},
  year={2021},
  doi = {10.1109/TTHZ.2020.3029949}
}

@article{Suzuki2014,
  title={Performance of {SAFARI} short-wavelength-band transition edge sensors {(TES)} fabricated by deep reactive ion etching},
  author={T. Suzuki and P. Khosropanah and R. A. Hijmering and M. Ridder and M. Schoemans and H. Hoevers and J. R. Gao},
   journal={IEEE Transactions on Terahertz Science and Technology},
  volume={4},
  number={2},
  pages={171-178},
  year={2014},
  doi = {10.1109/TTHZ.2014.2298376}
}

@article{Day2003,
  title={A broadband superconducting detector suitable for use in large arrays},
  author={P. K. Day and H. G. {LeDuc} and B. A. Mazin and A. Vayonakis and J. Zmuidzinas},
   journal={Nature},
  volume={425},
  number={6960},
  pages={917-821},
  year={2003},
  doi = {10.1038/nature02037}
}

@article{DeVisser2014,
  title={Fluctuations in the electron system of a superconductor exposed to a photon flux},
  author={P. J. {de Visser} and J. J. A. Baselmans and  J. Bueno and N. Llombart and T. M. Klapwijk},
   journal={Nature Communications},
  volume={5},
  number={},
  pages={3130},
  year={2014},
  doi = {10.1038/ncomms4130}
}

@article{Irwin1995,
  title={An application of electrothermal feedback for high resolution cryogenic particle detection},
  author={K. D. Irwin},
   journal={Applied Physics Letters},
  volume={66},
  number={15},
  pages={1998-2000},
  year={1995},
  doi = {10.1063/1.113674}
}

@article{Abramova2023,
  title={Investigation of blur kernel of terahertz images},
  author={Abramova, V and Abramov, S and Lukin, V and Grigelionis, I and Minkevicius, L and Valusis, G},
  journal={Lithuanian Journal of Physics},
  volume={63},
  number={3},
  pages={173-190},
  year={2023},
  doi = {10.3952/physics.2023.63.3.8}
}

@incollection{Malof2023,
  title={Forward and Inverse Design of Artificial Electromagnetic Materials},
  author={Malof, Jordan M. and Ren, Simiao and Padilla, Willie J.},
  booktitle={Advances in Electromagnetics Empowered by Artificial Intelligence and Deep Learning},
  pages={345--370},
  year={2023},
  publisher={IEEE Press, Wiley}
}

@article{smith2005gradient,
  title={Gradient index metamaterials},
  author={Smith, David Ryan and Mock, Jack J and Starr, AF and Schurig, David},
  journal={Physical Review E—Statistical, Nonlinear, and Soft Matter Physics},
  volume={71},
  number={3},
  pages={036609},
  year={2005},
  publisher={APS}
}

@article{schurig2006metamaterial,
  title={Metamaterial electromagnetic cloak at microwave frequencies},
  author={Schurig, David and Mock, Jack J and Justice, 1 BJ and Cummer, Steven A and Pendry, John B and Starr, Anthony F and Smith, David R},
  journal={Science},
  volume={314},
  number={5801},
  pages={977--980},
  year={2006},
  publisher={American Association for the Advancement of Science}
}

@article{cui2014coding,
  title={Coding metamaterials, digital metamaterials and programmable metamaterials},
  author={Cui, Tie Jun and Qi, Mei Qing and Wan, Xiang and Zhao, Jie and Cheng, Qiang},
  journal={Light: science \& applications},
  volume={3},
  number={10},
  pages={e218--e218},
  year={2014},
  publisher={Nature Publishing Group}
}

@article{ni2013metasurface,
  title={Metasurface holograms for visible light},
  author={Ni, Xingjie and Kildishev, Alexander V and Shalaev, Vladimir M},
  journal={Nature Communications},
  volume={4},
  number={1},
  pages={2807},
  year={2013},
  publisher={Nature Publishing Group UK London}
}

@article{arbabi2015dielectric,
  title={Dielectric metasurfaces for complete control of phase and polarization with subwavelength spatial resolution and high transmission},
  author={Arbabi, Amir and Horie, Yu and Bagheri, Mahmood and Faraon, Andrei},
  journal={Nature Nanotechnology},
  volume={10},
  number={11},
  pages={937--943},
  year={2015},
  publisher={Nature Publishing Group UK London}
}

@article{pendry2006controlling,
  title={Controlling electromagnetic fields},
  author={Pendry, John B and Schurig, David and Smith, David R},
  journal={Science},
  volume={312},
  number={5781},
  pages={1780--1782},
  year={2006},
  publisher={American Association for the Advancement of Science}
}

@article{Knap2004,
author = {Knap, W. and Lusakowski, J. and Parenty, T. and Bollaert, S. and Cappy, A. and Popov, V. V. and Shur, M. S.},
doi = {10.1063/1.1689401},
journal = {Applied Physics Letters},
number = {13},
pages = {2331--2333},
title = {Terahertz emission by plasma waves in 60 nm gate high electron mobility transistors},
volume = {84},
year = {2004}
}

@Article{Valusis2021,
AUTHOR = {Valušis, Gintaras and Lisauskas, Alvydas and Yuan, Hui and Knap, Wojciech and Roskos, Hartmut G.},
TITLE = {Roadmap of Terahertz Imaging 2021},
JOURNAL = {Sensors},
VOLUME = {21},
YEAR = {2021},
NUMBER = {12},
ARTICLE-NUMBER = {4092},
PubMedID = {34198603},
ISSN = {1424-8220},
DOI = {10.3390/s21124092}
}

@article{cybenko1989approximation,
  title={Approximation by superpositions of a sigmoidal function},
  author={Cybenko, George},
  journal={Mathematics of control, signals and systems},
  volume={2},
  number={4},
  pages={303--314},
  year={1989},
  publisher={Springer}
}

@ARTICLE{ZhangZL2023,
author={Zhang, Z. L. and Qi, P. F. and Guo, L. J. and Zhang, N. and Lin, L. and Liu, W. W.},
journal={Acta Optica Sinica},
title={Review on Super-Resolution Near-Field Terahertz Imaging Methods},
note ={(in Chinese)},
year={2023},
month={},
volume={43},
number={6},
pages={0600001},
doi={10.3788/AOS221632}
}

@article{Friederich2011,
author = {Friederich, F. and {von Spiegel}, W. and Bauer, M. and Meng, F. and  Thomson, M. D. and Boppel, S. and Lisauskas, A. and Hils, B. and Krozer, V. and Keil, A. and L\"{o}ffler, T. and Henneberger, R. and Huhn, A. K. and Spickermann, G. and {Haring Bolívar}, P. H. and Roskos, H. G.},
title = {{THz} Active Imaging Systems With Real-Time Capabilities},
journal = {IEEE Transactions on Terahertz Science and Technology},
year = {2011},
volume = {1},
number = {1},
pages = {183-200},
doi = {10.1109/TTHZ.2011.2159559}
}

@article{Spiegel2010,
author = {{von Spiegel}, W. and {am Weg}, C. and Henneberger, R. and Zimmermann, R. and  Roskos, H. G.},
title = {Illumination aspects in active terahertz imaging},
journal = {IEEE Transactions on Microwave Theory and Techniques},
year = {2010},
volume = {58},
number = {7},
pages = {2008-2013},
doi = {10.1109/TMTT.2010.2050247}
}

@article{Cooper2011,
author = {Cooper, K. B. and Dengler, R. J. and  Llombart, N. and Thomas, B. and Chattopadhyay, G. and Siegel, P. H.},
title = {{THz} Imaging Radar for Standoff Personnel Screening},
journal = {IEEE Transactions on Terahertz Science and Technology},
year = {2011},
volume = {1},
number = {1},
pages = {169-182},
doi = {10.1109/TTHZ.2011.2159556}
}

@article{Castilla2019,
author = {Castilla, S. and Terrés, B. and Autore, M. and Viti, L. and Li, J. and Nikitin, A.Y. and Vangelidis, I. and Watanabe, K. and Taniguchi, T. and Lidorikis, E. and Vitiello, M.S. and Hillenbrand, R. and Tielrooij, K.-J. and Koppens, F.H.L.},
journal = {Nano Letters},
pages = {2765–2773},
title = {Fast and Sensitive Terahertz Detection Using an Antenna-Integrated Graphene pn Junction},
volume = {19},
year = {2019},
doi = {10.1021/acs.nanolett.8b04171}
}

@article{Vicarelli2012,
author = {Vicarelli, L. and Vitiello, M. S. and Coquillat, D. and Lombardo, A. and Ferrari, A. C. and Knap, W. and Polini, M. and Pellegrini, V. and Tredicucci, A.},
doi = {10.1038/nmat3417},
journal = {Nature Materials},
number = {10},
pages = {865--871},
title = {{Graphene field-effect transistors as room-temperature terahertz detectors}},
volume = {11},
year = {2012}
}

@article{Koppens2014,
  title = {Photodetectors based on graphene, other two-dimensional materials and hybrid systems},
  author = {Koppens, F. H. L. and Mueller, T. and Avouris, Ph.  and Ferrari, A. C. and Vitiello,  M. S. and Polini,  M.},
  journal = {Nature Nanotechnology},
  volume = {9},
  issue = {},
  pages = {780–793},
  numpages = {},
  year = {2014},
  doi = {10.1038/nnano.2014.215},
}

@article{asgari_chip-scalable_2021,
	title = {Chip-{Scalable}, {Room}-{Temperature}, {Zero}-{Bias}, {Graphene}-{Based} {Terahertz} {Detectors} with {Nanosecond} {Response} {Time}},
	volume = {15},
	doi = {10.1021/acsnano.1c06432},
	language = {en},
	number = {11},
	urldate = {2022-06-09},
	journal = {ACS Nano},
	author = {Asgari, Mahdi and Riccardi, Elisa and Balci, Osman and De Fazio, Domenico and Shinde, Sachin M. and Zhang, Jincan and Mignuzzi, Sandro and Koppens, Frank H. L. and Ferrari, Andrea C. and Viti, Leonardo and Vitiello, Miriam S.},
	month = nov,
	year = {2021},
	pages = {17966--17976},
}

@article{Zak2014,
author = {Zak, Audrey and Andersson, Michael A. and Bauer, Maris and Matukas, Jonas and Lisauskas, Alvydas and Roskos, Hartmut G. and Stake, Jan},
doi = {10.1021/nl5027309},
journal = {Nano Letters},
number = {10},
pages = {5834--5838},
title = {Antenna-Integrated 0.6 {THz} {FET} Direct Detectors Based on {CVD} Graphene},
volume = {14},
year = {2014}
}

@article{Heinz2010,
author = {Heinz, E. and May, T. and Zieger, G. and Born, D. and Anders, S. and Thorwirth, G. and Zakosarenko, V. and Schubert, M. and Krause, T. and Starkloff, M. and  Kr\"{u}ger, A. and Schulz, M. and Bauer, F. and Meyer, H. G.},
title = {Passive Submillimeter-wave Stand-off Video Camera for Security Applications},
journal = {Journal of Infrared, Millimeter, and Terahertz Waves},
year = {2010},
volume = {31},
number = {11},
pages = {1355-1369},
doi = {10.1007/s10762-010-9716-y}
}

@article{Rowe2016,
author = {Rowe, S. and Pascale, E. and Doyle, S. and Dunscombe, C. and Hargrave, P. and Papageorgio, A. and Wood, K. and Ade, P. A. R. and Barry, P. and Bideaud, A. and Brien, T. and Dodd, C. and Grainger, W. and  House, J. and Mauskopf, P. and Moseley, P. and Spencer, L. and Sudiwala, R. and Tucker, C. and Walker, I.},
title = {A passive terahertz video camera based on lumped element kinetic inductance detectors},
journal = {Review of Scientific Instruments},
year = {2016},
volume = {87},
number = {3},
pages = {033105},
doi = {10.1063/1.4941661}
}

@article{Sheikh2016,
author = {Sheikh, F. and Zarifeh, N. and Kaiser, T.},
title = {Terahertz band: Channel modelling for
short-range wireless communications in the
spectral windows},
journal = {IET Microwaves, Antennas \& Propagation},
year = {2016},
volume = {10},
number = {13},
pages = {1435-1444},
doi = {10.1049/iet-map.2016.0022}
}

@article{Clochiatti2025,
author = {Clochiatti, S. and Grygoriev, A.and  Kress, R. and Mutlu, E. and Possberg, A. and Vogelsang, F. and {van Delden}, M. and Pohl, N. and Weimann, N. G.},
title = {Low-noise resonant tunneling diode terahertz detector},
journal = {IEEE Transactions on Terahertz Science and Technology},
year = {2025},
volume = {15},
number = {1},
pages = {107-119},
doi = {10.1109/TTHZ.2024.3505599}
}

@article{Suzuki2018,
author = {Suzuki, D. and Ochiai, Y. and Kawano, Y.},
title = {Thermal device design for a carbon nanotube terahertz camera},
journal = {ACS Omega},
year = {2018},
volume = {3},
number = {3},
pages = {3540-3547},
doi = {10.1021/acsomega.7b02032}
}

@article{Tokizane2021,
author = {Tokizane, Y. and Ejiri, H. and Minamikawa, T. and Suzuki, S. and  Asada, M. and Yasui, T.},
title = {Hybrid optical imaging with near-infrared, mid-infrared, and terahertz wavelengths for nondestructive inspection},
journal = {Applied Optics},
year = {2021},
volume = {60},
number = {10},
pages = {B100-B105},
doi = {10.1364/AO.415131}
}

@article{Takida2020,
author = {Takida, Y. and Suzuki, S. and Asada, M. and Minamide, H.},
title = {Sensitive terahertz-wave detector responses originated by negative differential conductance of resonant-tunneling-diode oscillator},
journal = {Applied Physics Letters},
year = {2020},
volume = {117},
number = {2},
pages = {021107},
doi = {10.1063/5.0012318}
}

@article{hillger_terahertz_2019,
	title = {Terahertz Imaging and Sensing Applications With Silicon-Based Technologies},
	volume = {9},
	doi = {10.1109/TTHZ.2018.2884852},
	number = {1},
	journal = {IEEE Transactions on Terahertz Science and Technology},
	author = {Hillger, Philipp and Grzyb, Janusz and Jain, Ritesh and Pfeiffer, Ullrich R.},
	year = {2019},
	pages = {1--19}
}

@ARTICLE{Rivenson2018,
author={Rivenson, Y. and Zhang, Y. B. and Günayd{\i}n, H. and Teng, D. and Ozcan, A.},
journal={Light-Science \& Application}, 
title={Phase recovery and holographic image reconstruction using deep learning in neural networks},
year={2018},
volume={7},
pages={17141},
doi={10.1038/lsa.2017.141}
}

@ARTICLE{Zhang2024,
author={Zhan, Z. K. and Wang, F. and Min, Q. X. and Jin, Y. and Situ, G. H.},
journal={Optics Letters}, 
title={Fourier phase retrieval using physics-enhanced
deep learning},
year={2024},
volume={49},
number={21},
pages={6129-6132},
doi={10.1364/OL.537792}
}

@ARTICLE{Llombart2015,
author={Nuria Llombart and Beatriz Blázquez and Angelo Freni and Andrea Neto},
journal={IEEE Transactions on Terahertz Science and Technology}, 
title={Fourier Optics for the Analysis of Distributed
Absorbers Under {THz} Focusing Systems},
year={2015},
volume={5},
number={4},
pages={573-583},
doi={10.1109/TTHZ.2015.2439511}
}

@ARTICLE{Dabironezare2021,
author={Dabironezare, S. O. and Carluccio, G. and  Freni, A. and Neto, A. and Llombart, N.},
journal={IEEE Transactions on Antennas and Propagation}, 
title={Coherent {Fourier} Optics Model for the Synthesis of Large Format Lens-Based Focal Plane Arrays},
year={2021},
volume={69},
number={2},
pages={734-746},
doi={10.1109/TAP.2020.3016501}
}

@ARTICLE{Zhang_Llombart2024,
author={Zhang, H. S. and Dabironezare, S. O. and Baselmans, J. J. A. and Llombart, N.},
journal={IEEE Transactions on Antennas and Propagation}, 
title={Focal Plane Array of Shaped Quartz Lenses for Wide Field-of-View Submillimeter Imaging Systems},
year={2024},
volume={72},
number={2},
pages={1263-1274},
doi={10.1109/TAP.2023.3334391}
}

@ARTICLE{Yuan2026,
author={Yuan, H. and Xiang, M. J. and Bandyopadhyay, A. and Roskos, H. G.},
journal={---}, 
title={{Phase} retrieval and object reconstruction in {Fourier Imaging} with physics-informed deep learning},
year = {unpublished}
}

@article{Javadi2021,
  title = {Sensitivity of field-effect transistor-based terahertz detectors. },
  author = {Javadi, E. and But, D. B. and Ikamas, K. and Zdanevičius, J. and Knap, W. and Lisauskas, A.},
  year = {2021},
  journal = {Sensors},
  volume = {21},
  number = {9},
  pages = {2909},
  doi = {10.3390/s21092909}
}

@article{PhysRevApplied.20.014003,
  title = {Nonlocal Superconducting Single-Photon Detector},
  author = {Paolucci, Federico},
  journal = {Physical Review Applied},
  volume = {20},
  issue = {1},
  pages = {014003},
  numpages = {11},
  year = {2023},
  month = {Jul},
  publisher = {American Physical Society},
  doi = {10.1103/PhysRevApplied.20.014003}
}

@article{gou2017spiral,
  title = {Spiral antenna-coupled microbridge structures for {THz} application},
  author = {Gou, Jun and Zhang, Tian and Wang, Jun and Jiang, Yadong},
  year = {2017},
  journal = {Nanoscale Research Letters},
  shortjournal = {Nanoscale Res Lett},
  volume = {12},
  number = {1},
  pages = {91},
  doi = {10.1186/s11671-017-1857-7}
}

@article{Pankratov2025,
  title = {Detection of single-mode thermal microwave photons using an underdamped {Josephson} junction},
  author = {A. L. Pankratov and A. V. Gordeeva and A. V. Chiginev and L. S. Revin and A. V. Blagodatkin and N. Crescini and L. S. Kuzmin},
  year = {2025},
  journal = {Nature Communications},
  volume = {16},
  number = {1},
  pages = {3457},
  doi = {10.1038/s41467-025-56040-4}
}

@article{MIAO2023112,
title = {A terahertz detector based on superconductor-graphene-superconductor Josephson junction},
journal = {Carbon},
volume = {202},
pages = {112-117},
year = {2023},
doi = {10.1016/j.carbon.2022.11.040},
author = {Miao, W. and Li, F. M. and Luo, Q. H. and Wang, Q. C. and Zhong, J. Q. and Wang, Z. and Zhou, K. M. and Ren, Y. and Zhang, W. and Li, J. and Shi, S. C.  and Yu, C. and He, Z. Z.  and Liu, Q. B. and Feng, Z. H.}
}

@article{Luo2018,
title = {Terahertz focal plane imaging array sensor based on {AlGaN/GaN} field effect transistors},
author = {Luo, M. C. and Sun, J. D. and Zhang, Z. P. and Li, X. and Shen, Z. H. and Wang Y. and Chen, H. B. and Dong X. F. and Zhang,  J. F. and Chen, Y. and Zhou, J. Y. and Qin, H.},
journal = {Infrared and Laser Engineering},
volume = {47},
number = {3},
pages = {0320001},
year = {2018},
doi = {10.3788/IRLA201847.0320001},
}

@article{ludwig2024terahertz,
  title = {Terahertz Detection with Graphene FETs: {{Photothermoelectric}} and resistive self-mixing contributions to the detector response},
  author = {Ludwig, Florian and Generalov, Andrey and Holstein, Jakob and Murros, Anton and Viisanen, Klaara and Prunnila, Mika and Roskos, Hartmut G.},
  year = {2024},
  journal = {ACS Applied Electronic Materials},
  volume = {6},
  number = {4},
  pages = {2197--2212},
  publisher = {American Chemical Society},
  doi = {10.1021/acsaelm.3c01511}
}

@article{ludwig2024model,
  title = {Modeling of antenna-coupled {Si} {MOSFETs} in the terahertz frequency range},
  author = {Ludwig, Florian and Holstein, Jakob and Krysl, A. and Lisauskas, A. and Roskos, Hartmut G.},
  year = {2024},
  journal = {IEEE Transactions on Terahertz Science and Technology},
  volume = {14},
  number = {3},
  pages = {414-423},
  doi = {10.1109/TTHZ.2024.3388254}
}

@article{Andree2022,
  title = {Broadband modeling, analysis, and characterization of {SiGe} {HBT} terahertz direct detectors},
  author = {Andree, M. and Grzyb, J. and Jain, R. and Heinemann, B. and Pfeiffer, U. R.},
  year = {2022},
  journal = {IEEE Transactions on Microwave Theory and Techniques},
  volume = {70},
  number = {2},
  pages = {1314-1333},
  doi = {10.1109/TMTT.2021.3134646}
}

@article{Sizov2010,
  title = {{THz} detectors},
  author = {Sizov, F. and Rogalski, A.},
  year = {2010},
  journal = {Progress in Quantum Electronics},
  volume = {34},
  number = {5},
  pages = {278-347},
  doi = {10.1016/j.pquantelec.2010.06.002}
}

@article{Rogalski2011,
  title = {{THz} detectors and focal plane arrays},
  author = {Rogalski, A. and Sizov, F.},
  year = {2011},
  journal = {{Opto-Electronics Review}},
  volume = {19},
  number = {3},
  pages = {346-404},
  doi = {10.2478/s11772-011-0033-3}
}

@article{Wei2025,
  title = {High-performance {Teraherz} photodetection in {2D} materials and topological materials},
  author = {Wei, Y. D. and Bao, Z. W. and Wu, H. F. and Zhang, Y. D. and Wen, Y. F. and Hu, Z. and Pan, X. K. and Lan, S. Q. and Zhang, L. B. and Wang, L. and Chen, X. S.},
  year = {2025},
  journal = {Journal of Physics D: Applied Physics},
  volume = {58},
  number = {7},
  pages = {073002},
  doi = {10.1088/1361-6463/ad93e2}
}

@article{Thomson2024,
  title = {Coherent terahertz detection via ultrafast dynamics of hot {Dirac} fermions in graphene},
  author = {Thomson, M. D. and Ludwig, F. and Holstein, J. and {Al-Mudhafar}, R. and {Al-Daffaie}, S. and Roskos, H. G.},
  year = {2024},
  journal = {ACS Nano},
  volume = {18},
  number = {6},
  pages = {4765-4774},
  doi = {10.1021/acsnano.3c08731}
}

@article{Jumaah2023,
  title = {Novel antenna-coupled terahertz photodetector with graphene nanoelectrodes},
  author = {Jumaah, A. J. and Roskos, H. G. and {Al-Daffaie}, S.},
  year = {2023},
  journal = {APL Photonics},
  volume = {8},
  number = {2},
  pages = {026103},
  doi = {10.1063/5.0127264}
}

@article{Delgado2024,
  title = {Room-temperature plasmon-assisted resonant {THz} detection in single-layer graphene transistors},
  author = {Caridad, J. M. and Castelló, O. and Baptista, S. M. L. and Taniguchi, T. and Watanabe, K. and Roskos, H. G. and Delgado-Notario, J. A.},
  year = {2024},
  journal = {Nano Letters},
  volume = {24},
  number = {3},
  pages = {935-942},
  doi = {10.1021/acs.nanolett.3c04300}
}

@article{Komiyama2011,
  title = {Single-photon detectors in the terahertz range},
  author = {Komiyama, S.},
  year = {2011},
  journal = {IEEE Journal of Selected Topics in Quantum Electronics},
  volume = {17},
  number = {1},
  pages = {54-66},
  doi = {10.1109/JSTQE.2010.2048893}
}

@article{Hashiba2006,
  title = {Isolated quantum dot in application to terahertz photon counting},
  author = {Hashiba, H and Antonov, V and Kulik, L. and Tzalenchuk, A. and Kleinschmid, P. and Giblin, S. and Komiyama, S.},
  year = {2006},
  journal = {Physical Review B},
  volume = {73},
  number = {8},
  pages = {081310},
  doi = {10.1103/PhysRevB.73.081310}
}

@article{Karasik2007,
  title = {Record-low {NEP} in hot-electron titanium nanobolometers},
  author = {Karasik, B. S. and Olaya, D. and Wei, J. and Pereverzev, S. and Gershenson, M. E. and Kawamura, J. H. and {McGrath}, W. R. and Sergeev, A. V.},
  year = {2007},
  journal = {IEEE Transactions on Applied Superconductivity},
  volume = {17},
  number = {2},
  pages = {293-297},
  doi = {10.1109/TASC.2007.897167}
}

@article{Hirakawa2023,
  title = {Terahertz detectors using microelectromechanical system resonators},
  author = {Li, C. and Zhang, Y. and Hirakawa, K.},
  year = {2023},
  journal = {Sensors},
  volume = {23},
  number = {13},
  pages = {5938},
  doi = {10.3390/s23135938}
}

@article{Hirakawa2025,
  title = {Uncooled, broadband terahertz bolometers using {SOI MEMS} beam resonators with piezoresistive readout},
  author = {Zhang, Y. and Ebata, K. and Iimori, M. and Liu, Q. and Zhao, Z. H. Takeuchi, R. and  Li, H. and Maenaka, K. and   Hirakawa, K.},
  year = {2025},
  journal = {Microsystems \& Nanoengineering},
  volume = {11},
  number = {1},
  pages = {132},
  doi = {10.1038/s41378-025-00996-2}
}

@article{Viti2021,
title = {Thermoelectric graphene photodetectors with sub-nanosecond response times at terahertz frequencies},
author = {Leonardo Viti and Alisson R. Cadore and Xinxin Yang and Andrei Vorobiev and Jakob E. Muench and Kenji Watanabe and Takashi Taniguchi and Jan Stake and Andrea C. Ferrari and Miriam S. Vitiello},
pages = {89--98},
volume = {10},
number = {1},
journal = {Nanophotonics},
doi = {doi:10.1515/nanoph-2020-0255},
year = {2021}
}

@article{Fatimy2006,
  title = {Terahertz detection by {GaN/AlGaN} transistors},
  author = {{El Fatimy}, A. and Tombet, S. B. and  Teppe, F. and Knap, W. and Veksler, D. B. and Rumyantsev, S. and Shur, M. S. and Pala, N. and Gaska, R. and Fareed, Q. and Hu, X. and Seliuta, D. and Valusis, G. and Gaquiere, C. and Theron, D. and Cappy, A.},
  year = {2006},
  journal = {Electronics Letters},
  volume = {42},
  number = {23},
  pages = {1342-1344},
  doi = {10.1049/el:20062452}
}

@article{Lin2020,
  title = {Heterodyne terahertz detection through electronic and optoelectronic mixers},
  author = {Lin, Y. J. and Jarrahi, M.},
  year = {2020},
  journal = {Reports on Progress in Physics},
  volume = {83},
  number = {6},
  pages = {066101},
  doi = {10.1088/1361-6633/ab82f6}
}

@article{Feng2022,
  title = {Heterodyne terahertz detection based on antenna-coupled {AlGaN/GaN} high-electron-mobility transistor},
  author = {Feng, W. and Zhu, Y. F. and Ding, Q. F. and Zhu, K. Q. and Sun, J. D. and Zhang, J. F. and Li, X. X. and Shangguan, Y. and Jin, L. and Qin, H.},
  year = {2022},
  journal = {Applied Physics Letters},
  volume = {120},
  number = {5},
  pages = {051103},
  doi = {10.1063/5.0063650}
}

@article{Lippmann1908,
  title = {Épreuves réversibles. {Photographies} intégrales},
  author = {Lippmann, G.},
  year = {1908},
  journal = {Comptes Rendues Hebdomadaires des Séances de l'Académie des Sciences},
  volume = {146},
  number = {9},
  pages = {446-451},
  doi = {}
}

@article{Adelson1992,
  title = {Single lens stereo with a plenoptic camera},
  author = {Adelson, E. H. and Wang, J. Y. A.},
  year = {1992},
  journal = {IEEE Transactions on Pattern Analysis and Machine Intelligence},
  volume = {14},
  number = {2},
  pages = {99-106},
  doi = {10.1109/34.121783}
}

@article{Ding2023,
  title = {A room-temperature, low-impedance and high-{IF}-bandwidth terahertz heterodyne detector based on bowtie-antenna-coupled {AlGaN/GaN} {HEMT}},
  author = {Ding, Q. F. and Zhu, Y. F. and Xiang, L. Y. and Zhang, J. F. and Li, X. X. and Jin, L. and Shangguan, Y. and Sun, J. D. and Qin, H.},
  year = {2023},
  journal = {Applied Physics Express},
  volume = {16},
  number = {2},
  pages = {024002},
  doi = {10.35848/1882-0786/acb4de}
}

@article{Jang2019,
  title = {Spectral characterization of a microbolometer focal plane array at terahertz frequencies},
  author = {Jang, D. and Kimbrue, M. and Yoo, Y. J. and Kim, K. Y.},
  year = {2019},
  journal = {IEEE Transactions on Terahertz Science and Technology},
  volume = {9},
  number = {2},
  pages = {150-154},
  doi = {10.1109/TTHZ.2019.2893573}
}

@incollection{Simoens2013,
  title = {{THz} bolometer detectors},
  booktitle = {{Physics and Applications of Terahertz Radiation}},
  Publisher = {{Springer Series in Optical Sciences}, vol. 173},
  editor = {Perenzoni, M. and Paul, D. J.},
  author = {Simoens, F.},
  year = {2013},
  pages = {35-75},
  doi = {10.1007/978-94-007-3837-9_2}
}

@article{Seifert2020,
  title = {Magic-angle bilayer graphene nanocalorimeters: toward broadband, energy-resolving single photon detection},
  author = {Seifert, P. and Lu, X. and Stepanov, P. and Retamal, J. R. D. and Moore, J. N. and Fong, K. C. and Principi, A. and Efetov, D. K.},
  year = {2020},
  journal = {Nano Letters},
  volume = {20},
  number = {5},
  pages = {3459-3464},
  doi = {10.1021/acs.nanolett.0c00373}
}

@article{Echternach2018,
  title = {Single photon detection of 1.5 {THz} radiation with the quantum capacitance detector},
  author = {Echternach, P. M. and Pepper, B. J. and Reck, T.  and Bradford, C. M.},
  year = {2018},
  journal = {Nature Astronomy},
  volume = {2},
  number = {1},
  pages = {90-97},
  doi = {10.1038/s41550-017-0294-y}
}

@article{Schwenson2025,
  title = {Sparse terahertz frequency-domain sensing with kilohertz measurement rate},
  author = {Schwenson, L. and Walter, F. and Jaeckel, A. and Wenzel, K. and Liebermeister, L. and Mach, C. and Castro-Camus, E. and Koch, M. and Schell, M. and Kohlhaas, R. B.},
  year = {2025},
  journal = {Journal of Infrared, Millimeter, and Terahertz Waves},
  volume = {46},
  number = {8},
  pages = {52},
  doi = {10.1007/s10762-025-01072-6}
}

@article{Sizov2018,
  title = {Terahertz radiation detectors: the state-of-the-art},
  author = {Sizov, F.},
  year = {2018},
  journal = {Semiconductor Science and Technology},
  volume = {33},
  number = {12},
  pages = {123001},
  doi = {10.1088/1361-6641/aae473}
}

@article{HE2024118886,
title = {Carbon-based self-powered terahertz detector with multi-walled carbon nanotubes-graphene heterostructure},
journal = {Carbon},
volume = {221},
pages = {118886},
year = {2024},
issn = {0008-6223},
doi = {10.1016/j.carbon.2024.118886},
author = {Yuan He and Nanxin Fu and Mengjie Jiang and Xuyang Lv and Shuguang Guo and Li Han and Libo Zhang and Bin Zhao and Gang Chen and Xiaoshuang Chen and Lin Wang}
}

@article{fan2020broadband,
	title = {Broadband {THz} absorption of microbolometer array integrated  with split-ring resonators},
	volume = {15},
	issn = {1556-276X},
	doi = {10.1186/s11671-020-03454-2},
	language = {en},
	number = {1},
	urldate = {2024-10-15},
	journal = {Nanoscale Research Letters},
	author = {Fan, Shuming and Gou, Jun and Niu, Qingchen and Xie, Zheyuan and Wang, Jun},
	month = dec,
	year = {2020},
	pages = {223},
}

@INPROCEEDINGS{5615383,
  author={Rui, Yao and Sheng-hui, Ding and Qi, Li and Qi, Wang},
  booktitle={2010 Academic Symposium on Optoelectronics and Microelectronics Technology and 10th Chinese-Russian Symposium on Laser Physics and Laser TechnologyOptoelectronics Technology (ASOT)}, 
  title={Real-time {THz} reflection imaging using a pyroelectric array camera}, 
  year={2010},
  volume={},
  number={},
  pages={125-127},
  doi={10.1109/RCSLPLT.2010.5615383}}

@article{hack2016comparison,
	title = {Comparison of {Thermal} {Detector} {Arrays} for {Off}-{Axis} {THz} {Holography} and {Real}-{Time} {THz} {Imaging}},
	volume = {16},
	copyright = {https://creativecommons.org/licenses/by/4.0/},
	doi = {10.3390/s16020221},
	language = {en},
	number = {2},
	urldate = {2024-10-15},
	journal = {Sensors},
	author = {Hack, Erwin and Valzania, Lorenzo and Gäumann, Gregory and Shalaby, Mostafa and Hauri, Christoph and Zolliker, Peter},
	month = feb,
	year = {2016},
	pages = {221}}

@INPROCEEDINGS{9365832,
  author={Jain, Ritesh and Hillger, Philipp and Grzyb, Janusz and Ashna, Eamal and Jagtap, Vishal and Zatta, Robin and Pfeiffer, Ullrich R.},
  booktitle={2021 IEEE International Solid-State Circuits Conference (ISSCC)}, 
  title={A 32×32 Pixel 0.46-to-0.75{THz} Light-Field Camera {SoC} in 0.13 micron {CMOS}}, 
  year={2021},
  volume={64},
  number={},
  pages={484-486},
  doi={10.1109/ISSCC42613.2021.9365832}}

@article{Kutaish2024,
author = {Kutaish, A. and Conde, M. H. and Pfeiffer, U.},
journal = {IEEE Sensors Letters},
pages = {536--538},
title = {Coarse-to-fine sparse {3-D} reconstruction in {THz} light field imaging},
volume = {8},
number = {10},
year = {2024},
doi = {10.1109/LSENS.2024.3454567}
}

@article{Jain2016,
author = {Jain, R. and Grzyb, J. and Pfeiffer, U. R.},
journal = {IEEE Transactions on Terahertz Science and Technology},
pages = {649-657},
title = {Terahertz light-field imaging},
volume = {6},
number = {5},
year = {2016},
doi = {10.1109/TTHZ.2016.2584861}
}

@article{Niu2025,
author = {Niu, Z. T. and Wang, Q. W. and Ren, Y. T. and He, M. J. and Gao, B. H. and Li, Z. H. and Qi, H. and Zhang, B.},
journal = {Measurement Science and Technology},
pages = {092002},
title = {A review on the latest development of light field imaging in flow field and temperature field measurement
},
volume = {36},
number = {9},
year = {2025},
doi = {10.1088/1361-6501/adfaf9}
}

@article{Lin2019,
author = {Lin, RJ and Su, VC and Wang, SM and Chen, MK and Chung, TL and Chen, YH and Kuo, HY and Chen, JW and Chen, J and Huang, YT and Wang, JH and Chu, CH and Wu, PC and Li, T and Wang, ZL and Zhu, SN and Tsai, DP},
journal = {Nature Nanotechnology},
pages = {227},
title = {Achromatic metalens array for full-colour light-field imaging},
volume = {14},
number = {3},
year = {2019},
doi = {10.1038/s41565-018-0347-0}
}

@article{Liu_Lightfield2024,
author = {Liu, F and Wang, YL and Yang, Q and Zhou, SB and Zhang, KB},
journal = {IET Computer Vision},
pages = {1269-1284},
title = {A comprehensive research on light field imaging: {Theory} and
application},
volume = {18},
number = {8},
year = {2024},
doi = {10.1049/cvi2.12321}
}

@inproceedings{Rodriguez2012,
author = {{Rodríguez-Ramos}, LF and Montilla, I and Lüke, JP and López, R and {Marichal-Hernández}, JG and {Trujillo-Sevilla}, J and Femenía, B and López, M and {Fernández-Valdivia}, JJ and Puga, M and Rosa, F and {Rodríguez-Ramos}, JM},
booktitle = {SPIE Proc. on Three-Dimensional Imaging, Visualization and Display},
volume = {8384},
title = {Atmospherical wavefront phases using the plenoptic sensor (real data)},
year = {2012},
doi = {10.1117/12.923662}
}

@inproceedings{Dansereau2013,
author = {Dansereau, D. G. and Pizarro, O. and Williams, S. B.},
booktitle = {2013 IEEE Conference on Computer Vision and Pattern  Recognition (CVPR)},
pages = {1027-1034},
title = {Decoding, calibration and rectification for lenselet-based plenoptic cameras},
year = {2013},
doi = {10.1109/CVPR.2013.137}
}

@article{Boppel:12,
author = {Sebastian Boppel and Alvydas Lisauskas and Alexander Max and Viktor Krozer and Hartmut G. Roskos},
journal = {Optics Letters},
number = {4},
pages = {536--538},
title = {{CMOS} detector arrays in a virtual 10-kilopixel camera for coherent terahertz real-time imaging},
volume = {37},
year = {2012},
doi = {10.1364/OL.37.000536}
}

@article{Jepsen1995,
author = {Jepsen, P. U. and Keiding, S. R.},
journal = {Optics Letters},
volume = {20},
number = {8},
pages = {807-809},
title = {Radiation patterns from lens-coupled terahertz antennas},
year = {1995},
doi = {10.1364/OL.20.000807}
}

@INPROCEEDINGS{8310362,
  author={Hillger, Philipp and Jain, Ritesh and Grzyb, Janusz and Mavarani, Laven and Heinemann, Bernd and Grogan, Gaetan Mac and Mounaix, Patrick and Zimmer, Thomas and Pfeiffer, Ullrich},
  booktitle={2018 IEEE International Solid-State Circuits Conference - (ISSCC)}, 
  title={A 128-pixel 0.56{THz} sensing array for real-time near-field imaging in 0.13 micron {SiGe} {BiCMOS}}, 
  year={2018},
  volume={},
  number={},
  pages={418-420},
  doi={10.1109/ISSCC.2018.8310362}}

@INPROCEEDINGS{9567181,
  author={Liebchen, T. and Dischke, E. and Rämer, A. and Müller, F. and Schellhase, L. and Chevtchenko, S. and Heinrich, W. and Krozer, V.},
  booktitle={2021 46th International Conference on Infrared, Millimeter and Terahertz Waves (IRMMW-{THz})}, 
  title={Compact 12×12-Pixel {THz} Camera using AlGaN/GaN HEMT Technology Operating at Room Temperature}, 
  year={2021},
  volume={},
  number={},
  pages={1-2},
  doi={10.1109/IRMMW-THz50926.2021.9567181}}

@INPROCEEDINGS{8538071,
  author={Jiang, Mei and Lu, Xunxun and Xuan, Xiaobo and Han, Rubing and Wang, Min},
  booktitle={2018 IEEE Asia-Pacific Conference on Antennas and Propagation (APCAP)}, 
  title={A {THz} Slotted-Waveguide Array Antenna Based on {MEMS} Technology}, 
  year={2018},
  volume={},
  number={},
  pages={238-240},
  doi={10.1109/APCAP.2018.8538071}}

@INPROCEEDINGS{Yokoyama2019,
  author={Yokoyama, S. and Ikebe, M. and Kanazawa, Y. and Ikegami, T. and Ambalathankandy, P. and Hiramatsu, S. and Sano, E. and Takida, Y. and Minamide, H.},
  booktitle={IEEE International Solid-State Circuits Conference (ISSCC)}, 
  title={A {32x32}-pixel {0.9 THz} imager with pixel-parallel 12b {VCO}-based {ADC} in 0.18$\mu$m {CMOS}}, 
  year={2019},
  volume={},
  number={},
  pages={238-240},
  doi={10.1109/ISSCC.2019.8662483}
  }

@misc{i2s,
  author = {{TeTechS Inc.}},
  url = {https://www.terahertzstore.com/store/products/terahertz-camera-for-thz-imaging-cea-leti-tzcam.htm},
  urldate = {March 17, 2026},
  title = {Terahertz Camera with 320$\times$240 pixel sensor}
}

@misc{Tydex,
  author = {Tydex},
  url = {https://www.tydexoptics.com/pdf/THz_AR_Coatings.pdf},
  urldate = {April 20, 2026},
  title = {{THz AR Coatings}}
}

@misc{swiss,
  author = {{Swiss Terahertz}},
  url = {https://www.swissterahertz.com/rigicamera},
  urldate = {March 17, 2026},
  title = {{RIGI} Camera}
}

@misc{Teraphysics,
  author = {{Teraphysics}},
  url = {https://teraphysics.com},
  urldate = {April 18, 2026},
  title = {{A mm-wave Enabling Amplifier Platform Technology}}
}

@article{Lisauskas2013,
author = {Lisauskas, Alvydas and Boppel, Sebastian and Matukas, Jonas and Palenskis, Vilius and Minkevi{\v{c}}ius, Linas and Valu{\v{s}}is, Gintaras and {Haring Bol{\'{i}}var}, Peter and Roskos, Hartmut G.},
doi = {10.1063/1.4802208},
journal = {Applied Physics Letters},
pages = {153505},
title = {{Terahertz responsivity and low-frequency noise in biased silicon field-effect transistors}},
volume = {102},
number = {15},
year = {2013}
}

@article{Zatta2021,
author = {Zatta, R. and Jain, R. and Grzyb, J. and Pfeiffer, U. R.},
doi = {10.25926/b8y6-3x02},
journal = {IEEE Transactions on Terahertz Science and Technology},
pages = {277-286},
title = {Resolution limits of hyper-hemispherical silicon lens-integrated {THz} cameras employing geometrical multiframe super-resolution imaging},
volume = {11},
number = {3},
year = {2021}
}

@article{Tauk2006,
author = {Tauk, R. and Teppe, F. and Boubanga, S. and Coquillat, D. and Knap,  W. and Meziani, Y. M. and Gallon, C.  and Boeuf, F. and Skotnicki, T. and  Fenouillet-Beranger, C. and Maude, D. K. Rumyantsev, S. and Shur, M. S.},
journal = {Applied Physics Letters},
pages = {253511},
title = {Plasma wave detection of terahertz radiation by silicon field effects
transistors: {R}esponsivity and noise equivalent power},
volume = {89},
year = {2006}
}

@misc{ino,
  author = {{INO}},
  url = {https://www.ino.ca/en/solutions/thz/microxcam-384i-thz/},
  urldate = {March 17, 2026},
  title = {{MICROXCAM-384i-THz} Terahertz Camera}
}

@misc{terascense,
  author = {{TeraSense Group}},
  year = {2008-2022},
  url = {https://terasense.com/products/sub-thz-imaging-cameras/},
  urldate = {October 15, 2024},
  title = { Terahertz cameras}
}

@misc{FBH,
  author = {{Ferdinand-Braun-Institut}},
  year = {2018},
  url = {https://www.fbh-berlin.de/forschung/forschungsnews/gan-hemt-based-thz-detectors},
  urldate = {March 17, 2026},
  title = {{GaN} {HEMT}-based {THz} detectors}
}

@misc{ohio,
  author = {{ElectroScience Laboratory, Ohio State University}},
  year = {2024},
  url = {https://electroscience.osu.edu/research/terahertz},
  urldate = {October 15, 2024},
  title = {Terahertz}
}

@misc{GeGa-QMC,
  author = {{THz - QMC Instruments}},
  url = {https://www.terahertz.co.uk},
  urldate = {April 17, 2026},
  title = {Gallium doped {Germanium (Ge:Ga)} photoconductive detector}
}

@misc{ticwave,
  author = {{TicWave Solutions}},
  url = {https://ticwave-solutions.com/index.php/products/manufacturer/tws},
  urldate = {March 17, 2026},
  title = {Cameras}
}

@misc{Teradar1,
  author = {{America's seed fund}},
  url = {https://www.sbir.gov/portfolio/1953529},
  urldate = {March 17, 2026},
  title = {TeraDar, Inc.}
}

@misc{Teradar2,
  author = {{IEEE Spectrum}},
  url = {https://spectrum.ieee.org/terahertz-radar},
  urldate = {November 20, 2025},
  title = {Could Terahertz Radar in Cars Save Lives? {Startup} {Teradar’s} tech combines lidar’s precision with radar’s simplicity}
}

@misc{Huebner,
  author = {{Hübner Photonics}},
  url = {https://hubner-photonics.com/products/terahertz/},
  urldate = {March, 2026},
  title = {Terahertz technology}
}

@misc{CamTHz,
  author = {{Cambridge Terahertz}},
  url = {https://www.thzcorp.com/},
  urldate = {March, 2026},
  title = {{Cambridge Terahertz uses MIT-developed sensor and AI to detect concealed weapons and alert security before entry.}}
}

@misc{Basler,
  author = {{Basler}},
  url = {https://www.baslerweb.com/en/innovation/microwave-imaging/},
  urldate = {March, 2026},
  title = {{Microwave Imaging - Automated, damage-free screening of packaging}}
}

@misc{Terakalis,
  author = {{Terakalis}},
  url = {https://www.terakalis.com/en/},
  urldate = {March, 2026},
  title = {{Waves with Unique Abilities - Safe, Contactless, Non-desctructive, Fast}}
}

@misc{Sikora,
  author = {{Sikora}},
  url = {https://sikora.net/en/},
  urldate = {March, 2026},
  title = {{Thickness measurement with radar - CENTERWAVE 6000 and PLANOWAVE 6000}}
}

@misc{Orteh,
  author = {{Orteh}},
  url = {https://www.orteh.pl/page/22/research-development},
  urldate = {March, 2026},
  title = {{Fast THz Mail Scanner}}
}

@misc{Nuctech1,
  author = {Nuctech},
  url = {https://www.nuctech.com/productdetail?id=b99ce138-1907-45bc-8a35-b2da0f86e105},
  urldate = {March, 2026},
  title = {{MW1000AA Body Inspection Device}}
}

@misc{Leidos,
  author = {{Leidos Inc.}},
  url = {https://www.leidos.com/markets/aviation/security-detection/aviation-checkpoint/people-screening},
  urldate = {March, 2026},
  title = {{People Screening}}
}

@misc{RohdeSchwarz,
  author = {{Rohde\& Schwarz}},
  url = {https://www.rohde-schwarz.com/products/aerospace-defense-security/security-scanner_254881.html?change_c=HQ},
  urldate = {March, 2026},
  title = {{Security Scanners}}
}

@misc{Canon2023,
  author = {{Canon Global}},
  url = {https://global.canon/en/news/2023/20230116.html},
  urldate = {Jan. 16, 2023},
  title = {Canon develops terahertz device with compact size, world-highest output and potential use cases in security, {6G} transmission and more}
}

@misc{yet2,
  author = {K. Ayers},
  url = {https://www.yet2.com/active-projects/canon-offering-high-power-semiconductor-terahertz-source-for-imaging-6g-and-radar/},
  urldate = {2026},
  title = {Canon Offering: High-Power Semiconductor Terahertz Source for Imaging, {6G}, and Radar}
}

@misc{ACST,
  author = {{ACST}},
  url = {https://acst.de},
  urldate = {March 26, 2026},
  title = {{THz} from components to system}
}

@INPROCEEDINGS{6105155,
  author={Bolduc, M. and Terroux, M. and Marchese, L. and Tremblay, B. and Savard, E. and Doucet, M. and Oulachgar, H. and Alain, C. and Jeronimek, H. and Bergeron, A.},
  booktitle={2011 International Conference on Infrared, Millimeter, and Terahertz Waves}, 
  title={{THz} imaging and radiometric measurements using a microbolometer-based camera}, 
  year={2011},
  volume={},
  number={},
  pages={1-2},
  doi={10.1109/irmmw-THz.2011.6105155}}

@ARTICLE{6557453,
  author={Han, Ruonan and Zhang, Yaming and Kim, Youngwan and Kim, Dae Yeon and Shichijo, Hisashi and Afshari, Ehsan and O, Kenneth K.},
  journal={IEEE Journal of Solid-State Circuits}, 
  title={Active Terahertz Imaging Using {Schottky} Diodes in {CMOS}: Array and 860-{GHz} Pixel}, 
  year={2013},
  volume={48},
  number={10},
  pages={2296-2308},
  doi={10.1109/JSSC.2013.2269856}}

@ARTICLE{Hu2019,
  author={Hu, Z. and Wang, C. and Han, R. N.},
  journal={IEEE Journal of Solid-State Circuits}, 
  title={A 32-Unit 240-{GHz} Heterodyne Receiver Array in 65-nm {CMOS} With Array-Wide Phase Locking}, 
  year={2019},
  volume={54},
  number={5},
  pages={1216-1227},
  doi={10.1109/JSSC.2019.2893231}
  }

@article{sarker2025prost,
  title={ProST: Progressive Sub-task Training for Pareto-Optimal Multi-agent Systems Using Small Language Models},
  author={Sarker Bijoy, Biddut and Saqib Hasan, Mohammad and Alipoormolabashi, Pegah and Sil, Avirup and Balasubramanian, Aruna and Balasubramanian, Niranjan},
  journal={arXiv e-prints},
  pages={arXiv--2509},
  year={2025}
}

@article{belcak2025small,
  title={Small language models are the future of agentic ai},
  author={Belcak, Peter and Heinrich, Greg and Diao, Shizhe and Fu, Yonggan and Dong, Xin and Muralidharan, Saurav and Lin, Yingyan Celine and Molchanov, Pavlo},
  journal={arXiv preprint arXiv:2506.02153},
  year={2025}
}

@INPROCEEDINGS{Xiang2026,
  author={Xiang, M. J. and Zhang, J. W. and Zhou, K. and Roskos, H. G.},
  booktitle={Proc. of SPIE 14079, Photonics Europe}, 
  title={Solving Ill-posed problems of inline terahertz holography: a unified physics-informed deep learning framework}, 
  year={2026},
  volume={},
  pages={},
  doi={}
  }

@INPROCEEDINGS{8048280,
  author={Song, Kiryong and Kim, Jungsoo and Kim, Doyoon and Seo, Myeong-Gyo and Rieh, Jae-Sung},
  booktitle={2017 IEEE International Symposium on Radio-Frequency Integration Technology (RFIT)}, 
  title={A {CMOS} 300-{GHz} 7 by 7 detector array for {THz} imaging}, 
  year={2017},
  volume={},
  number={},
  pages={31-33},
  doi={10.1109/RFIT.2017.8048280}}

@INPROCEEDINGS{Duelme2019,
  author={Dülme, S. and Grzeslo, M. and Morgan, J. and Steeg, M. and Lange, M. and Tebart, J. and Schrinski, N. and  Mohammad, I. and Neerfeld, T. and Lu, P. and Beling, A. and Stöhr, A.},
  booktitle={International Topical Meeting on Microwave Photonics (MWP2019)}, 
  title={300 {GHz} photonic self-mixing imaging-system with vertical illuminated triple-transit-region photodiode terahertz emitters}, 
  year={2019},
  number={},
  pages={30-33},  doi={10.1109/mwp.2019.8892098}
}

@phdthesis{Ludwig_diss2024,
author = {Florian A. Ludwig},
title = {Terahertz detection with field-effect transistors: experimental characterization and hydrodynamic modeling},
school = {Johann Wolfgang Goethe-Universität Frankfurt am Main},
year = {2024},
doi = {10.21248/gups.93123}
}

@article{Bauer2014,
author = {Bauer, M. and Venckevi{\v{c}}ius, R. and Ka{\v{s}}alynas, I. and Boppel, S. and Mundt, M. and Minkevi{\v{c}}ius, L. and Lisauskas, A. and Valu{\v{s}}is, G. and Krozer, V. and Roskos, H. G.},
doi = {10.1364/OE.22.01914119250},
journal = {Optics Express},
volume = {22},
number = {16},
pages = {19250-18256},
title = {Antenna-coupled field-effect transistors for multi-spectral terahertz imaging up to 4.25~{THz}},
year = {2014}
}

@ARTICLE{7404034,
  author={Kim, Dae Yeon and Park, Shinwoong and Han, Ruonan and Kenneth, K.O.},
  journal={IEEE Transactions on Terahertz Science and Technology}, 
  title={Design and Demonstration of 820-{GHz} Array Using Diode-Connected {NMOS} Transistors in 130-nm {CMOS} for Active Imaging}, 
  year={2016},
  volume={6},
  number={2},
  pages={306-317},
  doi={10.1109/TTHZ.2015.2513061}}

@article{Zda15,
author = {Zdanevi\v{c}ius, J. and Bauer, M. and Boppel, S. and Palenskis, V. and Lisauskas, A. and Krozer, V. and Roskos, H. G.},
journal = {Journal of Infrared, Millimeter, and Terahertz Waves},
number = {},
pages = {986--997},
publisher = {},
title = {Camera for High-Speed {THz} Imaging},
volume = {36},
month = {},
year = {2015},
url = {},
doi = {}
}

@article{Boppel2012,
author = {Boppel, S. and Lisauskas, A. and Mundt, M. and Seliuta, D. and Minkevi\v{c}ius, L. and Ka\v{s}alynas, I. and Valu\v{s}is, G. and Mittendroff, M. and Winnerl, S. and Krozer, V. and Roskos, H. G.},
journal = {IEEE Transactions on Microwave Theory and Techniques},
number = {12},
pages = {3834--3843},
title = {{CMOS} Integrated Antenna-Coupled Field-Effect Transistors for the Detection of Radiation From 0.2 to 4.3 {THz}},
volume = {60},
year = {2012},
doi = {10.1109/TMTT.2012.2221732}
}

@article{Minkevicius2011,
author = {Minkevi\v{c}ius, L. and Tamo\v{s}iunas, V. and Ka\v{s}alynas, I. and Seliuta, D. and Valu\v{s}is, G. and Lisauskas, A. and Boppel, S. and Roskos, H. G. and Köhler, K.},
journal = {Applied Physics Letters},
pages = {131101},
title = {Terahertz heterodyne imaging with InGaAs-based bow-tie diodes},
volume = {99},
number = {13},
year = {2011},
doi = {10.1063/1.3641907}
}

@article{Zhu2013,
author = {Zhu, Z. X. and Joshi, S. and Grover, S. and Moddel, G.},
journal = {Journal of Physics D: Applied Physics},
pages = {185101},
title = {Graphene geometric diodes for terahertz rectennas},
volume = {46},
number = {18},
year = {2013},
doi = {10.1088/0022-3727/46/18/185101}
}

@article{Sangare2013,
author = {Sangaré, P. and Ducournau, G. and Grimbert, B. and Brandli, V. and  Faucher, M. and Gaquière, C. and {Iñiguez-de-la-Torre}, A. and {Iñiguez-de-la-Torre}, I. and Millithaler, J. F. and  Mateos, J. and González, T.},
journal = {Journal of Applied Physics},
pages = {034305},
title = {Experimental demonstration of direct terahertz detection at room-temperature in {AlGaN/GaN} asymmetric nanochannels},
volume = {113},
number = {3},
year = {2013},
doi = {10.1063/1.4775406}
}

@article{Thumm2020,
author = {Thumm, M.},
journal = {Journal of Infrared, Millimeter, and Terahertz Waves},
pages = {1-140},
title = {State-of-the-art of high-power gyro-devices and free electron masers},
volume = {41},
number = {1},
year = {2020},
doi = {10.1007/s10762-019-00631-y}
}

@article{Lueck2011,
author = {Lueck, M. R. and Malta, D. M. and Gilchrist, K. H. and Kory, C. L. and Mearini, G. T. and Dayton, J. A.},
journal = {Journal of Micromechanics and Microengineering},
pages = {065022},
title = {Microfabrication of diamond-based slow-wave circuits for mm-wave and {THz} vacuum electronic sources},
volume = {21},
number = {6},
year = {2011},
doi = {10.1088/0960-1317/21/6/065022}
}

@article{Ulisse2022,
author = {Ulisse, G. and Schurch, P. and Hepp, E. and Koelmans, W. W. and Doerner, R. and Krozer, V.},
journal = {IEEE Transactions on Electron Devices},
pages = {6358-6361},
title = {A {3-D} printed helix to traveling-wave tubes},
volume = {69},
number = {11},
year = {2022},
doi = {10.1109/TED.2022.3209645}
}

@article{Xu-Song2008,
author = {Xu, K. Y. and Lu, X. F. and Song, A. M. and Wang, G.},
journal = {Journal of Applied Physics},
pages = {113708},
title = {Enhanced terahertz detection by localized surface plasma oscillations in a nanoscale unipolar diode},
volume = {113708},
number = {11},
year = {2008},
doi = {10.1063/1.2937175}
}

@article{Ito2017,
author = {Ito, H. and Ishibashi, T.},
journal = {Japanese Journal of Applied Physics},
volume = {56},
number = {},
pages = {014101},
title = {{InP/InGaAs} {Fermi}-level managed barrier diode for broadband
and low-noise terahertz-wave detection},
year = {2017},
doi = {10.7567/JJAP.56.014101}
}

@article{Kokkoniemi2017,
author = {Kokkoniemi, R. and Govenius, J. and Vesterinen, V. and Lake, R. E. and Gunyhó, A. M. and Tan, K. Y. and Simbierowicz, S. and Grönberg, L. and Lehtinen, J. and Prunnila, M. and Hassel, J. and Lamminen, A. and Saira, O. P. and Möttönen, M.},
journal = {Communications Physics},
volume = {2},
number = {},
pages = {124},
title = {Nanobolometer with ultralow noise equivalent power},
year = {2019},
doi = {10.1038/s42005-019-0225-6}
}

@article{Kokkoniemi2020,
author = {Kokkoniemi, R. and Girard, J. P. and Hazra, D. and Laitinen, A. and Govenius, J. and Lake, R. E. and Sallinen, I. and Vesterinen, V. and Partanen, M. and Tan, J. Y. and Chan, K. W. and Tan, K. Y. and Hakonen, P. and Möttönen, M.},
journal = {Nature},
volume = {586},
number = {7827},
pages = {47},
title = {Bolometer operating at the threshold for circuit quantum electrodynamics},
year = {2020},
doi = {10.1038/s41586-020-2753-3}
}

@ARTICLE{Batra2021,
  author={Batra, A. and Barowski, J. and Damyanov, D. and Wiemeler, M. and Rolfes, I. and Schultze, T. and Balzer, J. C. and Göhringer, D. and Kaiser, T.},
  journal={IEEE Journal of Microwaves}, 
  title={Short-Range {SAR} Imaging From {GHz} to {THz} Waves}, 
  year={2021},
  volume={1},
  number={2},
  pages={574-585},
  doi = {10.1109/JMW.2021.3063343}
}

@ARTICLE{Mehdi2017,
  author={Mehdi, I. and Siles, J. V. and Lee, C. and Schlecht, E.},
  journal={Proceedings of the IEEE}, 
  title={{THz} diode technology: {Status}, prospects, and applications}, 
  year={2017},
  volume={105},
  number={6},
  pages={990-1007},
  doi = {10.1109/JPROC.2017.2650235}
}

@ARTICLE{Hoefle2014,
  author={Hoefle, M. and Haehnsen, K. and Oprea, I. and Cojocari, O. and Penirschke, A. and Jakoby, R.},
  journal={Journal of Infrared, Millimeter, and Terahertz Waves}, 
  title={Compact and sensitive millimetre wave detectors based on low barrier {Schottky} diodes on impedance matched planar antennas}, 
  year={2014},
  volume={35},
  number={11},
  pages={891-908},
  doi = {10.1007/s10762-014-0090-z}
}

@ARTICLE{Tarasov2007,
  author={Tarasov, M. and  Svensson, J. and Kuzmin, L. and Campbell, E. E. B.},
  journal={Applied Physics Letters}, 
  title={Carbon nanotube bolometers}, 
  year={2007},
  volume={90},
  number={16},
  pages={163503},
  doi = {10.1063/1.2722666}
}

@article{Alekseyev2011,
author = {Alekseyev, L and Narimanov, E and Khurgin, J},
journal = {Optics Express},
pages = {22350-22357},
title = {Super-resolution imaging via spatiotemporal frequency shifting and coherent detection},
volume = {19},
number = {22},
year = {2011},
doi = {10.1364/OE.19.022350},
}

@article{Zhao2021,
author = {Zhao, J and Yin, LZ and Han, FY and Wang, YD and Huang, TJ and Du, CH and Liu, PK},
journal = {Optics Express},
pages = {36366-36378},
title = {Terahertz non-label subwavelength imaging with composite photonics-plasmonics structured illumination},
volume = {29},
number = {22},
year = {2021},
doi = {10.1364/OE.437544},
}

@article{chen2016review,
  title={A review of metasurfaces: physics and applications},
  author={Chen, Hou-Tong and Taylor, Antoinette J and Yu, Nanfang},
  journal={Reports on Progress in Physics},
  volume={79},
  number={7},
  pages={076401},
  year={2016},
  publisher={IOP Publishing}
}

@article{fan2022active,
  title={Active and tunable nanophotonic metamaterials},
  author={Fan, Kebin and Averitt, Richard D and Padilla, Willie J},
  journal={Nanophotonics},
  volume={11},
  number={17},
  pages={3769--3803},
  year={2022},
  publisher={De Gruyter}
}

@book{capolino2009theory,
  title={Theory and Phenomena of Metamaterials},
  author={Capolino, Filippo},
  year={2009},
  publisher={CRC Press},
  isbn={9781420054255}
}

@book{Ulaby,
  title={Microwave Radar and Radiometric Remote Sensing},
  author={F. T. Ulaby and D. Long},
  year={2015},
  publisher={Artech House},
  isbn={9780472119356}
}

@book{Karmakar,
  title={Ground-Based Microwave Radiometry and Remote Sensing: Methods and Applications},
  author={P. K. Karmakar},
  year={2017},
  publisher={Taylor \& Francis Group},
  isbn={9781138074521, 1138074527}
}

@article{Vittucci2026,
  author = {C. Vittucci and M. Picchiani},
  title = {Satellite microwave radiometry for the observation of land surfaces: {A} general review },
  journal = {Sensors},
  volume = {26},
  number = {5},
  pages = {1638},
  year = {2026},
  doi = {10.3390/s26051638}
}

@article{Santamaria2020,
  author = {G. Santamaría-Botello and Z. Popovic and K. A. Abdalmalak and D. {Segovia-Vargas} and E. R. Brown and L. E. {García Muñoz}},
  title = {Sensitivity and noise in {THz} electrooptic upconversion radiometers},
  journal = {Scientific Reports},
  volume = {10},
  number = {},
  pages = {9403},
  year = {2020},
  doi = {10.1038/s41598-020-65987-x}
}

@article{deng2025physics,
  author = {Deng, Y. and Fan, K. and Jin, B. and Malof, J. M. and Padilla, W. J.},
  title = {Physics-informed learning in artificial electromagnetic materials},
  journal = {Applied Physics Reviews},
  volume = {12},
  pages = {011331},
  year = {2025},
  doi = {10.1063/5.0232675}
}

@article{xu2024physics,
  author = {Xu, Yucheng and Yang, Jia-Qi and Fan, Kebin and Wang, Sheng and Wu, Jingbo and Zhang, Caihong and Zhan, De-Chuan and Padilla, Willie J. and Jin, Biaobing and Chen, Jian and Wu, Peiheng},
  title = {Physics-Informed Inverse Design of Programmable Metasurfaces},
  journal = {Advanced Science},
  volume = {11},
  pages = {2406878},
  year = {2024},
  doi = {10.1002/advs.202406878}
}

@article{rozman2024deep,
  author = {Rozman, N. and Peng, R. and Padilla, W. J.},
  title = {Deep Inverse Design of an Infrared Metasurface Diffuser},
  journal = {Advanced Optical Materials},
  volume = {12},
  pages = {2401462},
  year = {2024},
  doi = {10.1002/adom.202401462}
}

@article{peng2024transfer,
  author = {Peng, R. and Ren, S. and Malof, J. M. and Padilla, W. J.},
  title = {Transfer learning for metamaterial design and simulation},
  journal = {Nanophotonics},
  volume = {13},
  pages = {2323},
  year = {2024},
  doi = {10.1515/nanoph-2023-0691}
}

@article{li2024machine,
  author = {Li, W. and Sedeh, H. Barati and Tsvetkov, D. and Padilla, W. J. and Ren, S. and Malof, J. M. and Litchinitser, N. M.},
  title = {Machine Learning for Engineering Meta-Atoms with Tailored Multipolar Resonances},
  journal = {Laser \& Photonics Reviews},
  volume = {18},
  pages = {2300855},
  year = {2024},
  doi = {10.1002/lpor.202300855}
}

@article{yang2023normalizing,
  author = {Yang, Jia-Qi and Xu, YuCheng and Fan, Kebin and Wu, Jingbo and Zhang, Caihong and Zhan, De-Chuan and Jin, Biao-Bing and Padilla, Willie J.},
  title = {Normalizing flows for efficient inverse design of thermophotovoltaic emitters},
  journal = {ACS Photonics},
  volume = {10},
  pages = {1001},
  year = {2023},
  doi = {10.1021/acsphotonics.2c01803}
}

@article{deng2022deep,
  author = {Deng, Yang and Ren, Simiao and Malof, Jordan M. and Padilla, Willie J.},
  title = {Deep inverse photonic design: A tutorial},
  journal = {Photonics and Nanostructures - Fundamentals and Applications},
  volume = {52},
  pages = {101070},
  year = {2022},
  doi = {10.1016/j.photonics.2022.101070}
}

@article{khatib2022learning,
  author = {Khatib, O. and Ren, S. and Malof, J. M. and Padilla, W. J.},
  title = {Learning the Physics of All-Dielectric Metamaterials with Deep Lorentz Neural Networks},
  journal = {Advanced Optical Materials},
  volume = {10},
  pages = {2200097},
  year = {2022},
  doi = {10.1002/adom.202200097}
}

@article{ren2022inverse,
  author = {Ren, S. and Mahendra, A. and Khatib, O. and Deng, Y. and Padilla, W. J. and Malof, J. M.},
  title = {Inverse deep learning methods and benchmarks for artificial electromagnetic material design},
  journal = {Nanoscale},
  volume = {14},
  pages = {3958},
  year = {2022},
  doi = {10.1039/D1NR08346E}
}

@inproceedings{Appleby2002,
  author = {Sinclair, G. N. and Coward, P. R. and Anderton, R. N. and Appleby, R. and Seys, T. and Southwood, P.},
  title = {Detection of illegal passengers in lorries using a passive millimetre wave scanner. },
  booktitle = {IEEE proceedings of the 36th Annual International Carnahan Conference on Security Technology, Atlantic City, NJ, USA (20-24 Oct. 2002)},
  year = {2002},
  pages = {167–170}, 
  doi = {10.1109/CCST.2002.1049245}  
}

@inproceedings{deng2022benchmarking,
  author = {Deng, Y. and Dong, J. and Ren, S. and Khatib, O. and Soltani, M. and Tarokh, V. and Padilla, W. J. and Malof, J. M.},
  title = {Benchmarking data-driven surrogate simulators for artificial electromagnetic materials},
  booktitle = {Thirty-fifth Conference on Neural Information Processing Systems Datasets and Benchmarks Track (Round 2)},
  year = {2022},
}

@article{Ahmed2021,
  author = {Ahmed, S. S.},
  title = {Microwave imaging in security — two decades of innovation},
  journal = {IEEE Journal of Microwaves},
  volume = {1},
  number = {1},
  pages = {191-201},
  year = {2021},
  doi = {10.1109/JMW.2020.3035790}
}

@article{LiLiLi2019,
  author = {Li, R. and Li, C. and Li, H. W. and Wu, S. Y. and Fang, G. Y.},
  title = {Study of automatic detection of concealed targets in passive terahertz images for intelligent security screening},
  journal = {IEEE Transactions on Terahertz Science and Technology},
  volume = {9},
  number = {2},
  pages = {165-176},
  year = {2019},
  doi = {10.1109/TTHZ.2018.2889407}
}

@article{khatib2021deep,
  author = {Khatib, O. and Ren, S. and Malof, J. M. and Padilla, W. J.},
  title = {Deep Learning the Electromagnetic Properties of Metamaterials---A Comprehensive Review},
  journal = {Advanced Functional Materials},
  volume = {31},
  pages = {2101748},
  year = {2021},
  doi = {10.1002/adfm.202101748}
}

@article{deng2021neural,
  author = {Deng, Y. and Ren, S. and Fan, K. and Malof, J. M. and Padilla, W. J.},
  title = {Neural-adjoint method for the inverse design of all-dielectric metasurfaces},
  journal = {Optics Express},
  volume = {29},
  pages = {7526--7534},
  year = {2021},
  doi = {10.1364/OE.419138}
}

@inproceedings{Appleby1999,
  author = {R. Appleby and R. N. Anderton and S. Price and N. A. Salmon and G. N. Sinclair and J. R. Borrill and P. R. Coward and V. {Paraskevi Papakosta} and A. H. Lettington and D. A. Robertson},
  title = {Compact real-time (video rate) passive millimetre-wave imager},
  booktitle = {Proc. SPIE AeroSense '99 - Orlando, FL, Passive Millimeter-Wave Imaging Technology III},
  volume = {3703},
  pages = {13-19},
  year = {1999},
  doi = {10.1117/12.353003}
}

@inproceedings{Peichl2007,
  author = {M. Peichl and S. Dill and M. Jirousek and H. Süß},
  title = {Microwave radiometry - imaging technologies and applications},
  booktitle = {Proc. of the Wave Propagation in Communication, Microwave Systems and Navigation Conference (WFMN07), Chemnitz, Germany},
  volume = {},
  pages = {75-83},
  year = {2007},
  doi = {},
  url = {https://monarch.qucosa.de/api/qucosa%3A18847/attachment/ATT-12/}
}

@inproceedings{ren2020benchmarking,
  author = {Ren, Simiao and Padilla, Willie J. and Malof, Jordan M.},
  title = {Benchmarking Deep Inverse Models over time, and the Neural-Adjoint method},
  booktitle = {Advances in Neural Information Processing Systems},
  volume = {33},
  pages = {38},
  year = {2020}
}

@article{nadell2019deep,
  author = {Nadell, Christian C. and Huang, Bohao and Malof, Jordan M. and Padilla, Willie J.},
  title = {Deep learning for accelerated all-dielectric metasurface design},
  journal = {Optics Express},
  volume = {27},
  pages = {27523},
  year = {2019},
  doi = {10.1364/OE.27.027523}
}

@article{o2007properties,
  title={Properties of planar electric metamaterials for novel terahertz applications},
  author={O'Hara, John F and Smirnova, Evgenya and Chen, Hou-Tong and Taylor, Antoinette J and Averitt, Richard D and Highstrete, Clark and Lee, Mark and Padilla, Andwillie J},
  journal={Journal of Nanoelectronics and Optoelectronics},
  volume={2},
  number={1},
  pages={90--95},
  year={2007},
  publisher={American Scientific Publishers}
}

@article{padilla2022imaging,
  title={Imaging with metamaterials},
  author={Padilla, Willie J and Averitt, Richard D},
  journal={Nature Reviews Physics},
  volume={4},
  number={2},
  pages={85--100},
  year={2022},
  publisher={Nature Publishing Group UK London}
}

@article{pendry1999magnetism,
  title={Magnetism from conductors and enhanced nonlinear phenomena},
  author={Pendry, John B and Holden, Anthony J and Robbins, David J and Stewart, William J},
  journal={IEEE Transactions on Microwave Theory and Techniques},
  volume={47},
  number={11},
  pages={2075--2084},
  year={1999},
  publisher={IEEE}
}

@article{zhang2019transformation,
  title={Transformation optics from macroscopic to nanoscale regimes: a review},
  author={Zhang, Jingjing and Pendry, John B and Luo, Yu},
  journal={Advanced Photonics},
  volume={1},
  number={1},
  pages={014001},
  year={2019},
  publisher={SPIE}
}

@article{padilla2006spectroscopy,
  title={Spectroscopy of metamaterials from infrared to optical frequencies},
  author={Padilla, Willie J and Smith, David R and Basov, Dmitri N},
  journal={Journal of the Optical Society of America B},
  volume={23},
  number={3},
  pages={404--410},
  year={2006},
  publisher={Optica Publishing Group}
}

@article{achiam2023gpt,
  title={GPT-4 technical report},
  author={Achiam, Josh and Adler, Steven and Agarwal, Sandhini and Aflalo, Lama and Agrawal, Garima and Aleman, Chirag and Almeida, Diogo and Altenschmidt, Janko and Altman, Sam and Anadkat, Shyamal and others},
  journal={arXiv preprint arXiv:2303.08774},
  year={2023}
}

@article{lupoiu2025multiagentic,
  title={A multi-agentic framework for real-time, autonomous freeform metasurface design},
  author={Lupoiu, R. and Shao, Y. and Dai, T. and Mao, C. and Ed{\'e}e, K. and Fan, J. A.},
  volume = {11},
  number = {44},
  pages = {eadx8006},
  year={2025},
  month={10},
  journal={Science Advances},
  doi = {10.1126/sciadv.adx8006},
}

@article{huang2025mcpenabled,
  title={{MCP}-enabled {LLM} for meta-optics inverse design: leveraging differentiable solver without {LLM} expertise},
  author={Huang, Yi and others},
  journal={Nanophotonics},
  volume={14},
  number={27},
  pages={5589--5602},
  year={2025},
  publisher={De Gruyter}
}

@article{lu2025agentic,
  title={An agentic framework for autonomous metamaterial modeling and inverse design},
  author={Lu, Darui and Malof, Jordan M and Padilla, Willie J},
  journal={ACS Photonics},
  volume={12},
  number={11},
  pages={6071--6080},
  year={2025},
  publisher={ACS Publications}
}

@article{anil2023gemini,
  title={Gemini: a family of highly capable multimodal models},
  author={Anil, Rohan and Borgeaud, Sebastian and Wu, Yonghui and Alayrac, Jean-Baptiste and Yu, Jiahui and Soricut, Radu and Schalkwyk, Johan and Dai, Andrew M and Hauth, Anja and Millican, Katie and others},
  journal={arXiv preprint arXiv:2312.11805},
  year={2023}
}

@article{touvron2023llama,
  title={Llama: Open and efficient foundation language models},
  author={Touvron, Hugo and Lavril, Thibaut and Izacard, Gautier and Martinet, Xavier and Lachaux, Marie-Anne and Lacroix, Timoth{\'e}e and Rozi{\`e}re, Baptiste and Goyal, Naman and Hambro, Eric and Azhar, Faisal and others},
  journal={arXiv preprint arXiv:2302.13971},
  year={2023}
}

@article{anthropic2024claude,
  title={The Claude 3 model family: Opus, Sonnet, Haiku},
  author={{Anthropic}},
  journal={Model card},
  year={2024},
  url={https://www-cdn.anthropic.com/de8ba9b01c9ab7cbabf5c33b80b7bbc618857627/Model_Card_Claude_3.pdf}
}

@article{bommasani2021opportunities,
  title={On the opportunities and risks of foundation models},
  author={Bommasani, Rishi and Hudson, Drew A and Adeli, Ehsan and Altman, Russ and Arora, Simran and von Arx, Sydney and Bernardo, Michael S and Bernstein, Michael S and Bhagavatula, Chandra and others},
  journal={arXiv preprint arXiv:2108.07258},
  year={2021}
}

@article{vaswani2017attention,
  title={Attention is all you need},
  author={Vaswani, Ashish and Shazeer, Noam and Parmar, Niki and Uszkoreit, Jakob and Jones, Llion and Gomez, Aidan N and Kaiser, {\L}ukasz and Polosukhin, Illia},
  journal={Advances in neural information processing systems},
  volume={30},
  year={2017}
}

@article{Boppel2016,
author = {Boppel, S. and Ragauskas, M. and Hajo, A. and Bauer, M. and Lisauskas, A. and Chevtchenko, S. and Rämer, A. and Kasalynas, I. and Valusis, G. and Würfl, H. J. and Heinrich, W. and Tränkle, G. and Krozer, V. and Roskos, H. G.},
journal = {IEEE Transactions on Terahertz Science and Technology},
volume = {6},
number = {2},
pages = {348-350},
title = {0.25-$\mu$m {GaN} {TeraFETs} optimized as {THz} power detectors and intensity-gradient sensors},
year = {2016},
doi = {10.1109/TTHZ.2016.2520202}
}

@article{But2023,
author = {But, D. B. and Ikamas, K. and Kolacinski, C. and  Chernyadiev, A. V. and Vizbaras, D. and Knap, W. and Lisauskas, A.},
journal = {Scientific Reports},
volume = {13},
number = {1},
pages = {16161},
title = {Sub-terahertz feedback interferometry and imaging with emitters in 130 nm {BiCMOS} technology},
year = {2023},
doi = {10.1038/s41598-023-43194-8}
}

@article{Yuan2023,
author = {Yuan, H. and {ul-Islam}, Q. and Roskos, H. G.},
journal = {Lithuanian Journal of Physics},
volume = {63},
number = {4},
pages = {241-250},
title = {Superstrate-lens integration using paraffin wax on top of semiconductor -based {THz} detector chips},
year = {2023},
doi = {10.3952/physics.2023.63.4.6}
}

@article{Rutledge1982,
author = {Rutledge, D. B. and Muha, M. S.},
journal = {IEEE Transactions on Antennas and Propagation},
volume = {30},
number = {4},
pages = {535-540},
title = {Imaging antenna-arrays},
year = {1982},
doi = {10.1109/TAP.1982.1142856}
}

@article{Trichopoulos2013,
author = {G. C. Trichopoulos and H. L. Mosbacker and D. Burdette and K. Sertel},
journal = {IEEE Transactions on Antennas and Propagation},
volume = {61},
number = {4},
pages = {1733-1740},
title = {A broadband focal plane array camera for real-time
{THz} imaging applications},
year = {2013},
doi = {10.1109/TAP.2013.2242829}
}

@article{Appleby2007,
author = {Appleby, R. and Wallace, H. B.},
journal = {IEEE Transactions on Antennas and Propagation},
volume = {55},
number = {11},
pages = {2944-2956},
title = {Standoff detection of weapons and contraband in the 100~{GHz} to 1~{THz} region},
year = {2007},
doi = {10.1109/TAP.2007.908543}
}

@article{Ghosh2025,
author = {Ghosh, S. K. and Kepros, E. and Chahal, P.},
journal = {IEEE Transactions on Components, Packaging and Manufacturing Technology},
volume = {13},
number = {2},
pages = {244-252},
title = {Aerosol-jet printed high-{Q} quasi-optical {FSSs} on flex substrates using a novel parylene lift-off process},
year = {2025},
doi = {10.1109/TCPMT.2024.3521292}
}

@article{Chen_lens2025,
author = {Chen, J. and Huang, S. X. and Chan, K. F. and Wu, G.-B. and Wong, H. and Chan, C. H.},
journal = {Advanced Optical Materials},
volume = {13},
number = {36},
pages = {e02943},
title = {The equivalence of microlens and metalens for {THz} high spatial-spectral resolution hyperspectral imaging with a large field of view},
year = {2025},
doi = {10.1002/adom.202502943}
}

@inproceedings{Park_microlens2013,
author = {Park, K. Y. and Wiwatcharagoses, N. and Chahal, P.},
booktitle = {IEEE 63RD Electronic Components and Technology Conference (ECTC)},
pages = {1912-1919},
title = {Wafer-level integration of micro-lens for {THz} focal plane array application},
year = {2013},
doi = {10.1109/ECTC.2013.6575839}
}

@inproceedings{Zatta2021a,
author = {Zatta, R. and Pfeiffer, U. R.},
booktitle = {46th International Conference on Infrared, Millimeter and Terahertz Waves (IRMMW-THz)},
pages = {},
title = {Radial distortion in silicon lens-integrated {THz} cameras
},
year = {2021},
doi = {10.1109/IRMMW-THz50926.2021.9567497}
}

@article{But2023a,
author = {But, D. B. and  Chernyadiev, A. V. and Ikamas, K. and Kolacinski, C. and Krysl, A. and Roskos, H. G. and Knap, W. and Lisauskas, A.},
journal = {Opto-Electronics Review},
volume = {31},
number = {2},
pages = {e144599},
title = {Compact terahertz devices based on silicon in {CMOS} and {BiCMOS} technologies},
year = {2023},
doi = {10.24425/opelre.2023.144599}
}

@article{Manh2018,
author = {Manh, L. D. and Diebold, S. and Nishio, K. and Nishida, Y. and Kim, J. and Mukai, T. and Fujita, M. and Nagatsuma, T.},
journal = {IEEE Transactions on Terahertz Science and Technology},
volume = {8},
number = {4},
pages = {455-464},
title = {External feedback effect in terahertz resonant tunneling diode oscillators},
year = {2018},
doi = {10.1109/TTHZ.2018.2842209}
}

@article{Farrah2019,
author = {Farrah, D. and Smith, K. E. and Ardila, D. and Bradford, C. M. and Dipirro, M. and Ferkinhoff, C. and Glenn, J. and Goldsmith, P. and Leisawitz, D. and Nikola, T. and Rangwala, N. and Rinehart, S. A. and Staguhn, J. and Zemcov, M. and Zmuidzinas, J. and Bartlett, J. and Carey, S. and Fischer, W. J. and Kamenetzky, J. and Kartaltepe, J. and Lacy, M. and Lis, D. C. and Locke, L. and {Lopez-Rodriguez}, E. and {MacGregor}, M. and Mills, E. and Moseley, S. H. and Murphy, E. J. Rhodes, A. and Richter, M. and Rigopoulou, D. and Sanders, D. and Sankrit, R. and Savini, G. and Smith, J. D. and Stierwalt, S.},
journal = {Journal of Astronomical Telescopes, Instruments, and Systems},
volume = {5},
number = {2},
pages = {020901},
title = {Review: far-infrared instrumentation and technological development for the next decade},
year = {2019},
doi = {10.1117/1.JATIS.5.2.020901}
}

@article{Hoogelander2023,
  author={Hoogelander, M. and {van Berkel}, S. and Malotaux, E. S. and {Alonso-delPino}, M. and Cavallo, D. and Spirito, M. and Llombart, N.},
  title={Chessboard Focal Plane Array for a {CMOS}-Integrated Terahertz Camera}, 
  year={2023},
  journal = {IEEE Transactions on Terahertz Science and Technology},
  volume={13},
  number={6},
  pages={704-717},
  doi={10.1109/TTHZ.2023.3297072}
  }

@article{Hoogelander2026,
  author={Hoogelander, M. and Spirito, M. and Sutbas, B. and Carta, C. and Llombart, N. and Alonso-Delpino, M.},
  title={Chessboard {FPA} in 130-nm {SiGe} {BiCMOS} for high-resolution passive terahertz imaging}, 
  year={2026},
  journal = {IEEE Journal of Microwaves},
  volume={6},
  number={2},
  pages={334-349},
  doi={10.1109/JMW.2026.3660185}
  }

@article{Kanazawa2019,
  author={Kanazawa, Y. and Yokoyama, S. and Hiramatsu, S. and Sano, E. and Lkegami, T. and  Takida, Y. and Ambalathankandy, P. and Minamide, H. and Ikebe, M.},
  title={Wideband terahertz imaging pixel with a small on-chip antenna in 180 nm {CMOS}}, 
  year={2019},
  journal = {Japanese Journal of Applied Physics},
  volume={58},
  number={B},
  pages={SBBL06},
  doi={10.7567/1347-4065/ab03c9}
  }

@article{Tretyakov2025,
  author={Tretyakov, I. V. and Khudchenko, A. V. and Rudakov, K. I. and Ivashentseva, I. V. and Kaurova, N. S. and Voronov, B. M. and Kirsanova, M. S. and Larchenkova, T. I. and Goltsman, G. N. and Baryshev, A. M. and Hesper, R. and Khan, F. V. and Zhukova, E. S. and Chekushkin, A. M. and Melentev, A. V. and Zhivetev, K. V. and Terentiev, A. V. and Koshelets, V. P. and Likhachev, S. F.},
  title={Development of Mixers for the High-Resolution Spectrometer of the {Millimetron} Space Observatory}, 
  year={2025},
  journal = {IEEE Transactions on Terahertz Science and Technology},
  volume={15},
  number={2},
  pages={191-199},
  doi={10.1109/TTHZ.2024.3505592}
  }

@article{Risacher2016,
  author={Risacher, C. and Güsten, R. and Stutzki, J. and Hübers, H.-W. and Büchel, D. and Graf, U. U. and Heyminck, S. and Honingh, C. E. and Jacobs, K. and Klein, B. and Klein, T. and Leinz, C. and Pütz, P. and Reyes, N. and Ricken, O. and Wunsch, H. J. and Fusco, P. and Rosner, S.},
  title={First Supra-{THz} Heterodyne Array Receivers for Astronomy With the {SOFIA} Observatory}, 
  year={2016},
  journal = {IEEE Transactions on Terahertz Science and Technology},
  volume={6},
  number={2},
  pages={199-211},
  doi={10.1109/TTHZ.2015.2508005}
  }

@article{Graf2015,
  author={Graf, U. U. and Honingh, C. E. and Jacobs, K. and Stutzki, J.},
  title={Terahertz Heterodyne Array Receivers for Astronomy}, 
  year={2015},
  journal = {Journal of Infrared, Millimeter, and Terahertz Waves},
  volume={36},
  number={10},
  pages={896-921},
  doi={10.1007/s10762-015-0171-7}
  }

@article{Hu:95,
author = {B. B. Hu and M. C. Nuss},
journal = {Optics Letters},
number = {16},
pages = {1716--1718},
publisher = {Optica Publishing Group},
title = {Imaging with terahertz waves},
volume = {20},
month = {Aug},
year = {1995},
doi = {10.1364/OL.20.001716},
}

@article{Naftaly2019,
title = {Industrial applications of terahertz sensing: state of play},
author = {Naftaly, M. and Vieweg, N. and Deninger, A.},
journal = {Sensors},
volume = {19},
number = {19},
pages = {4303},
year = {2019},
doi = {10.3390/s19194203},
}

@article{Jepsen2011,
  author  = {P. U. Jepsen and D. G. Cooke and M. Koch},
  title   = {Terahertz spectroscopy and imaging—{M}odern techniques and applications},
  journal = {Laser \& Photonics Reviews},
  volume  = {5},
  number  = {1},
  pages   = {124--166},
  year    = {2011},
  doi     = {10.1002/lpor.201000011}
}

@article{Mittleman2018,
  author  = {D. M. Mittleman},
  title   = {Twenty years of terahertz imaging},
  journal = {Optics Express},
  volume  = {26},
  number  = {8},
  pages   = {9417--9431},
  year    = {2018},
  doi     = {10.1364/OE.26.009417}
}

@article{Ferguson2002,
  author  = {B. Ferguson and X.-C. Zhang},
  title   = {Materials for terahertz science and technology},
  journal = {Nature Materials},
  volume  = {1},
  number  = {1},
  pages   = {26--33},
  year    = {2002},
  doi     = {10.1038/nmat708}
}

@article{Akyildiz2022,
  author  = {F. Akyildiz and C. Han and Z. Hu and S. Nie and J. M. Jornet},
  title   = {Terahertz Band Communication: {A}n Old Problem Revisited and Research Directions for the Next Decade},
  journal = {IEEE Transactions on Communications},
  volume  = {70},
  number  = {6},
  pages   = {4250--4285},
  month   = jun,
  year    = {2022},
  doi     = {10.1109/TCOMM.2022.3171800}
}

@article{Siegel2002,
  author  = {P. H. Siegel},
  title   = {Terahertz technology},
  journal = {IEEE Transactions on Microwave Theory and Techniques},
  volume  = {50},
  number  = {3},
  pages   = {910-928},
  year    = {2002},
  doi     = {10.1109/22.989974}
}

@article{Nongkseh2025,
  author  = {Nongkseh, P and Sur, S. N. and Kandar, D.},
  title   = {Detection of concealed object using terahertz images: {A} comprehensive review},
  journal = {Engineering Applications of Artificial Intelligence},
  volume  = {148},
  number  = {},
  pages   = {110432},
  year    = {2025},
  doi     = {10.1016/j.engappai.2025.110432}
  }

@article{Yang2016,
  author  = {X. Yang and X. Zhao and K. Yang and Y. Liu and Y. Liu and W. Fu and Y. Luo},
  title   = {Biomedical applications of terahertz spectroscopy and imaging},
  journal = {Trends in Biotechnology},
  volume  = {34},
  number  = {10},
  pages   = {810--824},
  year    = {2016},
  doi     = {10.1016/j.tibtech.2016.04.008}
  }

@book{mitchell1997machine,
  title={Machine learning},
  author={Mitchell, Tom M},
  year={1997},
  publisher={McGraw-Hill},
  address={New York}
}

@article{10.1063/1.1469679,
    author = {Siebert, Karsten J. and Quast, Holger and Leonhardt, Rainer and Löffler, Torsten and Thomson, Mark and Bauer, Tobias and Roskos, Hartmut G. and Czasch, Stephanie},
    title = {Continuous-wave all-optoelectronic terahertz imaging},
    journal = {Applied Physics Letters},
    volume = {80},
    number = {16},
    pages = {3003-3005},
    year = {2002},
    month = {04},
    issn = {0003-6951},
    doi = {10.1063/1.1469679},
}

@article{10.1063/1.1616668,
    author = {Chen, Hou-Tong and Kersting, Roland and Cho, Gyu Cheon},
    title = {Terahertz imaging with nanometer resolution},
    journal = {Applied Physics Letters},
    volume = {83},
    number = {15},
    pages = {3009-3011},
    year = {2003},
    month = {10},
    issn = {0003-6951},
    doi = {10.1063/1.1616668},
}

@article{Hartwick:76,
author = {T. S. Hartwick and D. T. Hodges and D. H. Barker and F. B. Foote},
journal = {Applied Optics},
number = {8},
pages = {1919--1922},
publisher = {Optica Publishing Group},
title = {Far infrared imagery},
volume = {15},
month = {Aug},
year = {1976},
doi = {10.1364/AO.15.001919},
}

@article{Wang2026,
  author  = {Wang, M. and Qin, H. and Wu, D. M. and  Zhou, Q. and Yu, R. and Yang, S. G. and Cai, X. H. and Jin, L. and Sun, J. D. Zhang, J. F. and Li, X. X. and Hao, C. J. and Liu, D. F.},
  title   = {High-angular-resolution adaptive terahertz scanning imaging via dynamic zoom optics and beam steering metasurfaces},
  journal = {Optics Express},
  volume  = {34},
  number  = {1},
  pages   = {608-622},
  year    = {2026}
}

@article{Soeda2019,
  author  = {Kawada, Y. and Soeda, J. and Satozono, H. and Ikeda, Y. and  Takahashi, H.},
  title   = {High-transparency polymer-silicon nano-particle composites for broadband anti-reflection of terahertz waves},
  journal = {Applied Physics Express},
  volume  = {12},
  number  = {9},
  pages   = {092004},
  year    = {2019}, 
  doi = {10.7567/1882-0786/ab3470} 
}

@article{Chen2009,
  author  = {Chen, Y. W. and Han, P. Y. and Zhang, X.-C.},
  title   = {Tunable broadband antireflection structures for silicon at terahertz frequency},
  journal = {Applied Physics Letters},
  volume  = {94},
  number  = {4},
  pages   = {041106},
  year    = {2009}, 
  doi = {10.1063/1.3075059} 
}

@article{Shevchick2023,
  author  = {{Shevchik-Shekera}, A. V. and Sizov, F. F. and Golenkov, O. G. and Lysiuk, I. O. and Petriakov, V. O. and  Kovbasa, M. Y.},
  title   = {Silicon lenses with {HDPE} anti-reflection coatings for low {THz} frequency range},
  journal = {Semiconductor Physics, Quantum Electronics \& Optoelectronics},
  volume  = {26},
  number  = {1},
  pages   = {59-67},
  year    = {2023}, 
  doi = {10.15407/spqeo26.01.059} 
}

@article{Yao2018,
  author  = {Yao, H.-Y and Chen, Z.-Y and Chang, T.-H.},
  title   = {A design of broadband and low-loss multilayer antireflection coating in {THz} region},
  journal = {Progress In Electromagnetics Research C},
  volume  = {88},
  number  = {},
  pages   = {117-131},
  year    = {2018}, 
  doi = {} 
}

@article{Song2025,
  author  = {Song, J. and Kumar, P. and Raouf, I. and Kim, H. S.},
  title   = {Advancements and challenges in anti-reflective coatings: A comprehensive review},
  journal = {Journal of Materials Research and Technology},
  volume  = {39},
  number  = {},
  pages   = {2926-2938},
  year    = {2025}, 
  doi = {10.1016/j.jmrt.2025.09.268} 
}

@article{Gatesman2000,
  author  = {Gatesman, A. J. and Waldman, J. and Ji, M. Musante, C. and Yngvesson, S.},
  title   = {An anti-reflection coating for silicon optics at terahertz frequencies},
  journal = {IEEE Microwave and Guided Wave Letters},
  volume  = {10},
  number  = {7},
  pages   = {264-266},
  year    = {2000}, 
  doi = {10.1109/75.856983} 
}

@article{Hunsche1998,
  author  = {S. Hunsche and M. Koch and I. Brener and M. C. Nuss},
  title   = {{THz} near-field imaging},
  journal = {Optics Communications},
  volume  = {150},
  number  = {1--6},
  pages   = {22--26},
  month   = may,
  year    = {1998}
}

@article{Tao2020,
  author  = {Tao, Y. H. and Fitzgerald, A. J. and Wallace, V. P.},
  title   = {Non-contact, non-destructive testing in various industrial sectors with terahertz technology},
  journal = {Sensors},
  volume  = {20},
  number  = {3},
  pages   = {712},
  year    = {2020},
  doi     = {10.3390/s20030712}
}

@article{Koch2023,
  author  = {Koch, M. and Mittleman, D. M. and Ornik, J. and Castro-Camus, E.},
  title   = {Terahertz time-domain spectroscopy},
  journal = {Nature Reviews Methods Primer},
  volume  = {3},
  number  = {},
  pages   = {48},
  year    = {2023},
  doi     = {10.1038/s43586-023-00232-z}
}

@article{Zouaghi2013,
  author  = {Zouaghi, W. and Thomson, M. D. and Rabia, K. and Hahn, R. and Blank, V. and Roskos, H. G.},
  title   = {Broadband terahertz spectroscopy: {P}rinciples, fundamental research and potential for industrial applications},
  journal = {European Journal of Physics},
  volume  = {34},
  number  = {6},
  pages   = {S179-S199},
  year    = {2013},
  doi     = {10.1088/0143-0807/34/6/S179}
}

@article{Thomson2007,
  author  = {Thomson, M. D. and Kress, M. and Löffler, T. and Roskos, H. G.},
  title   = {Broadband {THz} emission from gas plasmas induced by femtosecond optical pulses: {From} fundamentals to applications},
  journal = {Laser \& Photonics Reviews},
  volume  = {1},
  number  = {4},
  pages   = {349-368},
  year    = {2007},
  doi     = {10.1002/lpor.200710025}
}

@article{Liu2025,
  author  = {Liu, Z. H. and Li, R. and Wu, Y. H. and Ye, D. D. and He, C. F.},
  title   = {Progress in terahertz nondestructive testing of thermal barrier coatings: a review},
  journal = {Nondestructive Testing and Evaluation},
  volume  = {40},
  number  = {10},
  pages   = {4509-4546},
  year    = {2025},
  doi     = {10.1080/10589759.2025.2452366}
}

@article{Shrekenhamer2013,
  author  = {D. Shrekenhamer and C. M. Watts and W. J. Padilla},
  title   = {Terahertz single pixel imaging with an optically controlled dynamic spatial light modulator},
  journal = {Optics Express},
  volume  = {21},
  number  = {10},
  pages   = {12507--12518},
  month   = may,
  year    = {2013}, 
  doi = {10.1364/OE.21.012507}
}

@article{Yeom2011,
  author  = {Yeom, S. and Lee, D.-S. and Lee, H. and Son, J.-Y. and Guschin, V. P.},
  title   = {Distance estimation of concealed objects with steroscopic passive millimeter-wave imaging},
  journal = {Progress in Electromagnetics Research},
  volume  = {115},
  number  = {},
  pages   = {399-407},
  year    = {2011},
  doi = {10.2528/PIER11030307}
}

@article{Luethi2005,
  author  = {Lüthi, T and Mätzler, C.},
  title   = {Stereoscopic Passive Millimeter-Wave Imaging and Ranging},
  journal={IEEE Transactions on Microwave Theory and Techniques},   
  volume  = {53},
  number  = {8},
  pages   = {2594-2599},
  year    = {2005},
  doi = {10.1109/TMTT.2005.852757}
}

@article{Chan2008,
  author  = {W. L. Chan and K. Charan and D. Takhar and K. F. Kelly and R. G. Baraniuk and D. M. Mittleman},
  title   = {A single-pixel terahertz imaging system based on compressed sensing},
  journal = {Applied Physics Letters},
  volume  = {93},
  number  = {12},
  pages   = {121105},
  month   = sep,
  year    = {2008}
}

@article{Rong2015,
  author  = {L. Rong and T. Latychevskaia and C. Chen and D. Wang and Z. Yu and X. Zhou and Z. Li and H. Huang and Y. Wang and Z. Zhou},
  title   = {Terahertz in-line digital holography of human hepatocellular carcinoma tissue},
  journal = {Scientific Reports},
  volume  = {5},
  pages   = {8445},
  month   = feb,
  year    = {2015}
}

@article{Stantchev2016,
  author  = {R. I. Stantchev and B. Sun and S. M. Hornett and P. A. Hobson and G. M. Gibson and M. J. Padgett and E. Hendry},
  title   = {Noninvasive, near-field terahertz imaging of hidden objects using a single-pixel detector},
  journal = {Science Advances},
  volume  = {2},
  number  = {6},
  pages   = {e1600190},
  doi = {10.1126/sciadv.1600190},
  year    = {2016}
}

@article{Stantchev2017,
  author  = {R. I. Stantchev and D. B. Phillips and P. Hobson and S. M. Hornett and M. J. Padgett and E. Hendry},
  title   = {Compressed sensing with near-field {THz} radiation},
  journal = {Optica},
  volume  = {4},
  number  = {8},
  pages   = {989-992},
  year    = {2017},
  doi = {10.1364/OPTICA.4.000989}
}

@article{Grant2013,
  author  = {Grant, J. and Escorcia-Carranza, I. and Li, C. and McCrindle, I. J. H. and Gough, J. and Cumming, D. R. S.},
  title   = {A monolithic resonant terahertz sensor element comprising a metamaterial absorber and micro-bolometer},
  journal = {Laser \& Photonics Reviews},
  volume  = {7},
  number  = {6},
  pages   = {1043-1048},
  year    = {2013}
}

@article{Valzania2018,
  author  = {Lorenzo Valzania and Thomas Feurer and Peter Zolliker and Erwin Hack},
  title   = {Terahertz ptychography},
  journal = {Optics Letters},
  volume  = {43},
  number  = {},
  pages   = {543--546},
  month   = {},
  year    = {2018}
}

@article{Yuan2019,
  author  = {H. Yuan and D. Vo{\ss} and A. Lisauskas and D. Mundy and H. G. Roskos},
  title   = {{3D} {F}ourier imaging based on {2D} heterodyne detection at {THz} frequencies},
  journal = {APL Photonics},
  volume  = {4},
  number  = {10},
  pages   = {106108},
  month   = oct,
  year    = {2019}
}

@article{Rakic2013,
  author = {A. D. Raki\'{c} and T. Taimre and K. Bertling and Y. L. Lim and P. Dean and D. Indjin and Z. Ikoni\'{c} and P. Harrison and A. Valavanis and S. P. Khanna and M. Lachab and S. J. Wilson and E. H. Linfield and A. G. Davies},
  title = {Swept-frequency feedback interferometry using terahertz frequency {QCLs}: a method for imaging and materials analysis},
  journal = {Optics Express},
  volume = {21},
  number = {22},
  pages = {22194--22205},
  year = {2013},
  month = sep,
  doi = {10.1364/OE.21.022194}
}

@article{Cocker2013,
  author = {T. Cocker and V. Jelic and M. Gupta and others},
  title = {An ultrafast terahertz scanning tunnelling microscope},
  journal = {Nature Photonics},
  volume = {7},
  number = {8},
  pages = {620--625},
  year = {2013},
  month = aug,
  doi = {10.1038/nphoton.2013.151}
}

@book{goodman2005introduction,
  title={Introduction to Fourier optics},
  author={Goodman, Joseph W},
  year={2005},
  publisher={Roberts and Company publishers}
}

@article{Zha10,
author = {Z. B. Zhang and X. Ma and J. G. Zhong},
journal = {Nature Communications},
number = {},
pages = {6225},
publisher = {},
title = {Single-pixel imaging by means of {Fourier} spectrum acquisition},
volume = {6},
month = {},
year = {2015},
url = {},
doi = {}
}

@article{Li2017,
  author = {Li, C. Z. and Peng, Z. Y. and Huang, T. Y. and Fan, T. L. and Wang, F. K. and Horng, T. S. and {Muñoz-Ferreras}, J. M. and {Gómez-García}, R. and Ran, L. X. and Lin, J. S.},
  title = {A Review on Recent Progress of Portable Short-Range Noncontact Microwave Radar Systems},
  journal = {IEEE Transactions on Microwave Theory and Techniques},
  volume = {65},
  number = {5},
  pages = {1692--1706},
  year = {2017},
  doi = {10.1109/TMTT.2017.2650911}
}

@inproceedings{Dobroiu2022,
  author = {A. Dobroiu and J. Ito and S. Suzuki and M. Asada and H. Ito},
  title = {Real-time subcarrier {FMCW} radar in the terahertz range based on a resonant-tunneling-diode oscillator: evaluation and demonstration},
  booktitle = {47th International Conference on Infrared, Millimeter and Terahertz Waves (IRMMW-{THz})},
  address = {Delft, Netherlands},
  year = {2022},
  doi = {10.1109/IRMMW-THz50927.2022.9895697}
}

@inproceedings{Taba2022,
  author = {Taba, M. T. and Naghavi, S. M. H. and Afshari, E.},
  title = {A review on the state-of-the-art {THz} {FMCW} radars implemented on silicon},
  booktitle = {ESSCIRC Conference},
  year = {2022},
  doi = {10.1109/ESSCIRC55480.2022.9911504}
}

@article{Pan2020,
  author = {Pan, M. M. and Chopard, A. and Fauquet, F. and Mounaix, P. and Guillet, J. P.},
  title = {Guided Reflectometry Imaging Unit Using Millimeter Wave {FMCW} Radars},
  journal = {IEEE Transactions on Terahertz Science and Technology},
  volume = {10},
  number = {6},
  pages = {647--655},
  year = {2020},
  doi = {10.1109/TTHZ.2020.3008330}
}

@article{Hu2022,
  author = {W. Hu and others},
  title = {High range resolution wideband terahertz {FMCW} radar},
  journal = {Applied Optics},
  volume = {61},
  number = {24},
  year = {2022},
  doi = {10.1364/AO.465647}
}

@article{Berkel2024,
  author = {{van Berkel}, S. and Khanal, S. and Rahiminejad, S. and {Jung-Kubiak}, C. and Maestrini, A. E. and Chattopadhyay, G.},
  title = {{MEMS} Phase Shifters for {THz} Beam-Scanning: Demonstration with a 500--600 {GHz} Phased Array with Leaky-Wave Feeds},
  journal = {IEEE Transactions on Terahertz Science and Technology},
  volume = {14},
  number = {6},
  pages = {830--842},
  year = {2024},
  doi = {10.1109/TTHZ.2024.3471898}
}

@article{Reyes2025,
  author = {{de los Reyes}, A. and Muldera, J. and Takida, Y. and Kajikawa, K. and Yadav, D. and Xu, Y. H. and Sato, M. and Otani, C. and Minamide, H.},
  title = {Swept-source optical coherence tomography based on a backward terahertz-wave parametric oscillator achieving depth super-resolution},
  journal = {Optics Express},
  volume = {33},
  pages = {40739--40754},
  year = {2025},
  doi = {10.1364/OE.565620}
}

@article{Loncarevic2025,
  author = {{Lončarević}, D. and  Ferrer, M. X. and Zhang, H. S. and Neto, A. and Llombart, N.},
  title = {A study of integrated low-permittivity lenses with strong shadow region illumination},
  journal = {IEEE Transactions on Antennas and Propagation},
  volume = {74},
  number = {1},
  pages = {86-98},
  year = {2025},
  doi = {10.1109/TAP.2025.3618091}
}

@article{Gezimati2023,
  author = {Gezimati, M. and Singh, G.},
  title = {Curved Synthetic Aperture Radar for Near-Field Terahertz Imaging},
  journal = {IEEE Photonics Journal},
  volume = {15},
  number = {3},
  pages = {1--13},
  year = {2023},
  note = {Art no. 5900113},
  doi = {10.1109/JPHOT.2023.3264747}
}

@article{Valles2020,
  author = {Adam Vallés and Jiahuan He and Seigo Ohno and Takashige Omatsu and Katsuhiko Miyamoto},
  title = {Broadband high-resolution terahertz single-pixel imaging},
  journal = {Optics Express},
  volume = {28},
  number = {20},
  pages = {28868--28881},
  year = {2020},
  month = {Sep},
  doi = {10.1364/OE.404143}
}

@article{Chan2009,
  author = {Wai Lam Chan and Hou-Tong Chen and Antoinette J. Taylor and Igal Brener and Michael J. Cich and Daniel M. Mittleman},
  title = {A spatial light modulator for terahertz beams},
  journal = {Applied Physics Letters},
  volume = {94},
  number = {21},
  pages = {213511},
  year = {2009},
  month = {May},
  doi = {10.1063/1.3147221}
}

@article{Bandyopadhyay2024,
  author = {Aparajita Bandyopadhyay and Jayshri Sabarinathan and Withawat Withayachumnankul},
  title = {Industrial sensing in the {IR}, {THz} and {GHz} range with advanced {AI}/{ML}: introduction},
  journal = {Applied Optics},
  volume = {64},
  number = {30},
  pages = {IS1},
  year = {2024},
  doi = {10.1364/AO.580621}
}

@article{Tai2023,
  author = {{Van Mai}, T. and Asada, M. and Namba, T. and Suzuki, Y. and },
  title = {Coherent Power Combination in a Resonant-Tunneling-Diode Arrayed Oscillator With Simplified Structure},
  journal = {IEEE Transactions on Terahertz Science and Technology},
  volume = {13},
  number = {4},
  pages = {405-414},
  year = {2023},
  doi = {10.1109/TTHZ.2023.3270672}
}

@article{Preu2025,
  author = {Preu, S.},
  title = {Continuous-wave terahertz photonics},
  journal = {APL Photonics},
  volume = {10},
  number = {11},
  pages = {110901},
  year = {2025},
  doi = {10.1063/5.0292242}
}

@article{Greenberg2026,
  author={Greenberg, James and Heffernan, Brendan M. and McGrew, William F. and Rolland, Antoine},
  title = {Terahertz Amplification by Injection Locking of Waveguide Resonant Tunneling Diode},
  journal = {IEEE Journal of Quantum Electronics},
  volume = {62},
  number = {1},
  pages = {8500107},
  year = {2026},
  doi = {10.1109/JQE.2026.3652530}
}

@article{Liu2024,
  author = {Liu, M. and Cai, Z. T. and Wang, Z. and Zhou, S. H. and Law, M. K. and Liu, J. and Ma, J. G. and Wu, N. J. and Liu, L. Y.},
  title = {A 3 {THz} {CMOS} Image Sensor},
  journal = {IEEE Journal of Solid-State Circuits},
  volume = {59},
  number = {9},
  pages = {2934-2947},
  year = {2024},
  doi = {10.1109/JSSC.2024.3381595}
}

@article{Koyama2022,
  author = {Koyama, Y. and Kitazawa, Y. and Yukimasa, K. and Uchida, T. and Yoshioka, T. and Fujimoto, K. and Sato, T. and Iba, J. and Sakurai, K. and Ichikawa, T.},
  title = {A High-Power Terahertz Source Over 10~{mW} at 0.45~{THz} Using an Active Antenna Array With Integrated Patch Antennas and Resonant-Tunneling Diodes},
  journal = {IEEE Transactions on Terahertz Science and Technology},
  volume = {12},
  number = {5},
  pages = {510-519},
  year = {2022},
  doi = {10.1109/TTHZ.2022.3180492}
}

@article{Oda2010,
  author = {Oda, N.},
  title = {Uncooled bolometer-type Terahertz focal plane array and camera for real-time imaging},
  journal = {Comptes Rendus Physique},
  volume = {11},
  number = {7-8},
  pages = {496-509},
  year = {2010},
  doi = {10.1016/j.crhy.2010.05.001}
}

@article{Nguyen2012,
  author = {Nguyen, D. T. and Simoens, F. and Ouvrier-Buffet, J. L. and Meilhan, J. and Coutaz, J. L.},
  title = {Broadband {THz} Uncooled Antenna-Coupled Microbolometer Array --Electromagnetic Design, Simulations and Measurements},
  journal = {IEEE Transactions on Terahertz Science and Technology},
  volume = {2},
  number = {3},
  pages = {299-305},
  year = {2012},
  doi = {10.1109/TTHZ.2012.2188395}
}

@article{Meng2024,
  author = {Meng, F. Q. and Tang, Z. L. and Ourednik, P. and Hazarika, J. and Feiginov, M. and Suzuki, S. and Roskos, H. G.},
  title = {High-power in-phase and anti-phase mode emission from linear arrays of resonant-tunneling-diode oscillators in the 0.4-to-0.8-{THz} frequency range
},
  journal = {APL Photonics},
  volume = {9},
  number = {8},
  pages = {086103},
  year = {2024},
  doi = {10.1063/5.0213695}
}

@article{Yuan2023b,
  author = {Hui Yuan and Alvydas Lisauskas and Mark D. Thomson and Hartmut G. Roskos},
  title = {600-{GHz} {Fourier} imaging based on heterodyne detection at the 2nd sub-harmonic},
  journal = {Optics Express},
  volume = {31},
  number = {24},
  pages = {40856-40870},
  year = {2023},
  doi = {10.1364/OE.487888}
}

@article{Yuan2025,
  author = {Hui Yuan and Aparajita Bandyopadhyay and Nan Zhang and M. J. Xiang and Hartmut G. Roskos},
  title = {Status of three-dimensional terahertz {Fourier} imaging for industrial application},
  journal = {Applied Optics},
  volume = {64},
  number = {30},
  pages = {F1-F12},
  year = {2025},
  doi = {10.1364/AO.569481}
}

@article{LiX2023,
  author = {X. R. Li and J. X. Li and Y. H. Li and A. Ozcan and M. Jarrahi},
  title = {High-throughput terahertz imaging: progress and challenges},
  journal = {Light: Science \& Applications},
  volume = {12},
  pages = {233},
  year = {2023},
  doi = {10.1038/s41377-023-01278-0}
}

@article{Pfeifer1998,
  author = {Pfeifer, T. and L\"{o}ffler, T. and Roskos, H. G. and Kurz, H. and Singer, M. and Biebl, E. M.},
  title = {Electro-optic near-field mapping of planar resonators},
  journal = {IEEE Transactions on Antennas and Propagation},
  volume = {46},
  pages = {284--291},
  doi = {10.1109/8.660974},
  year = {1998}
}

@article{Pfeifer1996,
  author = {Pfeifer, T. and Heiliger, H. M. and and L\"{o}ffler, T. and Ohlhoff, C. and Meyer, C. and Lüpke, G. and Roskos, H. G. and Kurz, H.},
  title = {Optoelectro-optic on-chip characterization of ultrafast electric devices: Measurement techniques and applications},
  journal = {IEEE Journal of Selected Topics in Quantum Electronics},
  volume = {2},
  pages = {586-604},
  doi = {10.1109/2944.571758},
  year = {1996}
}
\vfill
\onecolumn
 
\begin{center}
\begin{longtable}{m{2.7cm} m{9cm}}
 
\caption{Appendix: List of Abbreviations} \\[6pt]
\hline
\hline
\endfirsthead
 
\caption[]{Appendix: List of Abbreviations (continued)} \\[6pt]
\hline
\hline
\endhead
 
\hline
\multicolumn{2}{r}{\textit{Continued on next page}} \\
\endfoot
 
\hline
\hline
\endlastfoot
 
1D, 2D, 3D  & One-, two- or three-dimensional \\
AAE         & Adversarial autoencoder \\
ADM         & All-dielectric metamaterial \\
ADMM        & Alternating direction methods of multipliers \\
AEM         & Artificial electromagnetic material \\
AI          & Artificial intelligence \\
AIA         & Artificial intelligent agent \\
API         & Application programming interface \\
AR          & Anti-reflection \\
BiCMOS      & Bipolar CMOS \\
BWO         & Backward wave oscillator \\
CDI         & Coherent diffraction imaging \\
CMOS        & Complementary metal-oxide-semiconductor \\
CNN         & Convolutional neural network \\
CS          & Compressive sensing \\
DIM         & Deep inverse model \\
DL          & Deep learning \\
DNN         & Deep neural network \\
FET         & Field-effect transistor \\
FM          & Foundational model \\
FMCW        & Frequency-modulated continuous wave \\
FPA         & Focal plane array \\
FPI         & Focal plane imaging \\
FPM         & Fourier ptychographic microscopy \\
FoV         & Field of view \\
GAN         & Generative adversarial network \\
GFRP        & Glass-fibre-reinforced plastic \\
HBT         & Hetero-bipolar transistor \\
HEB         & Hot-electron bolometer \\
HIL         & Human in the loop \\
I\&Q        & In-phase and quadrature \\
IF          & Intermediate frequency \\
INN         & Invertible neural network \\
IPR         & Iterative phase retrieval \\
IR          & Infrared \\
ITU         & International Telecommunication Union \\
LLM         & Large language model \\
LO          & Local oscillator \\
LT-GaAs     & Low-temperature-grown GaAs \\
MCP         & Model context protocol \\
ML          & Machine learning \\
MLP         & Multi-layer perceptron \\
MOS         & Metal-oxide-semiconductor \\
MPA         & Multi-pixel array \\
MSE         & Mean squared error \\
NA          & Neural adjoint \\
NDT         & Non-destructive testing \\
NEP         & Noise-equivalent power \\
NHIL        & No human in the loop \\
NIR         & Near infrared \\
NP          & Non-deterministic polynomial \\
PCA         & Photoconductive antenna \\
PIDL        & Physics-informed deep learning \\
QCL         & Quantum cascade laser \\
RCS         & Radar cross-section \\
RTD         & Resonant tunneling diode \\
s-SNOM      & Scattering-type scanning near-field optical microscope \\
SAR         & Synthetic aperture radar \\
SCI         & Snapshot compressive imaging \\
SLM         & Spatial light modulator \\
SmLM        & Small language model \\
SNR         & Signal-to-noise ratio \\
TDS         & Time-domain spectroscopy \\
TeraFET     & Antenna-coupled FET as THz detector \\
TIE         & Transport of intensity equation \\
TPI         & Time-domain pulsed imaging \\
TV          & Total variation \\
VAE         & Variational autoencoder \\
VIS         & Visible spectral range \\
WRC-19      & World-Radiocommunication Conference 2019 \\
 
\end{longtable}
\end{center}
\twocolumn

\end{document}